\documentclass[fleqn,usenatbib]{mnras}

\usepackage{newtxtext,newtxmath}

\usepackage{relsize}
\usepackage[mathscr]{euscript}
\usepackage{nccmath}
\usepackage{hyperref}
\usepackage{mathtools}
\usepackage{caption}
\usepackage{multirow}
\usepackage[utf8]{inputenc}
\DeclareUnicodeCharacter{2212}{-}
\usepackage{multicol}        % Multi-column entries in tables
\usepackage[T1]{fontenc}
\usepackage{tablefootnote}
\usepackage{booktabs}
\usepackage{gensymb}
\usepackage[export]{adjustbox}
\usepackage{chngcntr}
\DeclareRobustCommand{\VAN}[3]{#2}
\let\VANthebibliography\thebibliography
\def\thebibliography{\DeclareRobustCommand{\VAN}[3]{##3}\VANthebibliography}

\usepackage{graphicx}	% Including figure files
\usepackage{amsmath}	% Advanced maths commands

\title[ELAIS-N1 Band-2 and broken power law]{ELAIS-N1 deep field uGMRT Band-2: constraints on diffuse Galactic synchrotron emission power spectrum}

\author[Sagar et al.]{
R. Sagar,$^{1}$\thanks{E-mail: rashmisagar777@gmail.com,
phd2101121003@iiti.ac.in} %abhirup.datta@iiti.ac.in (AD)}
A. Datta,$^{1}$ A. Chakraborty,$^{2}$ Nirupam Roy,$^{{4},{5}}$ A. Sinha,$^{3}$ A. Mazumder,$^{6}$ Prasun Dutta,$^{7}$ 
\newauthor
Khandakar Md Asif Elahi,$^{8}$ Kanan K. Datta,$^{9}$ Samir Choudhuri,$^{8}$ Somnath Bharadwaj,$^{10}$ Srijita Pal,$^{4}$ 
\newauthor
Anshuman Tripathi,$^{1}$ Suman Majumdar,$^{1}$ Tirthankar Roy Choudhury,$^{11}$ Sk. Saiyad Ali,$^{12}$ 
\\
$^{1}$Department of Astronomy, Astrophysics and Space Engineering, Indian Institute of Technology Indore, Indore 452020, India\\
$^{2}$Department of Physics and McGill Space Institute, McGill University, Montreal, QC H3A 2T8, Canada\\
$^{3}$Thüringer Landessternwarte, Sternwarte 5, D-07778 Tautenburg, Germany \\
$^{4}$Department of Physics, Indian Institute of Science, Bangalore 560012, India \\
$^{5}$Department of Physics, New Mexico Institute of Mining and Technology, Socorro, NM 87801, USA \\
$^{6}$Jodrell Bank Centre for Astrophysics, Department of Physics and Astronomy, The University of Manchester, Manchester M13 9PL, UK \\
$^{7}$Department of Physics, IIT (BHU) Varanasi, 221005 India \\
$^{8}$Department of Physics, Indian Institute of Technology Madras, Chennai 600036, India \\
$^{9}$Relativity and Cosmology Research Centre, Department of Physics, Jadavpur University, Kolkata 700032, India \\
$^{10}$Department of Physics, Indian Institute of Technology Kharagpur, Kharagpur 721302, India \\
$^{11}$National Centre For Radio Astrophysics, Post bag 3, Ganeshkhind, Pune 411007, India \\
$^{12}$Department of Physics, Jadavpur University, Kolkata 700032, India 
}

\date{Accepted XXX. Received YYY; in original form ZZZ}

\pubyear{\the\year{}}

\begin{document}
\label{firstpage}
\pagerange{\pageref{firstpage}--\pageref{lastpage}}
\maketitle

% Abstract of the paper
\begin{abstract}
We present high sensitivity, low radio frequency continuum observations of the ELAIS-N1 field with 32 hours of observations of the uGMRT Band-2 ($120-250$\,MHz) covering $5.86\,\text{deg}^2$ area, achieving a central off-source RMS noise of $237\,\mu\mathrm{Jy}/\mathrm{beam}$ with a resolution of $11.45''$ at the central frequency of 183 MHz. A radio source catalogue of 1027 sources statistically matches with similar observations at different frequencies within the sensitivity range of the uGMRT. The calibrated data is further used to characterise the dominant foreground, the diffuse Galactic synchrotron emission (DGSE), in angular scale and frequency regime. We derived the angular power spectrum (APS) of DGSE in two ways: image-based estimator (i-APS) and visibility-based Tapered Gridded Estimator (TGE; hereafter as t-APS). We assess the characteristics of DGSE with a power-law form of $C_{\ell} = A({1000}/{\ell})^{\beta}$. Combining data from Band-2 and earlier Band-3 observations, we derived a spectral variation of $C_{\ell}$ in the form of $C_{\ell} = A{\nu^{-2{\alpha}}}{\ell^{-{\beta}}}$. Our result indicates a spectral break at $\nu = 230\,{\pm}\,5$\,MHz, corresponding to a synchrotron age of $t_\text{syn} = 106\,{\pm}\,1$\,Myr for the cosmic-ray electrons (CRe). This break result suggests a low-energy cutoff in the CRe population, leading to spectral curvature at low frequencies. Using both of the techniques, i-APS and t-APS, we find that the mean spectral index $\alpha$ and power-law index $\beta$ are consistent within the frequency range $120-500$\,MHz.
\end{abstract}

% Select between one and six entries from the list of approved keywords.
% Don't make up new ones.
\begin{keywords}
instrumentation: interferometers -- methods: data analysis -- techniques: interferometric, image processing -- catalogues -- (cosmology:) diffuse radiation
\end{keywords}

%%%%%%%%%%%%%%%%%%%%%%%%%%%%%%%%%%%%%%%%%%%%%%%%%%

%%%%%%%%%%%%%%%%% BODY OF PAPER %%%%%%%%%%%%%%%%%%

\section{Introduction}
\label{sec:intro}

Deep and wide-field area surveys of the radio sky, when combined with high-sensitivity data over a wide range of wavelengths, offer promising physics to explore the dynamic nature of the universe. In the radio window of the spectrum, dust absorption has no impact on radio emission. Therefore, it provides an unbiased and unobscured view of galaxies undergoing active star formation with current radio telescopes. The following radio interferometers are the Very Large Array \citep[VLA,][]{2011ApJ...739L...1P}, upgraded Giant Meterwave Radio Telescope \citep[uGMRT,][]{2017CSci..113..707G}, upgraded Australia Telescope Compact Array \citep[ATCA,][]{2011MNRAS.416..832W}, the Murchison Wide-field Array \citep[MWA,][]{2013PASA...30....7T, 2013PASA...30...31B}, the Atacama Large Millimeter/Submillimeter Array \citep[ALMA,][]{2009IEEEP..97.1463W}, the Low Frequency Array \citep[LOFAR,][]{2013A&A...556A...2V}, Australian Square Kilometre Array Pathfinder\citep[ASKAP,][]{2008ExA....22..151J, 2020MNRAS.493.1662M}, SKA precursor telescope MeerKAT \citep{2016mks..confE...1J, 2020ApJ...888...61M, 2024MNRAS.528.2511T}, and anticipated future telescopes like the Square Kilometre Array \citep[SKA1-low,][]{2015aska.confE..10M, 2015aska.confE...1K}, \citep[SKA1-mid,][]{8105425}. \\
At lower radio frequencies, the predominant emission in galaxies arises from non-thermal synchrotron radiation \citep{2004MNRAS.355.1053D}, which emerges from processes such as supernovae \citep{1992ARA&A..30..575C}. This emission is generated by cosmic ray electrons (CRe) that are accelerated to high energies by supernova explosions and are observed throughout the Galaxy. It is also produced by young star-forming galaxies (SFGs), Galaxy clusters \citep{2014IJMPD..2330007B} and Active Galactic Nuclei (AGN). Thus, radio observations are valuable for investigating the evolving properties of the SFGs \citep{2018A&A...620A.192C, 2018A&A...611A..55K, 2020MNRAS.491.5911O, 2022arXiv220400831T, 2022ApJ...941...10V, 2023MNRAS.523.6082C} and provide information into the evolution of AGNs \citep{2009MNRAS.392..617D, 2020A&A...641A..85S, ocran2021evolution}, especially for the population with relatively low luminosities \citep{ 2021A&A...648A...3K, 2021ApJ...916..102A, 2022MNRAS.516..245W}. 
Hence, detailed surveys are important across all frequencies, probing lower radio frequencies and lower luminosities allow us to understand the properties of faint SFGs and AGNs \citep{2015MNRAS.453.1079B, 2017A&A...602A...3D, 2019A&A...622A..11G, 2019A&A...622A..10C, 2022MNRAS.514.4343S}. \\ 
Among the deep field surveys, the European Large-Area ISO Survey North-1 \citep[ELAIS-N1,][]{2000MNRAS.316..749O} field has been extensively studied at various frequencies due to its strategic location at high ecliptic latitude in the Northern sky and low far-infrared intensity \citep{2021A&A...648A...2S}. LOFAR results of ELAIS-N1 field \citep{2021A&A...648A...2S, 2025A&A...695A..80S} cover a large sky area and have higher sensitivity compared to previous low-frequency surveys. However, many research efforts have employed the GMRT \citep{1991CSci...60...95S} to explore the ELAIS-N1 field across different frequencies at 325/400\,MHz \citep{2009MNRAS.395..269S, 2019MNRAS.487.4102C, 2019MNRAS.490..243C}, 610/612\,MHz \citep{2008MNRAS.383...75G, 2016MNRAS.459L..36T, 2020MNRAS.497.5383I, 2020MNRAS.491.1127O} and 1.2\,GHz \citep{2023JApA...44...88S}. The recent upgrade of the GMRT has enabled access to Band-2 ($120-250$\,MHz) with its broad bandwidth, providing an opportunity to study the Northern sky at lower frequencies more profoundly. This work observed the ELAIS-N1 field at 150\,MHz with the GMRT software Backend \citep[GSB,][]{2010ExA....28...25R} and in the frequency range of $120-250$\,MHz with the GMRT wide-band Backend \citep[GWB,][]{2017JAI.....641011R}. This work marks the initial release of a series of papers focusing on the uGMRT Band-2 observations and presents independent results of uGMRT. This work is significant as low-frequency observations inherently face challenges related to RFI, foregrounds, systematics, and primary beam corrections, making it important to have results of the same sky patch with different telescopes. This survey provides valuable complementary data to the LOFAR survey ($115-177$\,MHz) on ELAIS-N1 within the survey capabilities of uGMRT observations. \\
However, deep field studies are not limited to the source population. There is an important domain of study i.e., foregrounds \citep{2008MNRAS.385.2166A, 2008MNRAS.389.1319J}. Foregrounds in radio observations encompass various emissions, including diffuse synchrotron emission \citep[DSE;][]{1999A&A...345..380S}, free-free emissions (galactic and extragalactic) \citep{2004ApJ...606L...5C}, faint radio-loud quasars \citep{2002ApJ...564..576D}, extragalactic point-sources and emission from supernova remnants. 
This paper covers the angular power spectrum (APS) approach in Band-2 ($120-250$\,MHz) and Band-3 ($300-500$\,MHz) wideband data and aims to study the Diffuse Galactic Synchrotron Emission (DGSE) characteristics in detail. This frequency range is of particular interest as it covers wide-range from Epoch of reionization (EoR) to post-EoR and synchrotron emission from CRe in the Galaxy dominates the sky in this frequency regime. We study the spectral evolution of the DGSE over a wide frequency range $120-500$\,MHz. We have compared visibility and image based estimation techniques to measure the APS. This is the first time we are presenting these two techniques to characterise the DGSE.\\
This paper provides the first release of the upgraded GMRT Band-2 Stokes I source catalogue, outlining the calibration methodology, catalogue process and detailed study of foreground in terms of APS. The paper is organised as follows: Section \ref{sec:observation} offers an overview of the uGMRT Band-2 observations used in this work. Section \ref{sec:calibration} describes the calibration technique and imaging, while Section \ref{sec:catalogue} presents the final catalogue and source classification. Section \ref{sec:comparison} discusses the comparison of this catalogue with previous available catalogues and surveys, confirming the accuracy of flux values. Section \ref{sec:count} analyses completeness, and source counts are presented, including raw and completeness-corrected counts. In Section~\ref{sec:APS}, we have discussed the power-law of diffuse synchrotron emission. In this section, we analysed the results using visibility- and image-based estimators. Section~\ref{sec:broken-PS} highlights the results of the spectral variation of DGSE power spectrum over the wideband range 120\,-\,500\,MHz with the conclusion in Section \ref{sec:conclusion}. 

\section{Observations} \label{sec:observation}

The GMRT is one of the biggest, large-bandwidth, and highly sensitive international facilities for radio astronomy operative in low frequency in the present day. 
Features of the updated GMRT (uGMRT) are as follows:\\
(i) Large frequency coverage $\sim$\,120\,-\,1450\,MHz; (ii) Availability of maximum usable bandwidth is 100\,MHz in Band-2; (iii) increased sensitivity and receivers with higher G/Tsys \citep{2017CSci..113..707G}. \\
This work used uGMRT observations of 32 hours of the ELAIS-N1 field ($\alpha_{2000}$  $=$ 16$^h$\,10.0$^m$\,1.00$^s$, $\delta_{2000}$ $=$54$^d$\,30.0$^m$\,36.00$^s$). The observations include 8 hours of a single night and 24 hours of three nights in the 34 and 41 observing cycles, respectively. The detailed information is in Table \ref{table:tab1}. Additionally, we used uGMRT data of the ELAIS-N1 deep field for 25 hours in observing cycle 32 in Band-3 (300-500 MHz) \citep{2019MNRAS.490..243C}. The ELAIS-N1 field is located at high galactic latitudes ($\ell = 86.95\degr$, $b = +44.48\degr$) and was at transit during the night in observing cycle 34 and in the morning in observing cycle 41. 
Observations were conducted exclusively during night to reduce the effect of Radio Frequency Interference (RFI) and the dynamic behaviour of the ionosphere. The radio frequency range designated for uGMRT Band-2 spans from $120-250$\,MHz (excluding $\approx$\,165\,-\,190\,MHz using a notch filter\footnote{\url{http://www.ncra.tifr.res.in:8081/~secr-ops/etc/etc_help.pdf}} to block persistent strong RFI) \citep{2022JAI....1150008B}. To reduce RFI impact significantly, we applied real-time broadband RFI filtering with a 4$\sigma$ cut-off and digital noise replacement \citep{doi:10.1142/S2251171719400063, 2022JAI....1150008B}. The usable bandwidth is $\sim$\,100\,MHz and is constrained by other factors within the band that are affected to some extent by RFI. The flux calibrator sources 3C286 and 3C48 were observed for calibration purposes at the beginning and end of each observation run, respectively, based on their availability during the observation period. Considering the temporal fluctuations in the system gain, we observed the phase calibrator (J1549+506) close to the target field every 25 min. The angular separation between the target and the phase calibrator is ${\sim}\,5^{\circ}$. The observation technique remained consistent for both Band-2 and Band-3 observations. In this work, we used Band-3 observation for the characterisation of DGSE in the angular scale and frequency domain only. For comprehensive details of Band-3 observations, see \cite{2019MNRAS.490..243C}.
%%%%%%%%%%%%%%%%%%%%%%%%%%%%%%%%%%%%%%%%%%%%%%%%%%%%%%%%%%%%%%%%%%%%%%%%%%%%%%%%%%%%%%%%%%%%%%
\begin{table}
\caption{Observations details of uGMRT Band-2 ($120-250$\,MHz).}
\centering
\begin{tabular}{ll}
\hline
Equinox & J2000 \\ [0.3ex]
\hline\hline
Project Code & 34\_141 \& 41\_099 \\ [0.3ex]
Observation Date & 2018 Jul 06 \\ 
                 & 2022 Feb 06, 07, 26 \\ [0.3ex]
Total Observation Time & 8 hours \& 24 hours \\ [0.3ex]
Central Frequency & 183 MHz \\ [0.3ex]
Bandwidth & 120 - 250 MHz \\ [0.3ex]
Integration Time & 5.37 sec \\ [0.3ex]
Number of Channels & 8192 \\ [0.3ex]
Frequency Resolution & 24 kHz \\ [0.3ex]
Working Antennas & 28, 27, 26, 27 \\ [0.3ex]
\hline
Flux Calibrators & 3C286 (27.308\,Jy) \\ 
                 & ($13^h 31^m 08^s + 30^\circ 30' 32''$) \\ 
                 & 3C48 (64.762\,Jy) \\ 
                 & ($01^h 37^m 41^s + 33^\circ 09' 35''$) \\ [0.3ex]
Phase Calibrator & J1549+506 \\
                 & ($15^h 49^m 17.47^s + 50^\circ 38' 05.79''$) \\ [0.3ex]
Pointing & ELAIS-N1 \\ 
         & ($16^h 10^m 01^s + 54^\circ 30' 36''$) \\ [0.3ex]
\hline
\end{tabular}
\label{table:tab1}
\end{table}

\section{Data Analysis} \label{sec:calibration}

Considering the importance of ionospheric phase delay in the low-frequency domain, the data reduction process has been done using Source Peeling and Atmospheric Modelling \citep[{\textsc{\tt\string SPAM}};][]{2009A&A...501.1185I, 2014ascl.soft08006I, 2014ASInC..13..469I} which is a semi-automated pipeline, based on the Astronomical Image Processing System \citep[{\textsc{\tt\string AIPS}};][]{Bridle_Greisen_1994, greisen1998recent, 2003ASSL..285..109G}. SPAM is composed of a series of Python scripts that use the \textsc{\tt\string ParselTongue} interface \citep{kettenis2006parseltongue} to access AIPS tasks, files, and tables using \textsc{\tt\string OBIT} \citep{2008PASP..120..439C} library.\\
SPAM is a direction-dependent (DD) calibration pipeline that mitigates ionospheric dispersive delay in calibration, modelling, and imaging. However, low-frequency observations face the challenge of large field-of-view (FoV), which encompasses several sources around the calibrator source. To overcome this, one can select a bright calibrator source (in this case, 3C286 or 3C48) with a flux that dominates over any other visible source. SPAM separates phase corrections to extract only the instrumental component which is later applied to the visibility of the target field to effectively separate instrumental errors from ionospheric effects \citep{2017A&A...598A..78I}. This also reduces the off-source RMS noise and increases the flux-scale precision. The SPAM data calibration follows these steps: \\

\textbf{Data handling $\&$ Pre-process:} For each night, data undergo calibration procedures for both GSB and GWB data separately. The bandwidth of GSB data is 16\,MHz and usable bandwidth of GWB is $\sim$100\,MHz. SPAM has a limitation when it comes to processing large fractional bandwidths (df/f\,$>{\sim}$\,0.2) in a single go. Thus, to process the large bandwidth, SPAM splits the GWB data into sub-bands, each with a 16\,MHz bandwidth covering the entire range from 120\,MHz to 250\,MHz (excluding $\approx$\,$165\,-\,190$\,MHz). Similarly, for Band-3 SPAM processed visibilities of bandwidth 32\,MHz of each sub-bands. SPAM calibrates the GWB data using a reference sky model extracted as a source catalogue from the calibrated GSB image, which was observed simultaneously with the GWB. The main pipeline performs this step.\\ 
In the pre-processing step, the flux-scale model by \cite{2012MNRAS.423L..30S} is used to determine the flux density of the flux calibrators. The data are then subjected to RFI flagging and bad antenna removal, followed by gain and bandpass solutions for each antenna per polarization. For instrumental calibration, SPAM selects the optimal scan of the primary calibrator.
The gain solutions of the dominant flux calibrator (i.e. 3C286) are applied to the target (ELAIS-N1). \\
 
\textbf{Calibration:} The main pipeline involves the process of direction-independent (DI) and DD calibration \citep{2009A&A...501.1185I, 2014ASInC..13..469I}. The final calibrated wideband data are averaged in time and frequency, resulting in the number of channels $=$ 45, channel width $=$ 0.366\,MHz, and an effective bandwidth $=$ 16\,MHz for each subband. The final output of the main pipeline is a map of the ELAIS-N1 region that has been corrected for primary-beam effects and calibrated visibilities for each subband of Band-2. However, there is no information on the primary beam coefficients used to correct the primary beam in the imaging process in SPAM for Band-2. Thus, we used the output visibilities resulting from the main pipeline of SPAM for each subband to combine for final wideband imaging, discussed in Section \ref{sec:imaging}. We have followed the same steps for the calibration of Band-3 data. \\ 
Table \ref{table:tab2} represents the percentage of visibility data flagged for each observation night in the two observing cycles. The flagging percentage of the target field after averaging and flagging in both time and frequency domain highlights the cause of significant variation in the RMS noise of the image each night. The flagging percentage depends not only on the RFI flagging percentage (approximately 35\% each night), but also on the instrumental and ionospheric conditions contributing further to the bad data. The detailed analysis of individual observational nights is beyond the scope of this paper and will be addressed in future work.
%%%%%%%%%%%%%%%%%%%%%%%%%%%%%%%%%%%%%%%%%%%%%%%%%%%%%%%%%%%%%%%%%%%%%%%%%%%%%%%%%%%%%%%%%%%%%%%%%%%%
\begin{figure*}
\centering
\makebox[\textwidth]{\includegraphics[width=0.8\paperwidth]{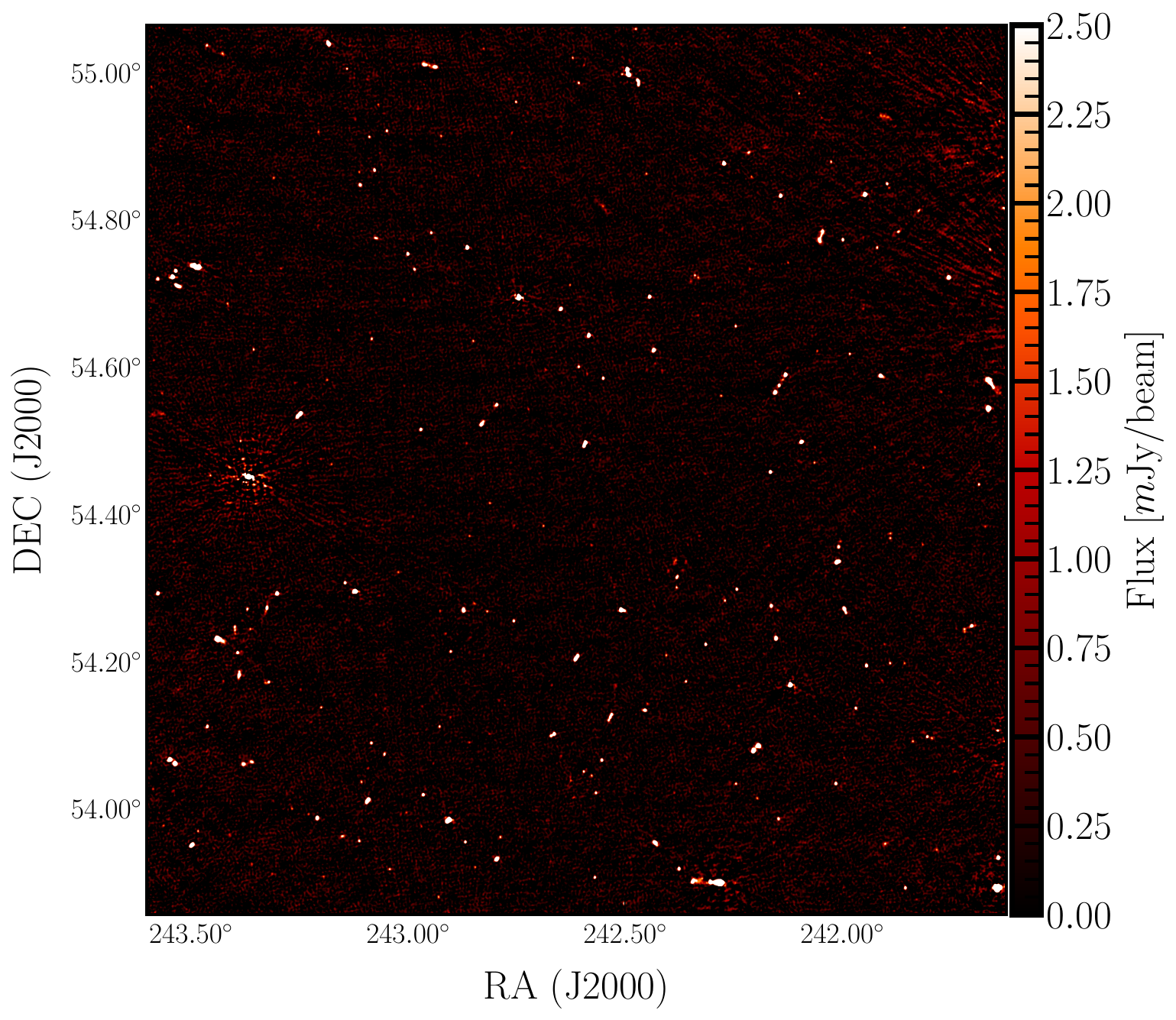}}
\caption{Primary-beam corrected 32 hours final image of ELAIS-N1 field after DD-calibration at 183\,MHz (usable bandwidth $\sim$\,100\,MHz). The final image covers an area of $5.86\,\text{deg}^2$ with the off-source RMS noise near the center is 237\,${\mu}$Jy/beam. Synthesized beam size is $11.45'' {\times} 9.02''$.}
\label{fig:Fig1}
\end{figure*}
%%%%%%%%%%%%%%%%%%%%%%%%%%%%%%%%%%%%%%%%%%%%%%%%%%%%%%%%%%%%%%%%%%%%%%%%%%%%%%%%%%%%%%%%%%

\subsection{Imaging} \label{sec:imaging}
Following the DD-calibration, we used {\tt\string WSClean}\footnote{\url{https://gitlab.com/aroffringa/wsclean}} \citep{2014MNRAS.444..606O} to create a combined continuum image from 32 hours of wideband data. The FoV of uGMRT at 183\,MHz is quite large ($3.5{^\circ}{\times}3.5{^\circ}$). For wideband imaging, w-stacking is used to minimise distortion due to non-coplanar sky curvature and to combine the visibilities at different frequencies. Considering the wide bandwidth and spectral variation of the field sources, we used the multi-scale (MS) deconvolution algorithm \citep{2017MNRAS.471..301O} with fit-spectral-pol $=$ 2. We selected the Briggs robust parameter to ${-1}$ \citep{1995PhDT.......238B}, which provides a favourable balance between sensitivity and resolution. 
The beam model was corrected using {\tt\string uGMRTprimarybeam}\footnote{\url{https://github.com/ruta-k/uGMRTprimarybeam}} with the primary-beam parameters\footnote{\url{http://www.ncra.tifr.res.in/ncra/gmrt/gmrt-users/observing-help/beam_shape_band_2.pdf}} of Band-2 ($120-250$\,MHz) of uGMRT. Here, we have implemented the threshold at 20$\%$ of the peak value to mitigate the effects of the primary-beam response. Figure \ref{fig:Fig1} presents the resulting combined final image of the ELAIS-N1 field at 183\,MHz with primary-beam correction. The off-source RMS noise achieved close to the center of the FoV is 237 $\mu$Jybeam$^{-1}$. The synthesised beam size is $11.45'' \times 9.02''$, and the position angle is $9.96^\circ$.\\
%%%%%%%%%%%%%%%%%%%%%%%%%%%%%%%%%%%%%%%%%%%%%%%%%%%%%%%%%%%%%%%%%%%%%%%%%%%%%%%%%%
% \begin{table}
% \caption{Effective time after flagging over observed target field for each observation night in the observing cycle 34 and 41, respectively.}
% \centering
% \begin{tabular}{ccccc}
% \hline
% Observing & Date & Total on-source & Flagging & \textbf{Usable on-source}  \\ 
% Night & ~ & Time & ~ & Time \\
%  ~& DD MM YYYY & (hours) & (\%) & (hours)  \\ 
% \hline\hline
% 1& 06 Jul 2018 & 2.4 & 32.2 & 2.0 \\ %[1ex]
% 2& 06 Feb 2022 & 5.1 & 34.3 & 3.2 \\ %[1ex]
% 3& 07 Feb 2022 & 6.2 & 36 & 4.5 \\ %[1ex]
% 4& 26 Feb 2022 & 5.1 & 34.3 & 3.1 \\ %[1ex]
% \hline
% \end{tabular}
% \label{table:tab2}
% \end{table}
\begin{table}
\caption{The approximate flagging percentage after averaging and flagging in both time and frequency domain over observed target field for each observation night in the observing cycle 34 and 41, respectively.}
\centering
\begin{tabular}{ccccc}
\hline
Observing & Date & Total on-source & Flagging  \\ 
Night &  & Time & ~ \\
 & YYYY-Mon-DD & (hours) & (\%)  \\ 
\hline\hline
Night 1 & 2018-Jul-06 & 2.31 & 32.6  \\ %[1ex]
Night 2 & 2022-Feb-06 & 5.16 & 34.6  \\ %[1ex]
Night 3 & 2022-Feb-07 & 6.08 & 37.6 \\ %[1ex]
Night 4 & 2022-Feb-26 & 5.59 & 35.1 \\ %[1ex]
\hline
\end{tabular}
\label{table:tab2}
\end{table}
%%%%%%%%%%%%%%%%%%%%%%%%%%%%%%%%%%%%%%%%%%%%%%%%%%%%%%%%%%%%%%%%%%%%%%%%%%%%%%%%%%

\section{The 183 MH\tiny{z}\,\, \normalsize{Source catalogue}} \label{sec:catalogue}

\begin{figure*}
    \centering
    \begin{minipage}{0.48\textwidth}
        \centering
        \includegraphics[width=\linewidth]{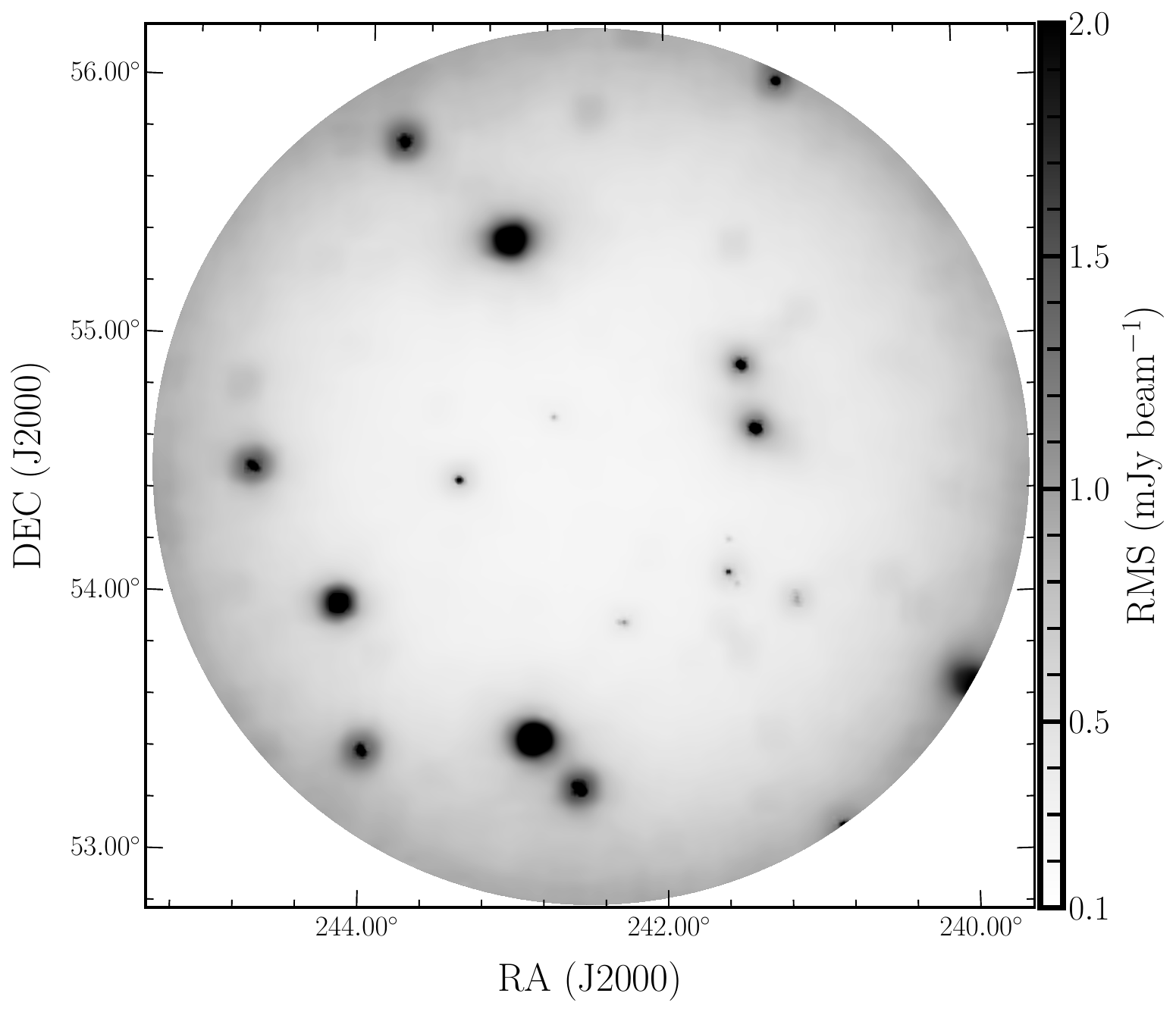}
    \end{minipage}\hfill
    \begin{minipage}{0.45\textwidth}
        \centering
        \includegraphics[width=\linewidth]{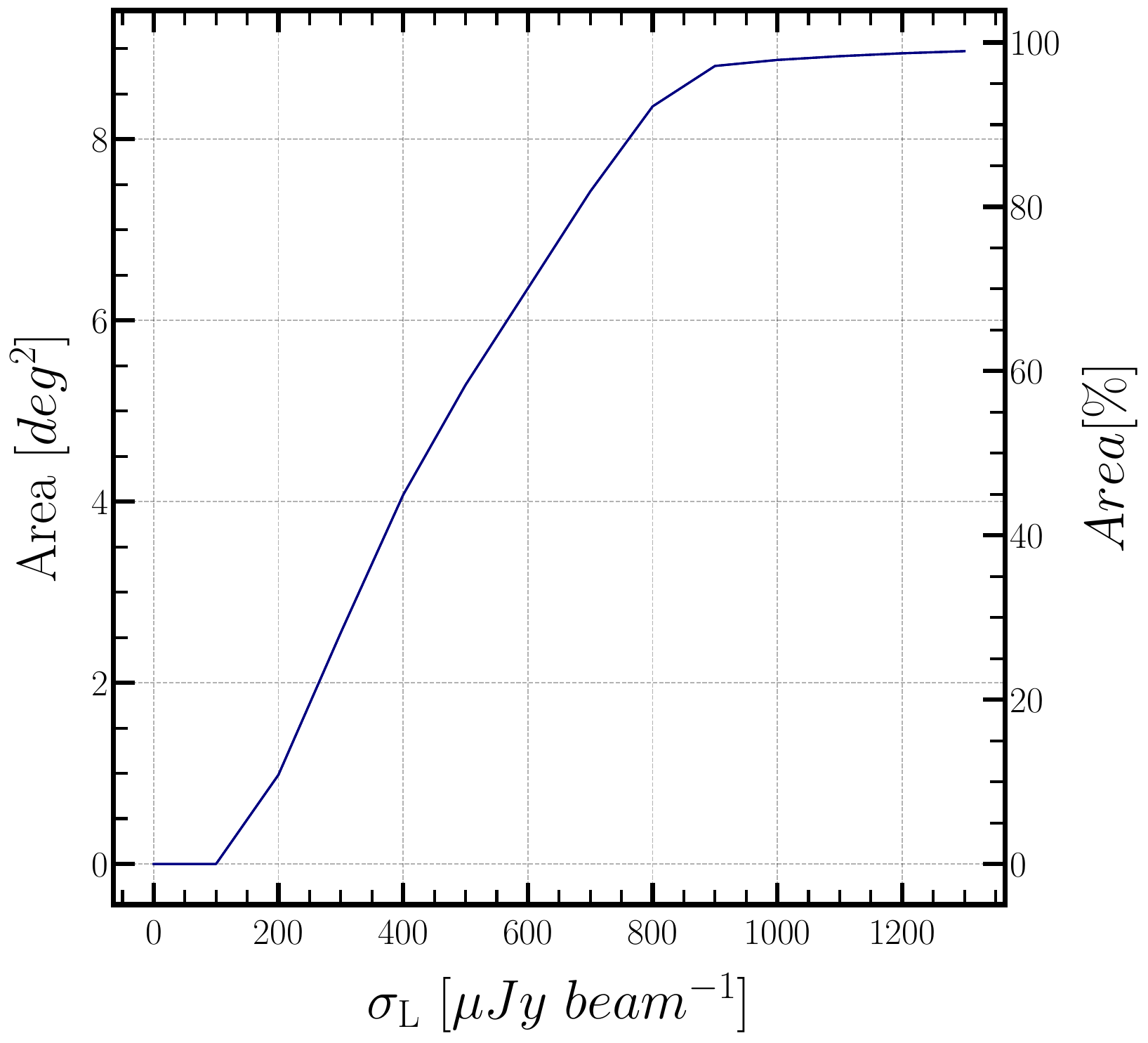}
    \end{minipage}
    \caption{Left Panel: The background RMS noise is high near the bright sources and at the edge of FoV in the 32 hours final image of the ELAIS-N1 field. Right Panel: Cumulative area of primary-beam corrected (final) image as a function of $\sigma_L$, local RMS noise.}
    \label{fig:Fig2}
\end{figure*}

We used Python-based Blob Detection and Source Finder\footnote{\url{https://www.astron.nl/citt/pybdsf/}} \citep[{\tt\string PyBDSF};][]{2015ascl.soft02007M}, source extraction software to generate a source catalogue from the final image corrected for the primary beam to characterise the sources present in the ELAIS-N1 field. The background RMS noise variation and background mean across the entire image were calculated using a 2-D sliding square box with 180 pixels width and a step size of 50 pixels (i.e. $\tt{rms\_box = (180,50)}$ equivalent to $\tt{(450'',125'')}$). To mitigate the influence of strong artefact sources on the SNR ratio, we identified regions with peak amplitudes exceeding an adaptive threshold of 150$\sigma$. To prevent identifying artefacts as real sources, we also applied a smaller box with dimensions of ${\tt rms\_box\_bright = (35,7)}$ near the bright sources. The PyBDSF software identified islands with contiguous source emissions above a pixel threshold, fitting each island as multiple Gaussian models. We set a detection threshold of 3${\sigma}_\text{rms}$ for island identification and a pixel threshold of 5${\sigma}_\text{rms}$ for source detection. To account for the variations of the PSF caused by ionospheric fluctuations during low-frequency observations, we utilised PyBDSF with the ``$\tt{psf\_vary\_do = True}$'' option to calculate and correct the PSF variation across the FoV. \\ 
We produced an RMS map that reveals the background noise variation throughout the field. Figure \ref{fig:Fig2} (left panel) shows that the background RMS is significantly higher close to the bright sources as well as towards the edges of the field. The right panel (Figure \ref{fig:Fig2}) displays the image area and the percentage of the field with a noise level lower than $5{\sigma}$ threshold. The former is due to the image artefacts, whereas the latter is an effect of the reduced amplitude of the primary beam away from the center of the FoV.  
In addition to resolution and thermal noise, the accuracy of any generated source catalogue is also constrained by confusion noise. This noise arises due to the fluctuation in the background sky brightness caused by the presence of multiple faint sources within the telescope beam. It can be estimated as :
%%%%%%%%%%%%%%%%%%%%%%%%%%%%%%%%%%%%%%%%%%%%%%%%%%%%%%%%%%%%%%%%%%%%%%%%%%%%%%%%%%%%%%%%%%%%%%%%%%%%%%
\begin{equation}
\label{eqn:1}
\begin{aligned}
\sigma_{CN} = 1.2\left(\frac{\nu}{3.02\,\text{GHz}}\right)^{-0.7} \left(\frac{\theta}{8\,\text{arcsec}}\right)^{10/3}\,\mu\text{Jy/beam}
\end{aligned}
\end{equation}
%%%%%%%%%%%%%%%%%%%%%%%%%%%%%%%%%%%%%%%%%%%%%%%%%%%%%%%%%%%%%%%%%%%%%%%%%
where $\nu$ is the observing frequency, and $\theta$ is the FWHM of the telescope beam \citep{2012ApJ...758...23C}. This provides a confusion noise $\sim$30\,$\mu$Jy/beam. Thus, the RMS noise in this work (i.e., 237$\mu$Jy/beam) is higher than the confusion noise and does not significantly impact the results. For uGMRT Band-2, we have generated the catalogue comprising 1027 sources of the ELAIS-N1 field covering an area of $5.86\,\text{deg}^2$ with a detection ${\geq}$\,5$\sigma_\text{rms}$ threshold. A sample of a few sources from the catalogue is shown in Table \ref{table:tab3} (the complete version of Table \ref{table:tab3}, including all columns with their respective uncertainties (covering both convolved and deconvolved source sizes and position angles), is available online as supplementary material).

%%%%%%%%%%%%%%%%%%%%%%%%%%%%%%%%%%%%%%%%%%%%%%%%%%%%%%%%%%%%%%%%%%%%%%%%%%%%%%%%%%%%%%%%%%

\begin{table*}
\caption{This table represents the sample source from comprises catalogue for uGMRT Band-2 final image of ELAIS-N1 field.}
\centering
\begin{tabular}{|ccccccccccc|}
\hline
Source Id & RA & DEC & Total Flux & Peak Flux & Maj & Min & DC\_Maj & DC\_Min & DC\_PA & RMS \\
 & (deg $\pm$ arcsec) & (deg $\pm$ arcsec) & (mJy) & (mJy beam$^{-1}$) & (arcmin) & (arcmin) & (arcmin) & (arcmin) & (deg) & (mJy beam$^{-1}$) \\
\hline
1 & 245.3040 $\pm$ 0.53 & 54.6526 $\pm$ 0.07 & 18.05 & 6.19 & 0.40 & 0.21 & 0.35 & 0.12 & 114.54 & 0.87 \\
2 & 245.2519 $\pm$ 0.08 & 54.4271 $\pm$ 0.36 & 26.16 & 9.82 & 0.38 & 0.20 & 0.33 & 0.09 & 65.65 & 0.88 \\
3 & 245.2438 $\pm$ 0.46 & 54.4203 $\pm$ 0.37 & 6.40 & 6.50 & 0.18 & 0.16 & 0. & 0. & 0. & 0.87 \\
4 & 245.2285 $\pm$ 0.76 & 54.8149 $\pm$ 0.70 & 24.27 & 17.34 & 0.21 & 0.19 & 0.12 & 0.03 & 40.88 & 0.84 \\
5 & 245.1558 $\pm$ 0.01 & 54.0549 $\pm$ 0.34 & 12.86 & 6.82 & 0.28 & 0.20 & 0.20 & 0.10 & 77.55 & 0.85 \\
\hline
\end{tabular}
\label{table:tab3}
\vspace{1ex}
\parbox{0.95\textwidth}{\footnotesize Note: The final catalogue (``FITS'' format) comprises columns for source Ids, positions, flux density, peak flux density, convolved source sizes, deconvolved source sizes and position angle and local RMS noise. The complete version of this table with all the columns and their respective uncertainties (including the convolved and deconvolved source sizes and their respective position angles) will be accessible with the online version of the paper.}
\end{table*}
%%%%%%%%%%%%%%%%%%%%%%%%%%%%%%%%%%%%%%%%%%%%%%%%%%%%%%%%%%%%%%%%%%%%%%%%%%%%%%%%%%%%%%%

\subsection{Source Classification} \label{sec:classification}
\begin{figure}
    \centering
    \includegraphics[width=3.35in]{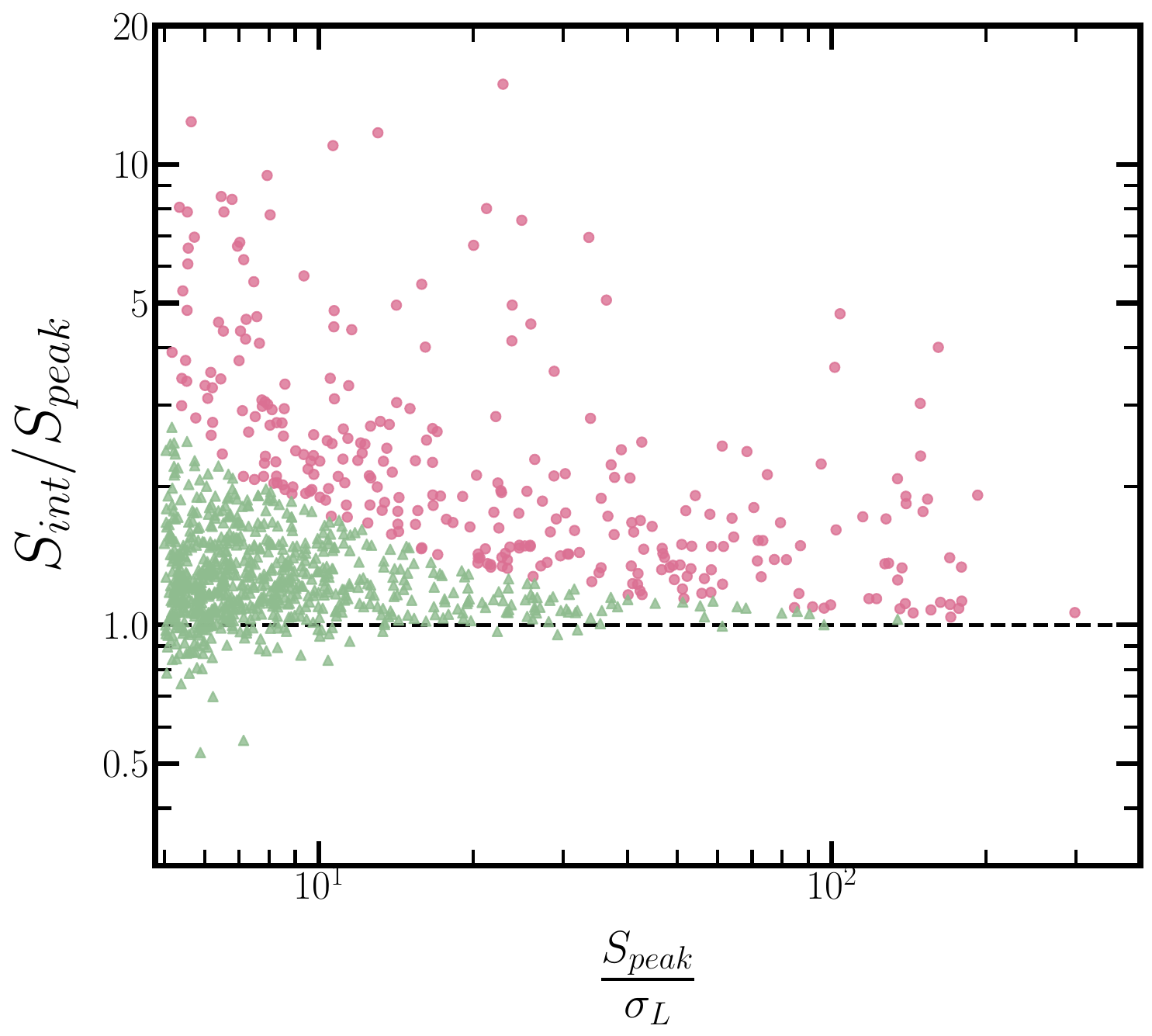}
    \caption{The ratio of integrated to peak flux density ($S_{\text{int}}/S_{\text{peak}}$) as a function of SNR ($S_{\text{peak}}/\sigma_L$) for uGMRT sources. Resolved sources are presented in red circles, and compact sources are in sea-green triangles.}
    \label{fig:Fig3}
\end{figure}

Time and bandwidth smearing affect artificially stretched image plane sources, making it difficult to classify them as resolved or compact. Typically, this classification is determined by the ratio of integrated to peak flux density (${S_{\text{int}}}/{S_{\text{peak}}}$). Bandwidth and time-average smearing reduce the peak flux densities of the sources but do not affect the integrated flux density. As a result, the integrated-to-peak flux density ratio is not equal to one for the sources that were initially unresolved. However, this classification is also affected by calibration errors and fluctuating noise, making it unreliable for calculations based solely on the (${S_{\text{int}}}/{S_{\text{peak}}}$)\,$>$\,1 criterion. The magnitude of this effect depends on the channel width (frequency resolution), integration time (time resolution), and radial distance from the pointing center. To address this, Figure \ref{fig:Fig3} illustrates a plot of (${S_{\text{int}}}/{S_{\text{peak}}}$) as a function of (${S_{\text{peak}}}/{\sigma_L}$), where ${\sigma_L}$ represents RMS noise. The plot reveals a skewed distribution at low SNR. 
%%%%%%%%%%%%%%%%%%%%%%%%%%%%%%%%%%%%%%%%%%%%%%%%%%%%%%%%%%%%%%%%%%%%%%%%%%%%%%%%%%%%%%%%%%%%%%%%%
%%%%%%%%%%%%%%%%%%%%%%%%%%%%%%%%%%%%%%%%%%%%%%%%%%%%%%%%%%%%%%%%%%%%%%%%%%%%%%%%%%%%%%%
%%%%%%%%%%%%%%%%%%%%%%%%%%%%%%%%%%%%%%%%%%%%%%%%%%%%%%%%%%%%%%%%%%%%%%%%%%%%%%%%%%%%%%%%%%%%%%%%%%%%
%%%%%%%%%%%%%%%%%%%%%%%%%%%%%%%%%%%%%%%%%%%%%%%%%%%%%%%%%%%%%%%%%%%%%%%%%%%%
\cite{2015MNRAS.453.4020F, 2019PASA...36....4F} describes an alternate way to identify the resolved sources at the 3$\sigma$ level, defined as follows:
%%%%%%%%%%%%%%%%%%%%%%%%%%%%%%%%%%%%%%%%%%%%%%%%%%%%%%%%%%%%%%%%%%%%%%%%%%%%%%%%%%%%%%%%%%%%%%%%%%%%
\begin{equation}
\begin{aligned}
\label{eqn:2}
\sigma_R &= \ln\left(\frac{S_{\text{int}}}{S_{\text{peak}}}\right) > 3\sqrt{\left(\frac{\sigma_{S_{\text{int}}}}{S_{\text{int}}}\right)^2 + \left(\frac{\sigma_{S_{\text{peak}}}}{S_{\text{peak}}}\right)^2}
\end{aligned}
\end{equation}
%%%%%%%%%%%%%%%%%%%%%%%%%%%%%%%%%%%%%%%%%%%%%%%%%%%%%%%%%%%%%%%%%%%%%%%%%%%%%%%%%%%%%%%%%%%%%%%%%%%%
where $\sigma_{S_{\text{int}}}$ and $\sigma_{S_{\text{peak}}}$ represent uncertainties in the integrated flux density ($S_{\text{int}}$) and peak flux density ($S_{\text{peak}}$), respectively. Based on the described criteria, we have found 301 resolved sources (represented by red circles) and 726 unresolved sources (represented by sea-green triangles), as shown in Figure \ref{fig:Fig3}. 

\section{COMPARISON WITH EXISTING RADIO CATALOGUES} \label{sec:comparison}

We provide a comparison of this work with other existing radio catalogues for the same field. We considered the study of uGMRT at 400\,MHz by \cite{2019MNRAS.490..243C} with a catalogue of 2528 sources. We also compared our results with the LOFAR Two-metre Sky Survey (LoTSS) of the ELAIS-N1 field at 150\,MHz by \cite{2021A&A...648A...2S}, which is one of the largest low-frequency deep field surveys with an available catalogue of 84862 sources and an RMS noise of $\lesssim$ 20\,$\mu$Jy/beam. Furthermore, we performed comparisons with the Faint Images of the Radio Sky at Twenty-centimeters (FIRST) survey catalogue at 1400\,MHz by \cite{1997ApJ...475..479W}, the TIFR GMRT Sky Survey (TGSS) catalogue at 150\,MHz by \cite{2017A&A...598A..78I}, and the GMRT 610\,MHz in Band-4 by \cite{2020MNRAS.497.5383I} and the uGMRT 1250\,MHz catalogue by \cite{2023JApA...44...88S}.\\
Ionospheric fluctuations at low frequencies can result in shift in the source positions and calibration errors can create flux density offsets. Hence, comparison with existing catalogues of the same part of the sky (in different frequencies) allows for determining potential systematic bias in source position offsets and flux densities (only when compared to observations at the same frequency). In order to find counterparts of the sources detected in the present catalogue, we applied a %$7''$ 
search radius of 1$\times$PSF in other catalogues, while for TGSS, the threshold is %$12''$ 
2$\times$PSF due to its low resolution. Each catalogue has a specific limit on flux density based on the sensitivity and completeness of the observations. Each catalogue at a different frequency was scaled to 183\,MHz using the relation $S_\nu \propto \nu^{\alpha}$ with $\alpha = -0.7$. Only sources with flux densities above this cutoff, scaled to 183\,MHz were considered. The flux limit at 183\,MHz, indicated as $S_{\text{cut,183MHz}}$, was employed as a constraint for selecting sources. Table \ref{table:tab4} provides details on the resolution and flux limits of each chosen catalogue, as well as their equivalent 183\,MHz cut-off.
%%%%%%%%%%%%%%%%%%%%%%%%%%%%%%%%%%%%%%%%%%%%%%%%%%%%%%%%%%%%%%%%%%%%%%%%%%%%%%%%%%%%

\subsection{Flux Density Offset} \label{sec:fluxoffset} 

\begin{figure}
\includegraphics[width=3.35in]{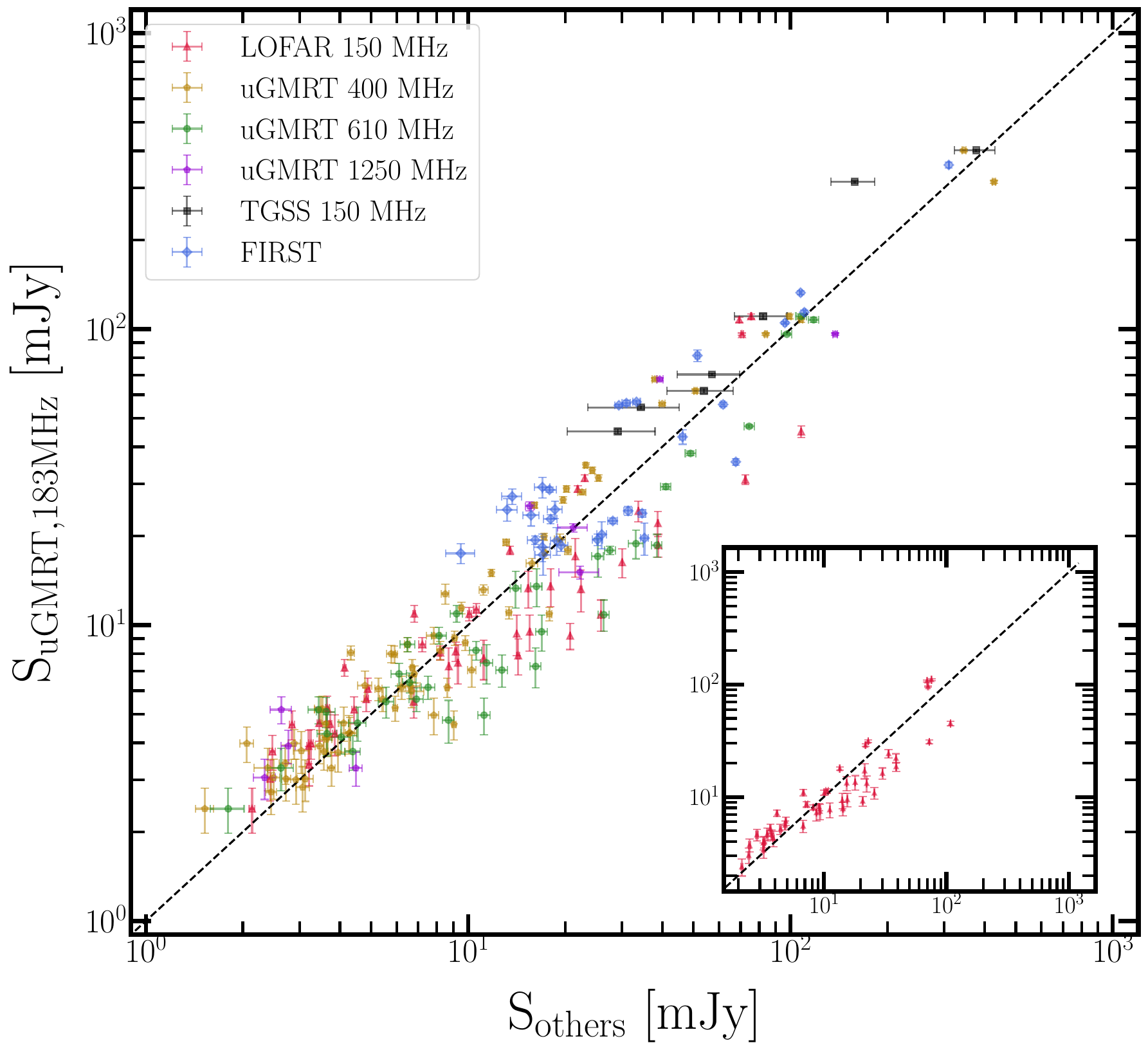}
\caption{Comparing the integrated flux densities of compact and isolated sources estimated with uGMRT at 183\,MHz with other available catalogues after scaling them at 183\,MHz frequency, including 150\,MHz LoTSS (red, also in the sub-plot), 400\,MHz GMRT (golden), 610\,MHz GMRT (green), uGMRT 1250\,MHz (purple), 150\,MHz TGSS (black) and FIRST survey (blue). The black dashed line in both plots represents $S_{\text{uGMRT}}/S_{\text{others}} = 1$.}
\label{fig:Fig4}
\end{figure}

Variations in the absolute flux density scale can introduce systematic offsets in the estimated flux density values. Different catalogues can be derived using different flux density scales, resulting in an offset in the measured flux density of sources. In this study, we used the \cite{2012MNRAS.423L..30S} flux scale which has also been utilised in the LoTSS 150\,MHz data from \cite{2021A&A...648A...2S} and the TGSS 150\,MHz catalogue \cite{2017A&A...598A..78I}. The flux density of the sources in this catalogue has been compared with their counterparts in other catalogues to check for flux offset.
%%%%%%%%%%%%%%%%%%%%%%%%%%%%%%%%%%%%%%%%%%%%%%%%%%%%%%%%%%%%%%%%%%%%%%%%%%%%%%%%%%%%%%
\begin{table}
\caption{Flux limit ($\alpha = -0.7$) of prior available catalogues with their frequency and resolution, respectively.}
\centering
\begin{tabular}{|ccccc|}
\hline
Catalogue & Frequency & Resolution & $S_\text{limit}^{\diamond}$ & $S_\text{{cut},{183MHz}}$ \\ [0.5ex]
 & (MHz) & (arcsec) & (mJy) & (mJy) \\ [1ex]
\hline\hline
uGMRT & 183 & 11.45 & 1.1 & 1.1 \\ [0.5ex]
FIRST & 1400 & 5.4 & 1.0 &  4.15 \\ [0.5ex]
uGMRT & 400 & 4.6 &  0.10 & 0.17 \\ [0.5ex]
GMRT & 610 & 6 & 0.2 &  0.46 \\ [0.5ex]
uGMRT & 1250 & 2.0 & 0.06 &  0.23 \\ [0.5ex]
LoTSS & 150 & 6 & 0.1 &  0.087 \\ [0.5ex]
TGSS & 150 & 25 & 17.5 &  15.22 \\ [0.5ex]
\hline
\end{tabular}
%\vspace{0.5em}
\parbox{\linewidth}{\footnotesize \hspace{1.8em} $S_\text{limit}^{\diamond}$ -- Flux density limit for the corresponding catalogue.}
\label{table:tab4}
\end{table}
%%%%%%%%%%%%%%%%%%%%%%%%%%%%%%%%%%%%%%%%%%%%%%%%%%%%%%%%%%%%%%%%%%%%%%%%%%%%%%%%%%%%
To ensure a fair comparison, we have selected only high-SNR sources with peak flux densities above 10$\sigma$ and applied compact source condition where the size of the sources must be below the resolution limit at a higher frequency. In addition, we have applied a flux limit to our sample, choosing only sources above a certain flux threshold. For example, for the \cite{2020MNRAS.497.5383I} catalogue, this threshold is 0.2\,mJy, which corresponds to 0.46\,mJy at 183\,MHz (assumption of spectral index, $\alpha=-0.7$). Only sources with flux density above the limits specified in Table \ref{table:tab4} were considered for analysis. Flux density ratios ($S_{183 \text{MHz}}/S_{\text{others}}$) were calculated after proper flux scaling. Here, ``others'' corresponds to the catalogues used for comparison. The median values of the flux ratio are found to be: 1.07$^{+0.50}_{-0.31}$ (LoTSS 150\,MHz), 0.89$^{+0.32}_{-0.40}$ (FIRST 1400\,MHz), 1.14$^{+0.23}_{-0.27}$ (uGMRT 400\,MHz), 0.84$^{+0.28}_{-0.34}$ (GMRT 610\,MHz),  1.30$^{+0.60}_{-0.40}$ (uGMRT 1250\,MHz) and 1.34$^{+0.20}_{-0.28}$ (TGSS 150\,MHz) with 16th and 85th percentile uncertainities. Figure \ref{fig:Fig4} illustrates these results, highlighting the reliability of the flux values obtained. Furthermore, the subplot provides a clear comparison of the 183\,MHz catalogue with LoTSS at 150\,MHz, including consideration of associated errors. This comparison provides the median flux density ratios with nearly one of the other catalogues.
%%%%%%%%%%%%%%%%%%%%%%%%%%%%%%%%%%%%%%%%%%%%%%%%%%%%%%%%%%%%%%%%%%%%%%%%%%%%%%%%%%
%%%%%%%%%%%%%%%%%%%%%%%%%%%%%%%%%%%%%%%%%%%%%%%%%%%%%%%%%%%%%%%%%%%%%%%%%%%%%%%%%%%%%%%

\subsection{Positional Accuracy} \label{sec:pos-off} 

The position accuracy of the sources in the catalogue can be affected by various factors, such as the phase calibration error, ionosphere and pointing error. To verify the astrometric precision of the obtained source positions, a comparison is made with the 1400\,MHz FIRST, LoTSS 150\,MHz, uGMRT 400\,MHz, GMRT 610\,MHz and uGMRT 1250\,MHz catalogues. The source selection criteria are similar, as in Section \ref{sec:fluxoffset}. The FIRST catalogue \citep{1997ApJ...475..479W} has a higher frequency (1400\,MHz) and a better resolution of $5.4''$, making it less affected by ionospheric fluctuations and hence providing better positional accuracy ($<$\,1 arcsec). The offsets follow the convention of \cite{2016MNRAS.460.2385W}:
%%%%%%%%%%%%%%%%%%%%%%%%%%%%%%%%%%%%%%%%%%%%%%%%%%%%%%%%%%%%%%%%%%%%%%%%%
\begin{equation}
\begin{aligned}
\label{eqn:3}
\delta_\mathrm{RA} & = \mathrm{RA_{uGMRT_{183}} - RA_{FIRST}} \\
\delta_\mathrm{DEC} & = \mathrm{DEC_{uGMRT_{183}} - DEC_{FIRST}}
\end{aligned}
\end{equation}
%%%%%%%%%%%%%%%%%%%%%%%%%%%%%%%%%%%%%%%%%%%%%%%%%%%%%%%%%%%%%%%%%%%%%%%%%

\begin{table}
\caption{Median deviations of the Right Ascension (RA) and Declination (Dec) of the uGMRT 183\,MHz source catalogue from those of other catalogues, along with their corresponding 16th and 85th percentile uncertainties.}
\centering
\begin{tabular}{ccccc}
\hline
catalogue & $\nu$ & $\delta_\text{RA,median}$ & $\delta_\text{DEC,median}$ & Match \\ [0.5ex]
& (MHz) & (arcsec) & (arcsec)  &  \\ [0.8ex]
\hline\hline
FIRST & 1400 & -0.027$^{+1.04}_{-1.96}$ & 0.368$^{+0.95}_{-0.61}$ & 30 \\ [1ex]
LoTSS & 150 & 0.637$^{+1.74}_{-1.46}$ & 0.255$^{+0.85}_{-0.34}$ & 46 \\ [1ex]
uGMRT & 400 & 0.912$^{+0.78}_{-1.18}$ & 0.011$^{+0.80}_{-0.50}$ & 72 \\ [1ex]
GMRT & 610 & 0.644$^{+1.66}_{-1.89}$ & 0.259$^{+0.76}_{-0.44}$ & 36 \\ [1ex]
uGMRT & 1250 & 0.740$^{+0.93}_{-0.88}$ & 0.221$^{+0.89}_{-0.26}$ & 9 \\ [1ex]
TGSS & 150 & -0.123$^{+0.98}_{-0.56}$ & 0.432$^{+1.07}_{-0.65}$ & 7 \\ [1ex]
\hline
\end{tabular}
\label{table:tab5}
\end{table}

%%%%%%%%%%%%%%%%%%%%%%%%%%%%%%%%%%%%%%%%%%%%%%%%%%%%%%%%%%%%%%%%%%%%%%%%%%%%%%%%%%%%
%%%%%%%%%%%%%%%%%%%%%%%%%%%%%%%%%%%%%%%%%%%%%%%%%%%%%%%%%%%%%%%%%%%%%%%%%%%%%%%%%%%%
\begin{figure}
\includegraphics[width=3.35in]{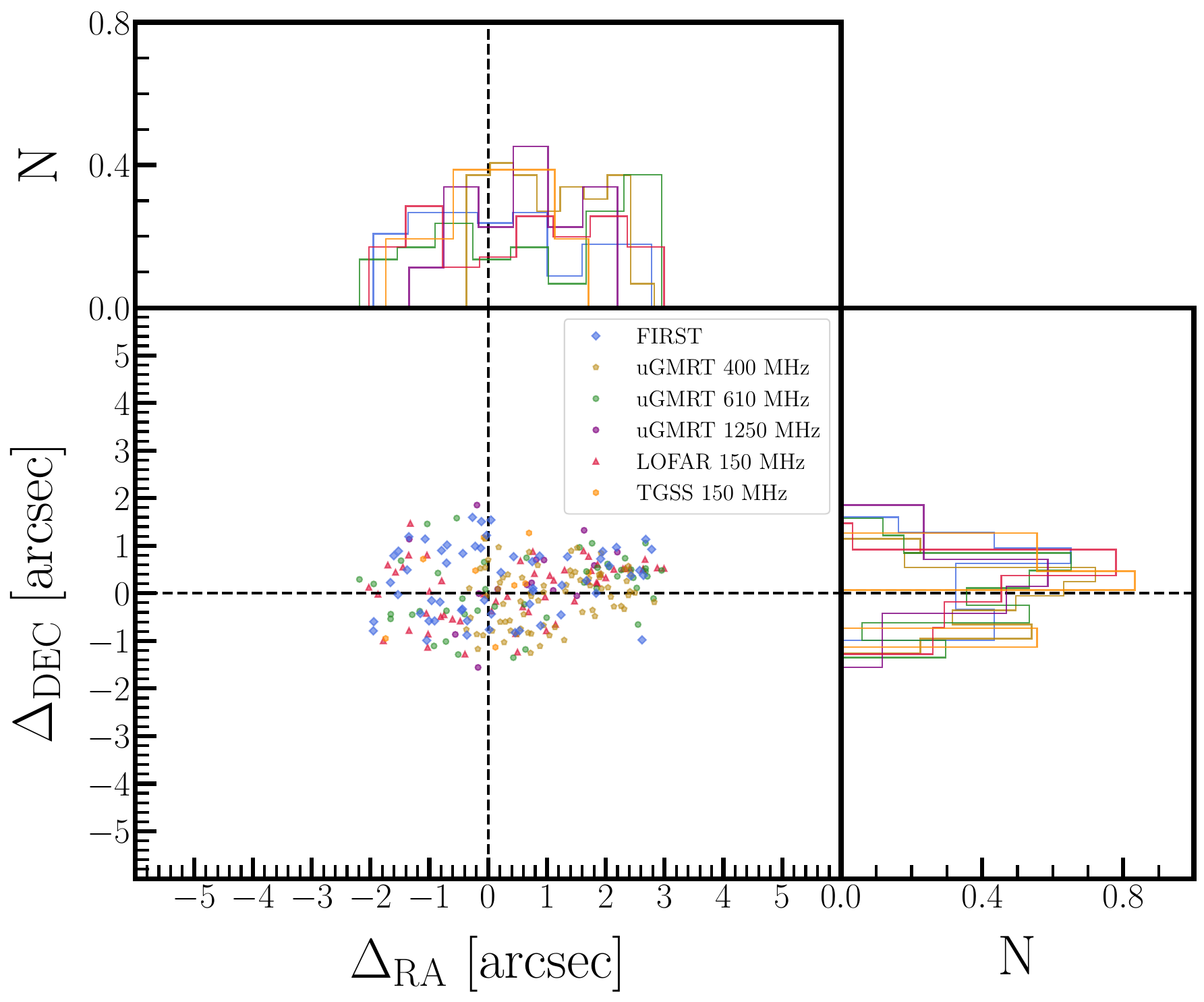}
\caption{Deviation in positions (RA $\&$ DEC) of the uGMRT sources at 183\,MHz compared with the FIRST (blue), uGMRT 400\,MHz (golden), GMRT 610\,MHz (green), uGMRT 1250\,MHz (purple), LoTSS (red) and TGSS 150MHz (orange) catalogue observations.}
\label{fig:Fig5}
\end{figure}
%%%%%%%%%%%%%%%%%%%%%%%%%%%%%%%%%%%%%%%%%%%%%%%%%%%%%%%%%%%%%%%%%%%%%%%%%%%%%%%%%%%%

Table \ref{table:tab5} displays the median offset values with errors, and Figure \ref{fig:Fig5} presents a histogram of the positional offset estimation with other compared observations. Although the offsets are less than the cell size of the image ($2.5''$), we have corrected it for any possible astrometric error. The constant value derived from the median offset value with respect to FIRST catalogue has been applied to correct the final catalogue source positions.

\subsection{Spectral Index Distribution}

\begin{figure}
    \includegraphics[width=3.35in]{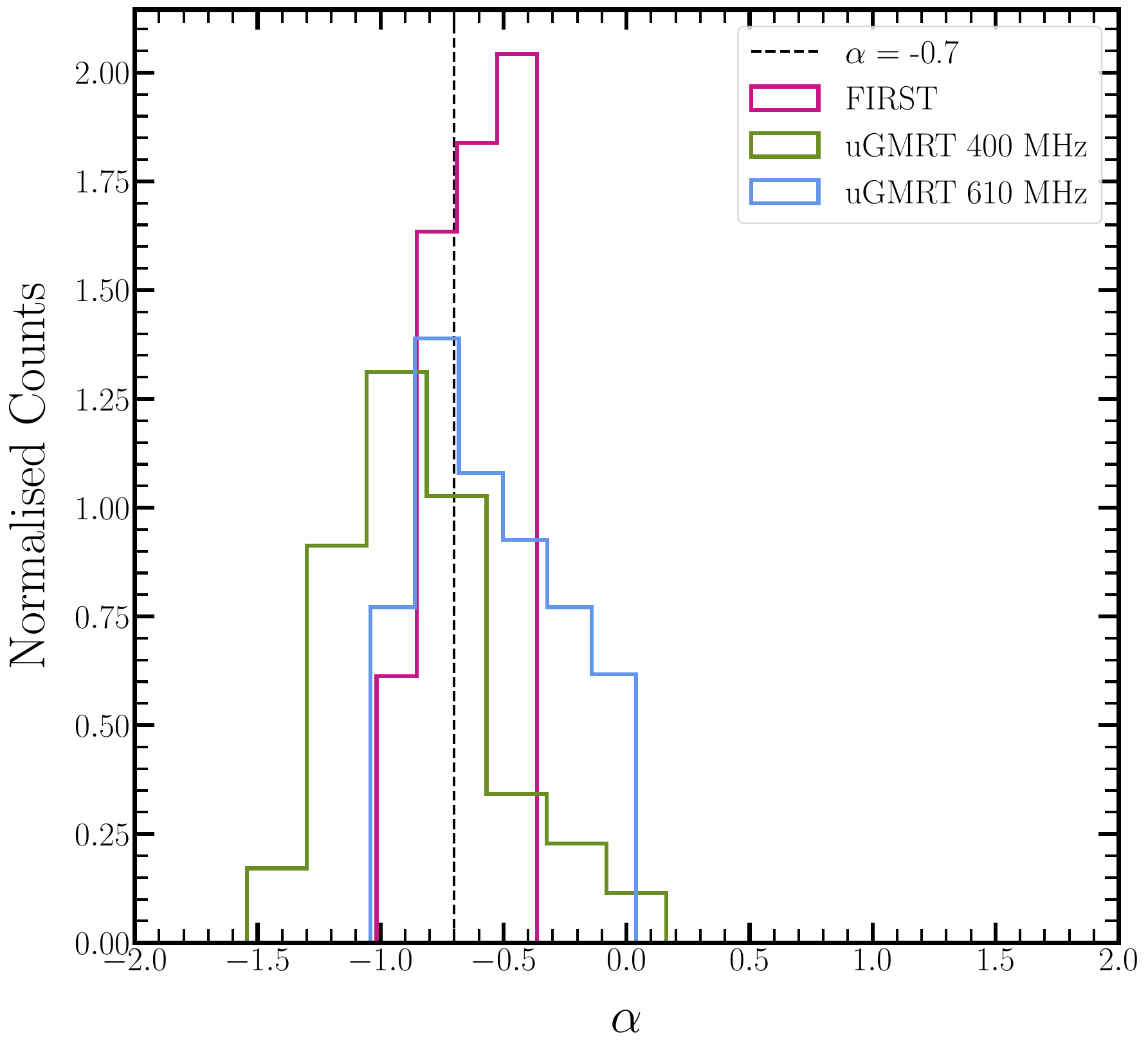}
     \caption{The normalised count of spectral indices measured in the field after matching with FIRST (deep pink), uGMRT 400\,MHz (green) and GMRT 610\,MHz catalogue (blue) has been shown using a solid line with a $7''$ match radius. The spectral index of power-law {$-0.7$} is represented by the black dashed line.}
     \label{fig:Fig6}
 \end{figure}

The data covers deep field studies in the ELAIS-N1 region, consisting of various sources with different spectral properties. To characterise these properties, the flux density of sources calculated in this work is compared with other available catalogues at higher frequencies, specifically uGMRT 400\,MHz \citep{2019MNRAS.490..243C}, GMRT 610\,MHz \citep{2020MNRAS.497.5383I} and 1400\,MHz FIRST \citep{1997ApJ...475..479W} catalogue. Source selection criteria are consistent with those used in previous sections \ref{sec:fluxoffset} and \ref{sec:pos-off}, namely compact, isolated, and high SNR. The spectral index, $\alpha$ is estimated from the flux densities of 30 matched sources from FIRST catalogue with a power-law distribution where $S_\nu \propto \nu^{\alpha}$. We have derived the spectral index of matched sources and plotted the histograms of $\alpha$ for the uGMRT 183\,MHz matched sources with the other two catalogues, shown in Figure  \ref{fig:Fig6}. The calculated median spectral values, ${\alpha}_\text{median}$ are -0.87$^{+0.26}_{-0.31}$ (uGMRT 400\,MHz), -0.64$^{+0.18}_{-0.23}$ (FIRST 1400\,MHz) and -0.56$^{+0.28}_{-0.34}$ (GMRT 610\,MHz) with uncertainties for the 16th and 85th percentile. Thus, after source matching, the ${\alpha}_\text{median} {\sim}\,{-0.7}$ for this catalogue.\\
Radio observations at other frequencies for the ELAIS-N1 region typically have a median spectral index value with FIRST catalogue: -0.81$^{+0.28}_{-0.32}$ \citep[at 400MHz;][]{2019MNRAS.490..243C}, -0.85 $\pm$ 0.05 and -0.83 $\pm$ 0.31 at 610MHz for \cite{2020MNRAS.497.5383I} and \cite{2020MNRAS.491.1127O} respectively. Similar distributions have been observed for other deep fields such as the Cosmic Evolution Survey \citep[COSMOS;][]{2017A&A...602A...1S}, Lockman Hole \citep{2016MNRAS.463.2997M, 2020MNRAS.495.4071M} and Bo\"{o}tes \citep{2016MNRAS.460.2385W, 2023MNRAS.525.5311S}, with few exceptions \citep{2011MNRAS.415.1597M, 2017MNRAS.464.3357W, 2021MNRAS.508.5259M}. However, the distribution of $\alpha$ in this work is broader than typically observed, including LOFAR at 150\,MHz. We acknowledge that a detailed study of the spectral index necessitates deeper observations, which will be communicated in future work.
%%%%%%%%%%%%%%%%%%%%%%%%%%%%%%%%%%%%%%%%%%%%%%%%%%%%%%%%%%%%%%%%%%%%%%%%%%%%%%%%%%%%%%%%%%%%%%%%%%%%
%%%%%%%%%%%%%%%%%%%%%%%%%%%%%%%%%%%%%%%%%%%%%%%%%%%%%%%%%%%%%%%%%%%%%%%%%%%%%%%%%%%%

\section{Source Count} \label{sec:count}

The relation between the population of radio galaxies and their associated flux density provides an understanding of the evolutionary properties of galaxies \citep{2020MNRAS.491.1127O}. The distribution of source counts characterises the abundance of radio sources. The observations of a plateau or flattening in the source counts below 1\,mJy has been firmly established and is widely accepted as evidence for the presence of large number of low flux SFGs in the radio sky \citep{2009ApJ...694..235P} except for the LoTSS deep fields study at 150\,MHz \citep{2021A&A...648A...5M}, which observed a dip at the sub-mJy level. However, in deep and wide fields, comprehensive studies are required to develop a standard model for low-frequency sources. These differential source counts can provide information about the nature of sources and contribute to constrained models.\\
Precise source counting is beneficial in assessing the population of radio galaxies, particularly at low frequencies, where SFGs and radio-quiet (RQ) AGNs dominate over AGNs and radio galaxies due to their low flux densities \citep{2015MNRAS.452.1263P, 2023MNRAS.523.1729B}. Previous studies, both observations and simulations \cite{2008MNRAS.388.1335W, 2015MNRAS.448.2665W}, have highlighted the importance of characterising the distribution of sources in faint flux densities. However, observational data on sub-mJy source populations are still limited. Therefore, radio interferometers such as MWA, uGMRT, LOFAR or upcoming SKA, require accurate characterisation of source population to allow better modelling and subtraction of source confusion noise when flux densities vary from sub-mJy to \textit{${\mu}$}Jy.\\
We present the differential source counts using flux densities of sources of wideband data derived from the PYBDSF catalogue. We measured the differential source counts at 183\,MHz down to 0.9\,mJy. Nevertheless, relying entirely on PYBDSF output to directly obtain source counts can lead to inaccurate quantification of the true extragalactic source distribution. The presence of catalogue incompleteness, resolution bias \citep{2006A&A...457..517P}, false detection and Eddington bias \citep{1913MNRAS..73..359E, 2024ApJ...964..158B} can cause errors that significantly impact the precision of source counts, particularly in low-frequency and the faint end of flux density bins. We calculated two corrections for the source counts: one for false detections of sources and another for completeness, which considers the resolution and visibility area effects on source detection. The subsequent sections elaborate on the corrections applied to the source count distribution in detail below.
%%%%%%%%%%%%%%%%%%%%%%%%%%%%%%%%%%%%%%%%%%%%%%%%%%%%%%%%%%%%%%%%%%%%%%%%%%%%%%%%%%%
\begin{figure}
\includegraphics[width=3.35in]{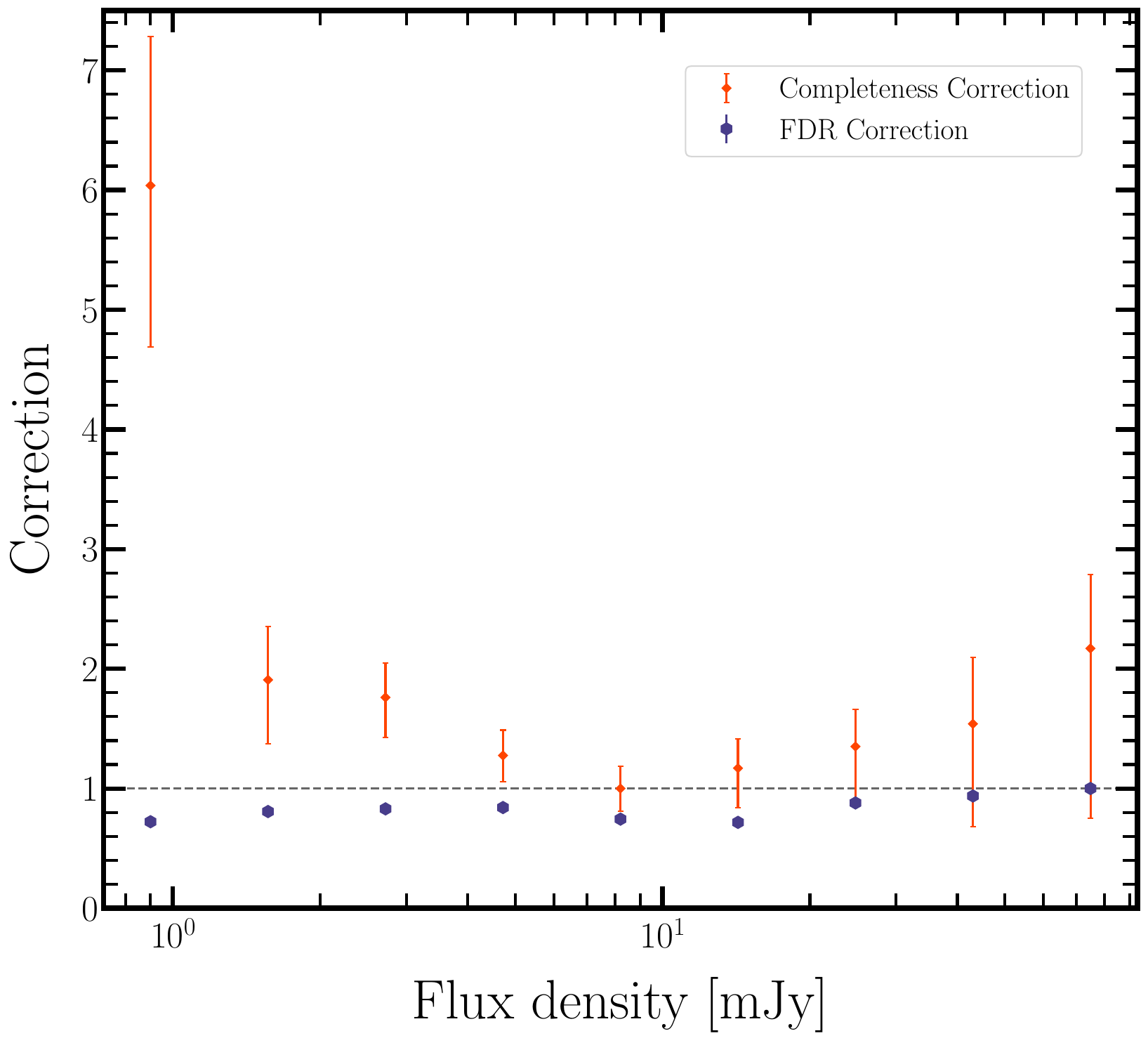}
\caption{Variation of correction factors for false detection rate (FDR) (indigo) and completeness (orange) with integrated flux density.}
\label{fig:Fig7}
\end{figure}
%%%%%%%%%%%%%%%%%%%%%%%%%%%%%%%%%%%%%%%%%%%%%%%%%%%%%%%%%%%%%%%%%%%%%%%%%%%%%%%%%%%%

\begin{table*}
\caption{Euclidean normalised differential source counts of the ELAIS-N1 field at 183\,MHz. The columns below correspond to the flux density bins, central value of the flux density bin, number of sources in a bin N, normalised source counts (S$^{2.5}$(dN/dS)), FDR, completeness, and corrected-normalised source counts.}
\centering
\begin{tabular}{|ccccccc|}
\hline
S$_\text{range}$ & S$_\text{mid}$ & N & Counts & FDR & Completeness & Corrected Counts \\ [0.5ex]
(mJy) & (mJy) & & (Jy$^{1.5}$sr$^{-1}$) & & & (Jy$^{1.5}$sr$^{-1}$) \\ [1.2ex]
\hline
0.900\,--\,1.564 & 1.232 & 17 & 7.316 $\pm$ 1.775 & 0.71 $\pm$ 0.03 &  6.04$^{+1.24}_{-1.35}$  & 31.365 $\pm$ 7.607 \\ [0.5ex]
1.564\,--\,2.718 & 2.141 & 134 & 27.705 $\pm$ 2.393 & 0.81 $\pm$ 0.01 & 1.91$^{+0.45}_{-0.55}$ & 42.818 $\pm$ 3.699 \\[0.5ex]
2.718\,--\,4.722 & 3.719 & 232 & 48.203 $\pm$ 3.165 & 0.83 $\pm$ 0.01 & 1.76$^{+0.29}_{-0.34}$ & 70.454 $\pm$ 4.626 \\[0.5ex]
4.722\,--\,8.206 & 6.464 & 222 & 79.424 $\pm$ 5.301 & 0.86 $\pm$ 0.00 & 1.27$^{+0.21}_{-0.22}$ & 87.156 $\pm$ 5.850 \\[0.5ex]
8.206\,--\,14.259 & 11.232 & 172 & 138.115 $\pm$ 10.533 & 0.77 $\pm$ 0.01 & 1.00$^{+0.19}_{-0.19}$ & 106.348 $\pm$ 8.109 \\[0.5ex]
14.259\,--\,24.777 & 19.518  & 106 & 194.327 $\pm$ 18.875 & 0.69 $\pm$ 0.01 & 1.17$^{+0.25}_{-0.33}$ & 156.88 $\pm$ 15.238 \\[0.5ex]
24.777\,--\,43.054 & 33.915  & 60 & 251.778 $\pm$ 32.504 & 0.88 $\pm$ 0.01 & 1.35$^{+0.31}_{-0.51}$ & 299.334 $\pm$ 38.644 \\[0.5ex]
43.054\,--\,74.813 & 58.933 & 47 & 451.706 $\pm$ 65.888 & 0.96 $\pm$ 0.00 & 1.54$^{+0.56}_{-0.86}$ & 667.803 $\pm$ 97.409 \\[0.5ex]
74.813\,--\,130.000 & 102.407 & 12 & 264.171 $\pm$ 76.260 & 0.83 $\pm$ 0.02  & 2.17$^{+0.62}_{-1.42}$ & 475.799 $\pm$ 137.351 \\[0.5ex]
\hline
\end{tabular}
\label{table:tab6}
\end{table*}
%%%%%%%%%%%%%%%%%%%%%%%%%%%%%%%%%%%%%%%%%%%%%%%%%%%%%%%%%%%%%%%%%%%%%%%%%%%%%%%%%%%%

\subsection{False Detection Rate}

FDR is a critical metric of spurious source detection in a PyBDSF radio catalogue, which can be adversely affected by incorporating false detections arising from noise spikes and strong artefacts, a source finding package, detected as real. To account for FDR, we leveraged the symmetry of noise spikes in the flipped (negative) image to detect sources. Furthermore, we optimised PyBDSF on this flipped (negative) image using the parameters described in Section \ref{sec:catalogue}. We have detected 184 sources with negative peaks less than ${-5}{\sigma}$. We binned the number of negative sources identified in the flipped image in 10 logarithmic bins to modify for FDR in flux density bins and then compared it with the positive sources found in the original image.
The fraction of real sources in each bin introduced by \cite{2019A&A...622A...4H} as follows:
%%%%%%%%%%%%%%%%%%%%%%%%%%%%%%%%%%%%%%%%%%%%%%%%%%%%%%%%%%%%%%%%%%%%%%%%%%%%%%%%%%%%
\begin{equation}
\begin{aligned}
\label{eqn:4}
f_{\text{real}, j} = \frac{N_{\text{catalogue}, j} - N_{\text{inv}, j}}{N_{\text{catalogue}, j}}
\end{aligned}
\end{equation}
%%%%%%%%%%%%%%%%%%%%%%%%%%%%%%%%%%%%%%%%%%%%%%%%%%%%%%%%%%%%%%%%%%%%%%%%%%%%%%%%%%%%
Here, N$_\text{inv,j}$ and N$_\text{catalogue,j}$ are the number of identified sources in each flux density bin of the flipped and original image, respectively. Poissonian fluctuations are used to calculate the uncertainties. The resulting fraction, obtained using Equation \ref{eqn:4}, is then multiplied by the number of sources in the corresponding flux density bin to acquire the correction factors for the original catalogue, presented in Figure \ref{fig:Fig7} for each bin.

\subsection{Completeness}

A catalogue shows incompleteness when it is unable to identify sources with flux densities greater than its detection threshold (5$\sigma$; in this case), primarily due to the noise fluctuations in the image. The PyBDSF employed in generating the catalogue has limited accuracy in measuring flux. This drawback causes the completeness of a source catalogue to be limited, possibly leading to either underestimation or overestimation of source counts derived from the image plane. There can be two types of bias: Eddington bias and Resolution bias, while 
%%%%%%%%%%%%%
\begin{figure*} 
\centering
 \makebox[\textwidth]{\includegraphics[width=0.65\paperwidth]{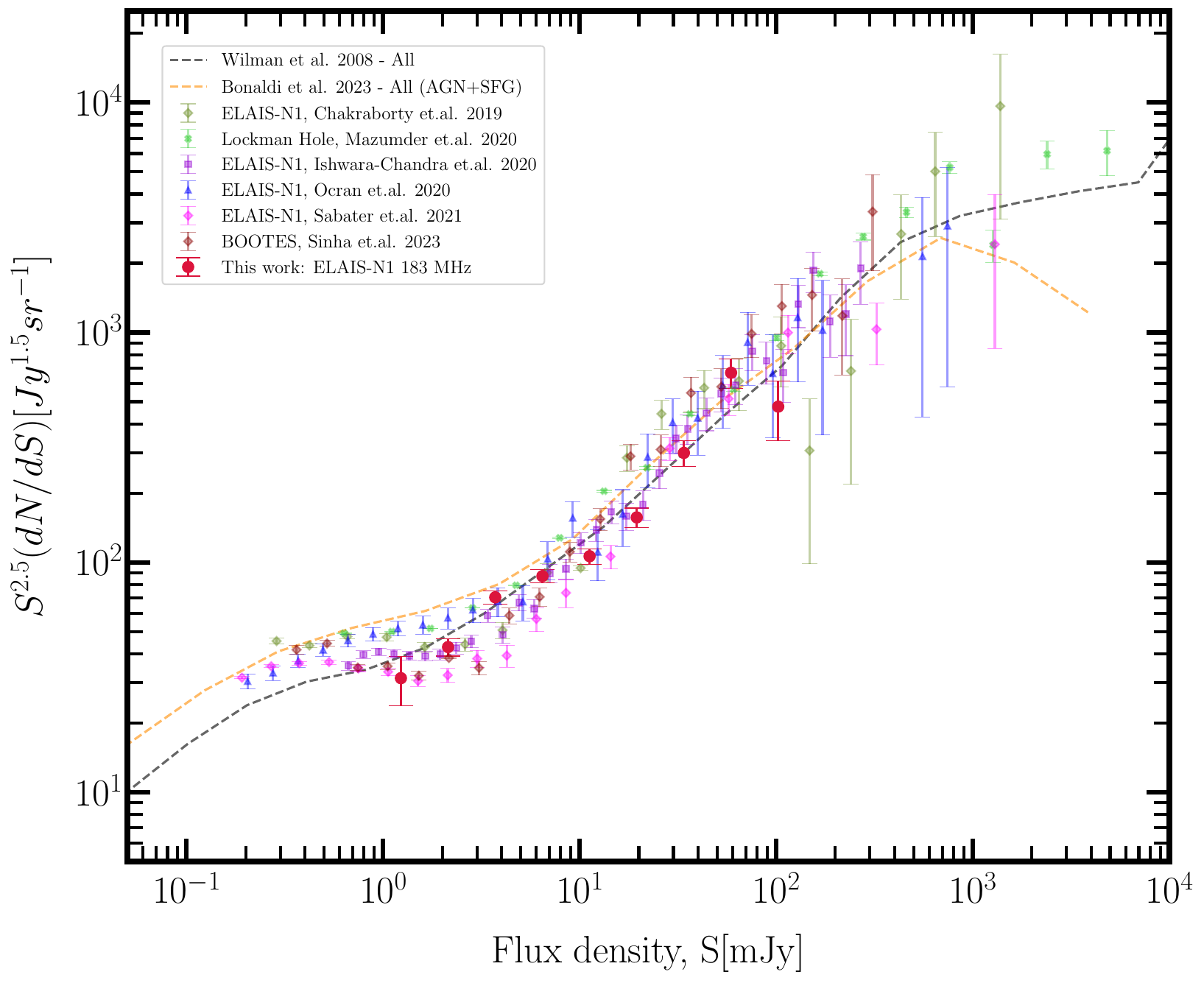}}
\caption{Euclidean normalised 183\,MHz differential source counts (red circles) of the ELAIS-N1 field after FDR and incompleteness corrections. Modelled source counts from simulations S$^3$ \citep{2008MNRAS.388.1335W} and T-RECS II \citep{2023MNRAS.524..993B} are shown as black-dashed and orange-dashed curves, respectively. The observations are also compared with other data observations covering the same sky region, such as uGMRT 400\,MHz \citep{2019MNRAS.490..243C} (olive diamonds), Lockman Hole \citep{2020MNRAS.495.4071M} (lime green triangles), GMRT 610\,MHz (\protect\cite{2020MNRAS.497.5383I} (dark violet squares) and \protect\cite{2020MNRAS.491.1127O} (blue triangles), LoTSS 150\,MHz \citep{2021A&A...648A...5M} (fuchsia diamonds) and Bo\"{o}tes 400\,MHz \citep{2023MNRAS.525.5311S} (maroon diamonds).}
\label{fig:Fig8}
\end{figure*}
%%%%%%%%%%%%%%%%%%%%%%%%%%%%%%%%%%%%%%%%%%%%%%%%%%%%%%%%%%%%%%%%%%%%%%%%%%%%%%%%%%%%
Eddington bias scatters more high-count bins into lower-count ones than vice versa, thereby falsely increasing the lower source counts. Resolution bias arises due to the lower detection probability of resolved sources compared to compact ones, leading to an underestimation of source counts. To correct for biases in source counts caused by incompleteness and resolution effects, simulations were conducted on the image plane using open source software {\tt\string AEGEAN}\footnote{\url{https://github.com/PaulHancock/Aegean}} \citep{2012ascl.soft12009H, 2018PASA...35...11H}. A total of 500 sources were injected artificially into the residual image \citep[following similar criteria as in ][]{2000A&AS..146...41P, 2016MNRAS.460.2385W}, with 150 being extended sources ($\theta_{\text{major}}, \theta_{\text{minor}} \geq 11.45''$) and the remainder being unresolved point sources, following the existing source classification in the catalogue. The flux density distribution followed a power-law distribution $dN/dS \propto S^{-1.6}$ \citep[see:][]{2011A&A...535A..38I, 2013A&A...549A..55W}, and the flux values were constrained between 0.9\,mJy and 130\,mJy. The source positions were chosen randomly across the whole RA and DEC range of the catalogue. One hundred different realisations of these simulations were carried out to account for visibility area impacts and source confusion constraints \citep{2016MNRAS.460.2385W, 2019PASA...36....4F, 2019A&A...622A...4H}. Sources were obtained from each image separately with similar parameters as discussed in Section \ref{sec:catalogue}. The resulting sources were organised into an equal number of bins as the original catalogue. The completeness correction factor in each flux density bin is given by:

%%%%%%%%%%%%%%%%%%%%%%%%%%%%%%%%%%%%%%%%%%%%%%%%%%%%%%%%%%%%%%%%%%%%%%%%%%%%%%%%%%%%%%%%%%%%%%%%%%%%
\begin{equation}
\begin{aligned}
\label{eqn:5}
\text{Correction}{j} = \frac{N_{\text{injected}, j}}{N_{\text{recovered}, j}}
\end{aligned}
\end{equation}
%%%%%%%%%%%%%%%%%%%%%%%%%%%%%%%%%%%%%%%%%%%%%%%%%%%%%%%%%%%%%%%%%%%%%%%%%%%%%%%%%%%%

where N$_\text{injected,j}$ indicates the number of injected sources and N$_\text{recovered,j}$ stands for the number of restored sources after removing original pre-simulation sources within the jth bin \citep{2019A&A...622A...4H}. Figure \ref{fig:Fig7} displays the correction factors of FDR and completeness for each flux bin. The median value of each flux bin from 100 simulations serves as the correction factor, whereas the associated uncertainties are estimated for the 16th and 85th percentile.

\subsection{Differential Source Count} \label{sec:dN/dS}

To estimate the Euclidean normalised differential source counts (in Jy$^{-1.5}$sr$^{-1}$), the generated source catalogue by PyBDSF was corrected for FDR and incompleteness. For each bin, the correction factor was multiplied with the uncorrected source counts, and the effective area per bin was also considered, as noise varies across the image (Figure \ref{fig:Fig2}). We fixed the effective area per bin over which sources can be found by finding the fraction of area (f) where a source with a certain flux density can be seen \citep{1985ApJ...289..494W}. We have binned the flux densities into 10 logarithmic bins down to 0.9\,mJy, and evaluated the Poisson fluctuations on the source counts. We first computed the Poisson fluctuation and then proceeded with the area correction per bin. Table \ref{table:tab6} lists the resulting source counts and fluctuations. After incorporating the necessary corrections, we plotted the normalised differential source counts in Figure \ref{fig:Fig8} and compared them with other observed counts and simulations. We obtain the source counts by considering the total flux densities of both point and extended sources. 
For cross-validation, we have compared the differential source counts against those of the 400\,MHz uGMRT data \citep{2019MNRAS.490..243C} and 610\,MHz GMRT data \citep{2020MNRAS.497.5383I, 2020MNRAS.491.1127O}, which were scaled to 183\,MHz using a spectral index, ${\alpha} = {-0.7}$. Similar comparisons were made with differential source counts at 1400\,MHz \citep{1997ApJ...475..479W, 2015ApJ...801...26H}, LoTSS 150\,MHz \citep{2021A&A...648A...5M}, Lockman Hole at 325\,MHz \citep{2020MNRAS.495.4071M} and Bo\"{o}tes at 400\,MHz \citep{2023MNRAS.525.5311S} using the same spectral index. We have also used simulated catalogues, namely $S^3$-SKADS \citep{2008MNRAS.388.1335W} and Tiered - Radio Extragalactic Continuum Simulation (T-RECS) II at 150\,MHz \citep{2023MNRAS.524..993B} for comparison.\\ 
The obtained source counts in red circles in Figure \ref{fig:Fig8}, matched well with the previous surveys and simulations before sub-mJy flux density sources. This may be primarily due to the limited number of sources below 1\,mJy after selecting sources above 5$\sigma$. This limits us to investigate the results from the LoTSS survey, where there is a ``drop and bump'' in source counts at flux densities ${\leqslant}$\,2\,mJy \citep{2021A&A...648A...5M} with spectral index ${\alpha}\,{=}\,{−0.7}$. This initial drop in source count and subsequent rise that produces a bump feature is neither predicted by existing simulation models nor observed in available high and low-frequency observations, except in the LoTSS 150MHz survey. The GLEAM survey conducted in frequency range 72\,–\,231\,MHz using MWA \citep{2016MNRAS.459.3314F, 2019PASA...36....4F}, is also limited to a few mJy/beam in sensitivity with a few arc-min resolution to reach the sub-mJy radio population.
%%%%%%%%%%%%%%%%%%%%%%%%%%%%%%%%%%%%%%%%%%%%%%%%%%%%%%%%%%%%%%%%%%%%%%%%%%%%%%

\section{Characterization of Diffuse Synchrotron Emission}
\label{sec:APS}
At low frequencies, the radio sky is dominated by non-thermal diffuse synchrotron radiation from the Milky Way Galaxy and extra-galactic radio sources associated with Active Galactic Nuclei (AGN), supernova remnants, and star-forming galaxies (SFGs). Thus, study of the physical properties of Galactic and extra-galactic foregrounds emission is important for the detection of a redshifted $21$-cm signal from Cosmic Dawn and EoR \citep{2004MNRAS.355.1053D}. DGSE refers to the radio emission produced by high-energy electrons spiralling around magnetic field lines in the Galaxy \citep{1992ARA&A..30..575C}. The study of DGSE in frequency dependence and angular scale introduces the properties of the magnetic field and CRe spectrum \citep{2011A&A...534A..54S, 2013JCAP...03..036D}.
%%%%%%%%%%%
\begin{figure*}
\centering
\begin{minipage}{0.45\textwidth}
    \centering
    \includegraphics[width=\linewidth]{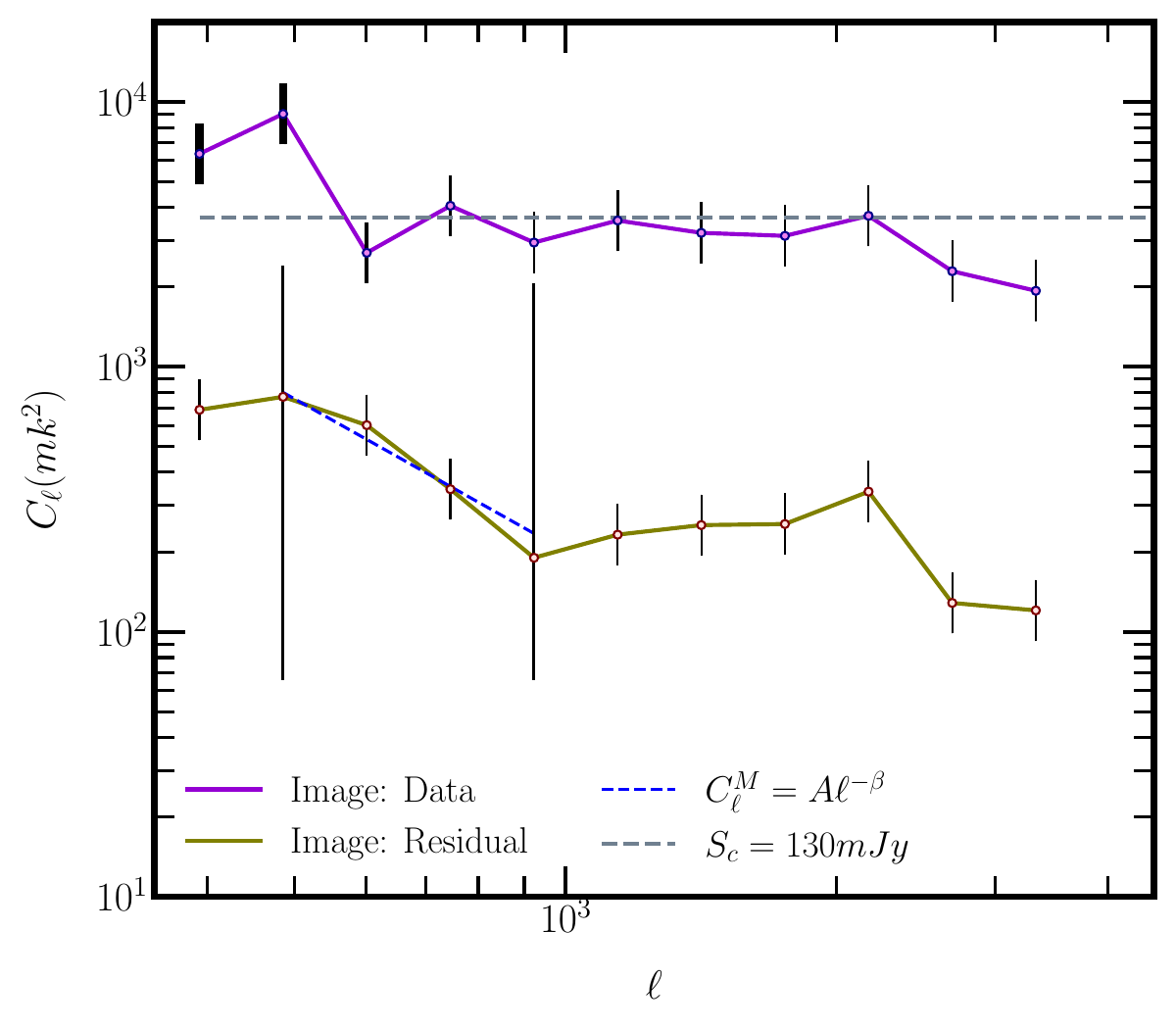}
\end{minipage}\hfill
\begin{minipage}{0.45\textwidth}
    \centering
    \includegraphics[width=\linewidth]{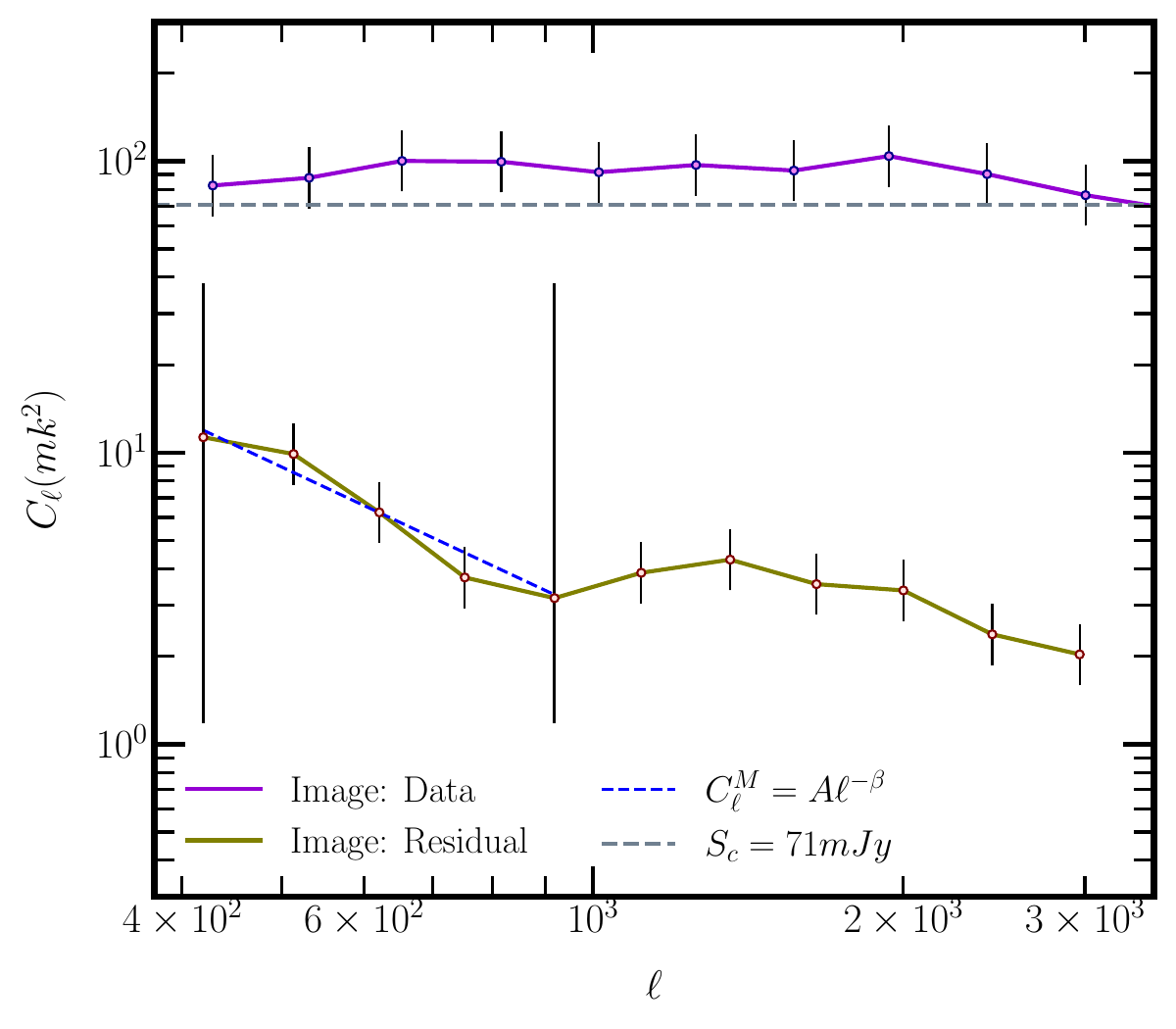}
\end{minipage}\vfill
\caption{i-APS of data containing point-source (in violet) and its corresponding residual (olive) after calibration of uGMRT data at 191\,MHz (left) and at 316\,MHz (right) central frequency. The blue dashed line shows the best-fit power law model used to characterise the DGSE and the black vertical solid lines indicate the ${\ell}_\text{min}-{\ell}_\text{max}$ range used to fit the power law. The dashed line in the upper curve (grey) represents the foreground contribution resulting from discrete point sources where the brightest source has a flux density of $S_c =$ 130\,mJy (left) and of $S_c =$ 71\,mJy (right) for Band-2 and Band-3 respectively. This is based on the model prediction developed by \protect\cite{2008MNRAS.385.2166A}.} 
\label{fig:Fig9}
\end{figure*}
%%%%%%%%%%%%%%%%%%%%%%%%%%%%%%
Here we study the ELAIS-N1 field with uGMRT to characterise DGSE in the frequency range of $120-250$\,MHz and $300-500$\,MHz.
The power spectrum of DSE is commonly described by a power-law function that depends on both frequency and angular scale. This model has been used in several studies \citep{2005ApJ...625..575S, 2008MNRAS.385.2166A}, and can be expressed mathematically as
%%%%%%%%%%%%%%%%%%%%%%%%%%%%%%%%%%%%%%%%%%%%%%%%%%%%%%%%%%%%%%%%%%%%%%%%%%%%%%%%%%%%%%
\begin{equation}
\label{eqn:6}
\begin{aligned}
C_{\ell}(\nu) = A \left( \frac{\ell}{{\ell}_0} \right)^{-\beta} \left( \frac{\nu}{{\nu}_0} \right)^{-2\alpha}
\end{aligned}
\end{equation}
%%%%%%%%%%%%%%%%%%%%%%%%%%%%%%%%%%%%%%%%%%%%%%%%%%%%%%%%%%%%%%%%%%%%%%%%%%%%%%%%%%%%%%%
where ${\beta}$ representing the power-law index and ${\alpha}$ representing the mean spectral index.
The study of diffuse emission using a radio interferometer has been the subject of several investigations aimed at measuring the APS in different frequency ranges \citep{2012MNRAS.426.3295G, 2013A&A...558A..72I, 2017MNRAS.470L..11C, 2020MNRAS.495.4071M, 2022A&A...662A..97G}. \cite{2019MNRAS.490..243C} studied DGSE at 400\,MHz with uGMRT and resulted in a broken power law with a possible break frequency at 405\,MHz (${\nu}_\text{break}$).

\subsection{Image-based Estimator vs Visibility-based Estimator} \label{sec:IME}
%%%%%%%%%%%%%%%%%%%%%%%%%%%%%%%%%%%%%%%%%%%%%%%%%%%%%%%%%%%%%%%%%%%%%%%%%%%%%%
Two different classes of estimators are used to estimate the two-point statistics of the sky brightness distribution. The first class, known as image-based estimators (Hereafter i-APS), involves using the reconstructed image to calculate either the structure or auto-correlation function in the image plane or estimate the power spectrum in its Fourier conjugate plane \citep{2010ApJ...724..526D, 2019RAA....19...60D, 2022A&A...662A..97G}. The second class, known as visibility-based estimators \citep{2016MNRAS.463.4093C, 2019MNRAS.490..243C, 2020MNRAS.495.4071M}, involves implementing visibilities directly to estimate the power spectrum of the sky brightness distribution. \\

\begin{figure*}
  \centering
  \begin{minipage}{0.9\textwidth}
    \includegraphics[width=\linewidth]{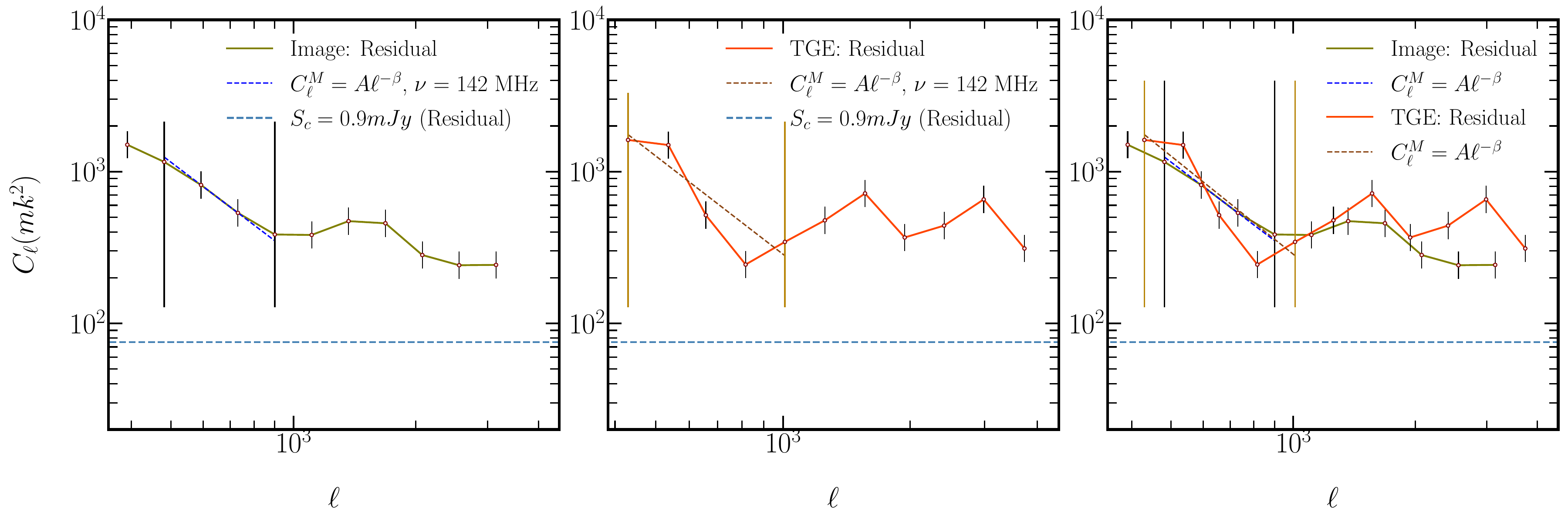}
  \end{minipage}
  \begin{minipage}{0.9\textwidth}
    \includegraphics[width=\linewidth]{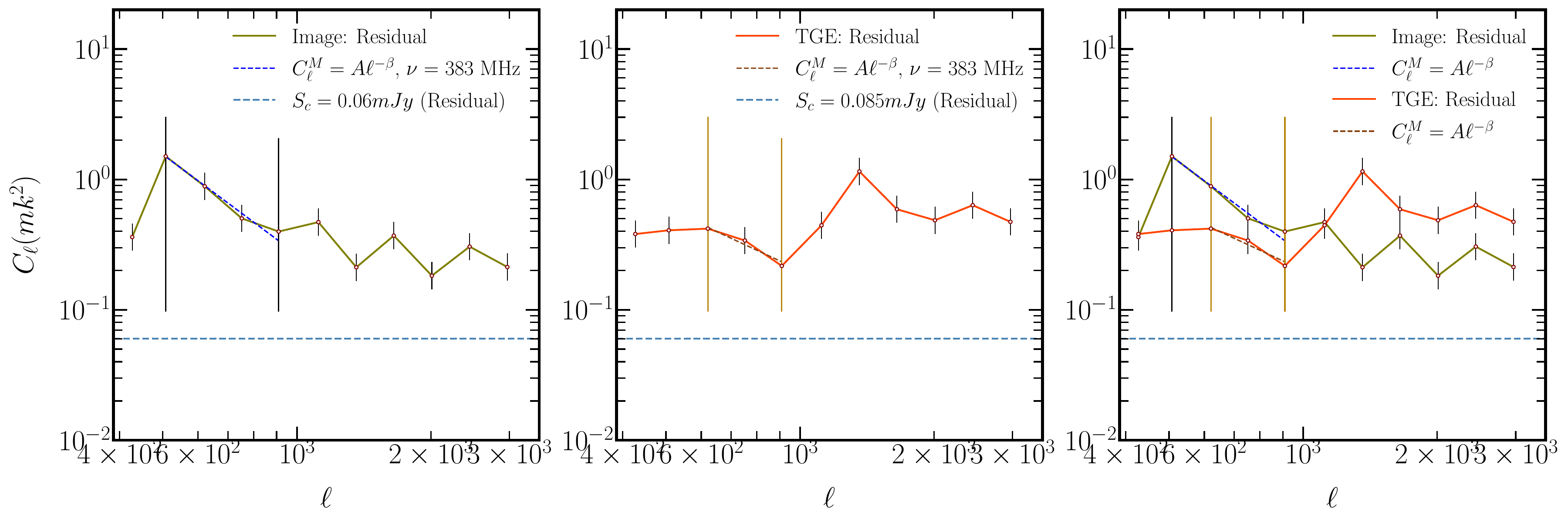}
  \end{minipage}
  \caption{Angular power spectrum ($C_{\ell}$) with $1{-}{\sigma}$ error bars (in olive and orange-red) plotted as a function of angular multipole $\ell$ for the Image-based estimator (i-APS) and the Visibility-based estimator (t-APS) at frequency sub-band in Band-2 and Band-3: 142\,MHz (top) and 383\,MHz (bottom). The vertical solid lines (in black and golden) indicate the range of ${\ell}$ values used to fit a power-law model shown in Table \ref{table:tab7}, while the dashed line (in blue and brown) shows the best-fitting model $C_{\ell} = A(1000/{\ell})^{{\beta}}$. The values of magnitude A and power-law index ${\beta}$ for i-APS and t-APS are given in Table \ref{table:tab7} for Band-2 and Band-3. The dashed curves (sky blue) in the plots represent the model fitting predictions based on \protect\cite{2008MNRAS.385.2166A}}. 
  \label{fig:Fig10}
\end{figure*}

\textbf{Tapered Gridded Estimator :} 
%\label{sec:TGE}
The Tapered Gridded Estimator\footnote{\url{https://github.com/samirchoudhuri/TGE}} \citep[TGE; hereafter t-APS,][]{2016MNRAS.459..151C} technique estimates the APS from gridded visibility data using correlation, effectively reducing the contribution from sidelobes and the outer primary beam by tapering the sky response with a carefully selected window function. 
In this work, to estimate the APS of the DGSE, t-APS is applied to the residual visibilities, which are obtained by subtracting the point source at the 5$\sigma$ threshold from the calibrated visibilities in the reduction procedure mentioned in Section \ref{sec:observation}. t-APS's ability to taper the sky response well before the first null of the primary beam is quantified by the parameter f \citep{2016MNRAS.463.4093C}. A smaller value of f indicates higher tapering of the primary beam, reducing the impact of bright point sources in the outer region of the FoV. 
%%%%%%%%%%%%%%%%%%%%%%%%%%%%%%%%%%%%%%%%%%%%%%%%%%%%%%%%%%%%%%%%%%%%%%%%%%%%%%%%%%%%%%%%%%
\begin{table*}
\centering
\caption{Comparison of angular power spectrum between i-APS and t-APS technique in the frequency range $120-500$\,MHz. In this table, we present 12 sub-bands with their respective central frequencies in this range.}
\label{table:tab7}
\begin{tabular}{cccccccc}
\hline
\multirow{3}{*}{$\nu$ (MHz)} & \multicolumn{3}{c}{i-APS} & \multicolumn{3}{c}{t-APS} \\ [1ex]
%\cline{2-7}
&  A\,(mK$^2$) & $\beta$ & ${\ell}_\text{min}-{\ell}_\text{max}$ & A\,(mK$^2$)  & $\beta$ & ${\ell}_\text{min}-{\ell}_\text{max}$ \\ [0.8ex]
\hline
142  & $283.002\,{\pm}\,63.509$ & $2.028\,{\pm}\,0.265 $ & $390-900$ & $285.641\,{\pm}\,140.413$ & $2.147\,{\pm}\,0.664$ & $429-1010$\\ [0.5ex]
%158 & $243.802\,{\pm}\,48.148$ & $1.891\,{\pm}\,0.457$ & $407-955$ & $260.794\,{\pm}\,63.639$ & $1.901\,{\pm}\,0.464$ & $533-1010$  \\ [0.5ex]
158 & $243.802\,{\pm}\,48.148$ & $1.744\,{\pm}\,0.347$ & $407-955$ & $260.794\,{\pm}\,63.639$ & $1.901\,{\pm}\,0.464$ & $533-1010$  \\ [0.5ex]
191  & $201.547\,{\pm}\,3.615$ & $2.253\,{\pm}\,0.265 $ & $507-955$ & $201.570\,{\pm}\,25.139$ & $1.852\,{\pm}\,0.369$ & $659-1561$ \\ [0.5ex]
%208 & $185.439\,{\pm}\,10.333$ & $2.133\,{\pm}\,0.169$ & $530-1007$ & $112.145\,{\pm}\,10.333$ & $1.713\,{\pm}\,0.169$ & $517-1012$  \\ [0.5ex]
208 & $185.439\,{\pm}\,10.333$ & $1.712\,{\pm}\,0.984$ & $530-1007$ & $112.145\,{\pm}\,10.333$ & $1.713\,{\pm}\,0.169$ & $517-1012$  \\ [0.5ex]
225  & $138.221\,{\pm}\,19.730$ & $2.737\,{\pm}\,0.744$ & $572-1088$ & $183.368\,{\pm}\,16.955$ & $2.476\,{\pm}\,0.489$ & $806-1255$\\ [0.5ex]
241  & $90.197\,{\pm}\,24.168$ & $2.179\,{\pm}\,0.499$ & $640-1206$ & $123.004\,{\pm}\,32.174$ & $1.932\,{\pm}\,0.499$ & $533-1556$\\ [0.5ex]
\hline
316  & $2.186\,{\pm}\,0.461$ & $2.244\,{\pm}\,0.275  $ & $512-918$ & $1.087\,{\pm}\,0.029$ & $1.779\,{\pm}\,0.120$ & $751-1114$ \\ [0.5ex]
350 & $0.366\,{\pm}\,0.026$ & $2.388\,{\pm}\,0.325 $ & $503-917$ & $0.472\,{\pm}\,0.039$ & $1.888\,{\pm}\,0.464$ & $621-755$  \\ [0.5ex]
383  & $0.266\,{\pm}\,0.030$ & $2.543\,{\pm}\,0.189 $ & $507-909$ & $0.174\,{\pm}\,0.028$ & $2.353\,{\pm}\,0.366$ & $620-909$\\ [0.5ex]
416  & $0.102\,{\pm}\,0.020$ & $1.716\,{\pm}\,0.557$ & $483-919$ & $0.144\,{\pm}\,0.020$ & $1.562\,{\pm}\,0.341$ & $609-919$ \\ [0.5ex]
450 & $0.038\,{\pm}\,0.005$ & $2.959\,{\pm}\,0.742$ & $514-919$ & $0.328\,{\pm}\,0.019$ & $1.574\,{\pm}\,0.179$ & $1105-1993$  \\ [0.5ex]
483  & $0.133\,{\pm}\,0.021$ & $2.839\,{\pm}\,0.984  $ & $517-1121$ & $0.471\,{\pm}\,0.009$ & $1.935\,{\pm}\,0.071$ & $1121-2001$\\ [0.5ex]
\hline
\end{tabular}
\vspace{1ex}
\parbox{0.85\textwidth}{\footnotesize
The parameters $A$ and $\beta$ correspond to the amplitude and spectral index in the power law, respectively. The ${\ell}_{\text{min}}$ and ${\ell}_{\text{max}}$ represent the multipole range in which GDSE dominates.}
\end{table*}
%%%%%%%%%%%%%%%%%%%%%%%%%%%%%%%%%%%%%%%%%%%%%%%%%%%%%%%%%%%%%%%%%%%%%%%%%%%%%%%%%%%%%%%%
However, as the tapering increases, the information on low multipoles ($\ell$) decreases. Here, we utilized a tapering parameter of f = 0.5 with a $U_{min}$ threshold of 60 $\lambda$, corresponding to a ${\ell}_{min} = $\,377 for the uGMRT baseline. These parameters were used for both the estimators, as shown in the figures for Band-2 and Band-3, respectively. \\

\textbf{Image based Estimator :} 
%\label{sec:i-APS}
This work demonstrates that for low-frequency observations with diffuse emission, using the reconstructed residual image from calibrated visibility data can enable the estimation of both the large and small-scale distribution of the intensity, allowing for the tracing of $C_{\ell}$ values at low multipole (${\ell}$) which is not possible with visibility-based methods like t-APS due to tapering.\\
The diffuse emission in the image was modelled using WSClean as mentioned in Section~\ref{sec:imaging}. We utilised the APS to characterise the behaviour of the diffuse emission by fitting it to a power-law model of the form $C_{\ell} = A(1000/{\ell})^{{\beta}}$. 
We utilised the publicly accessible LOFAR pipeline\footnote{\url{https://gitlab.com/flomertens/ps_eor}} to generate an i-APS \citep[in detail:][]{2022MNRAS.509..114O} at low-frequency data obtained from uGMRT.
We repeated the imaging process described in Section \ref{sec:observation} using the same parameters for Band-2 and Band-3.
In Figure \ref{fig:Fig9}, we have used the i-APS technique; the upper curves (violet) illustrate $C_{\ell}$ values with $1-{\sigma}$ error bars prior to the subtraction of point sources. We have selected sub-bands at 191\,MHz and 316\,MHz, the central frequency of Band-2 (left) and Band-3 (right). The measured $C_{\ell}$  values range nearly $10^4$\,mK$^2$ and $10^2$\,mK$^2$ for Band-2 and Band-3, respectively. Here, both of the plots (in violet) appear mostly flat (threshold in grey) and are broadly consistent with the theoretical model prediction \citep{2008MNRAS.385.2166A}. This dominant signal in the sky originates from the Poisson fluctuations of the point source distribution. We extracted the calibrated visibilities and subtracted the point source contribution to create a residual primary beam corrected image. Using these residual images, we estimated the APS and fitted the power-law ($C_{\ell} = A(1000/{\ell})^{{\beta}}$), shown in Figures \ref{fig:Fig9} and \ref{fig:Fig10}, for their respective ${\ell}_\text{min}-{\ell}_\text{max}$ range where power-law behaviour of DGSE dominates. However, at $\ell \gtrsim 1000$, the Poisson contribution from residual point sources, instrumentation noise and calibration/beam effects dominates the DGSE and results in flattening of APS \citep{2012MNRAS.426.3295G, 2017MNRAS.470L..11C, 2019MNRAS.487.4102C}. In Figures \ref{fig:Fig9} and \ref{fig:Fig10}, the power law fit is represented with the blue and brown dashed lines for the selected sub-bands for i-APS and t-APS, respectively. The ${\ell}_\text{min}-{\ell}_\text{max}$ range has been shown by the vertical solid lines, black for i-APS (in Figures \ref{fig:Fig9} and \ref{fig:Fig10}) and golden for t-APS (in Figure \ref{fig:Fig10}). Removing point sources reduces $C_{\ell}$ values for large ${\ell}$ scales. In Figure \ref{fig:Fig9}, the residual $C_{\ell}$ data shows a power-law behaviour with values around $10^2$\,mK$^2$ and $1$\,mK$^2$ for ${\ell} \lesssim 10^3$ at 191\,MHz and 316\,MHz, respectively. The dashed sky blue line in both of the figures indicates the value of $C_{\ell}$ resulting from the residual point sources as predicted by \cite{2008MNRAS.385.2166A}, below a threshold of $5{\sigma}$.
Using residual images for other sub-bands, we estimated the APS as shown in Figure \ref{fig:Fig10} for the Band-2 and Band-3. In this figure, we have fit the power-law  $C_{\ell} = A(1000/{\ell})^{{\beta}}$ with two sub-band at 142\,MHz (Band-2, upper panel) and at 383\,MHz (Band-3, lower panel) central frequency for comparison of both estimation techniques (i.e. i-APS and t-APS). In the figure, the i-APS (left), t-APS (middle), and their comparison (right) are shown. 
The best-fitting values of the amplitude A and power-law index ${\beta}$ for all sub-bands are listed in the Table \ref{table:tab7} for Band-2 (top panel) and Band-3 (bottom panel). The range of multipole values (${\ell}$) used for fitting is also provided in Table~\ref{table:tab7}. \\
From both the APS of Band-2 and Band-3 (Table \ref{table:tab7}), it is evident that in this work diffuse emission follows a power law over the selected multipole range ($400 \lesssim \ell \lesssim 1000$) probed by the fit, and its power sharply decreases with increasing frequency. The diffuse emission is dominating on large-scales (low multipole $\ell$) and has less power on small-scales (large multipole $\ell$), making i-APS more intriguing for detailed characterisation of DGSE. The value of $\beta$ sharply increases in Band-2, making it steeper, especially near 225\,MHz (Table \ref{table:tab7}) for both the i-APS and t-APS. For detailed analysis, we have fitted a multi-frequency angular power spectrum (MF-APS) in terms of single and broken power spectrum as discussed in section \ref{sec:broken-PS}.

\section{Results and Discussion : Broken Power Law}
\label{sec:broken-PS}
A spectral energy distribution with a broken power law occurs when the spectral index changes at a given frequency, called the ``break frequency''. This break often arises from the ageing of synchrotron-emitting electrons, which lose their energy over time through synchrotron radiation.
%%%%%%%%%%%%%%%%%%%%%%%%%%%%%%%%%%%%%%%%%%%%%%%%%%%%%%%%%%%%%%%%%%%%%%%%%%%%%%%%%%%%%%%%%%%%%%%%%
\begin{figure*}
\centering
\begin{minipage}{0.45\textwidth}
    \centering
    \includegraphics[width=\linewidth]{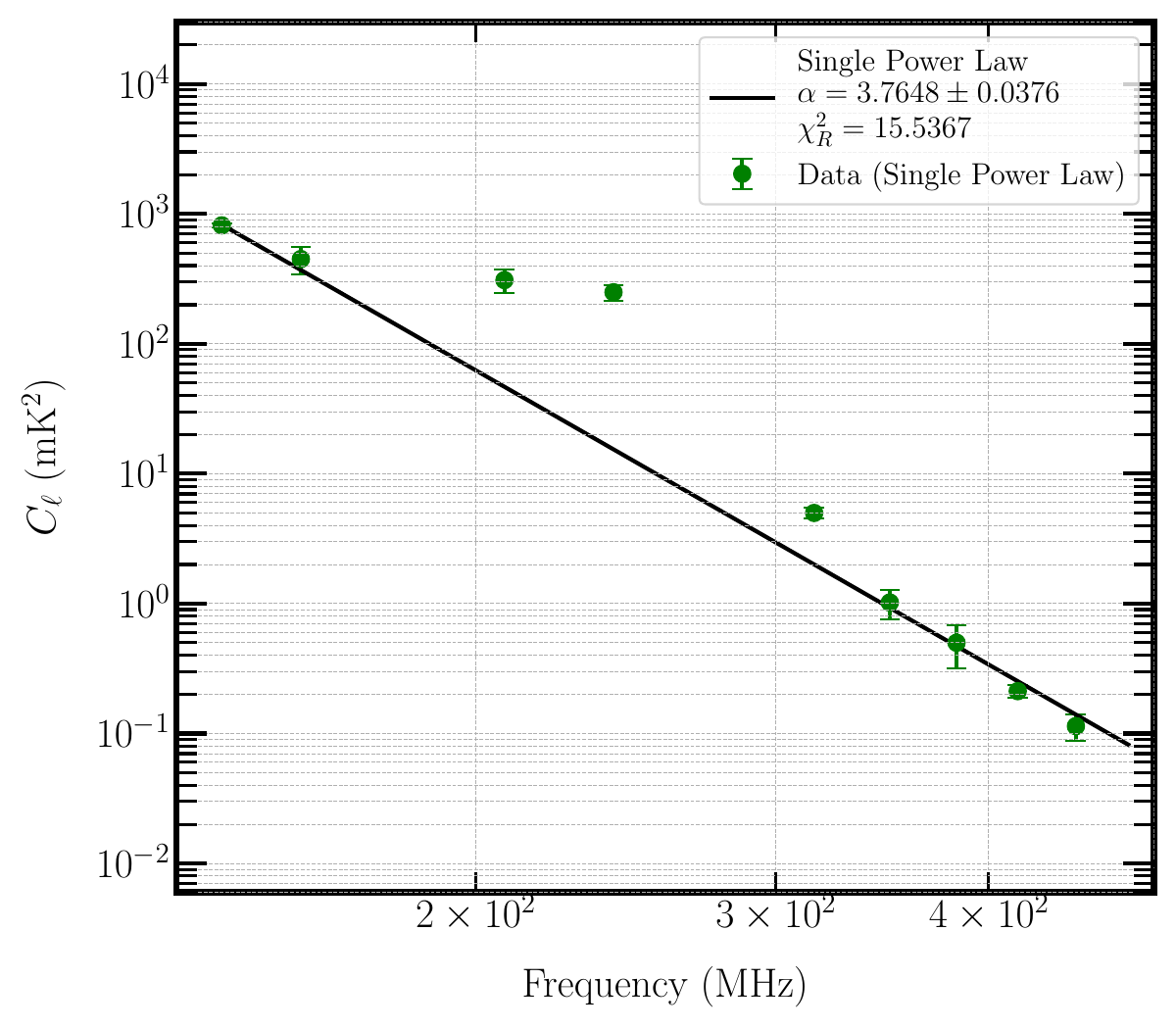}
\end{minipage}\hfill
\begin{minipage}{0.45\textwidth}
    \centering
    \includegraphics[width=\linewidth]{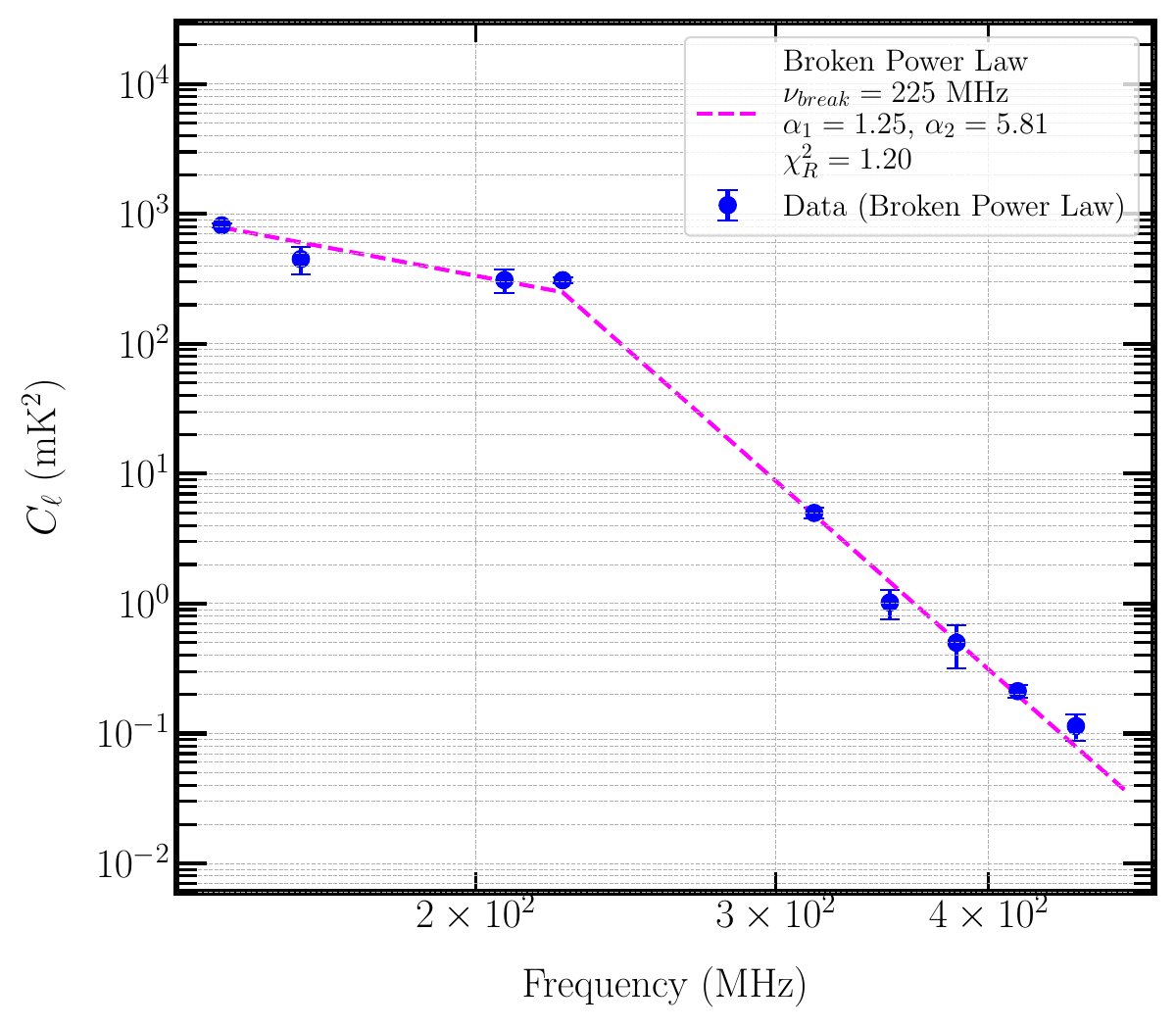}
\end{minipage}\vfill
\begin{minipage}{0.45\textwidth}
    \centering
    \includegraphics[width=\linewidth]{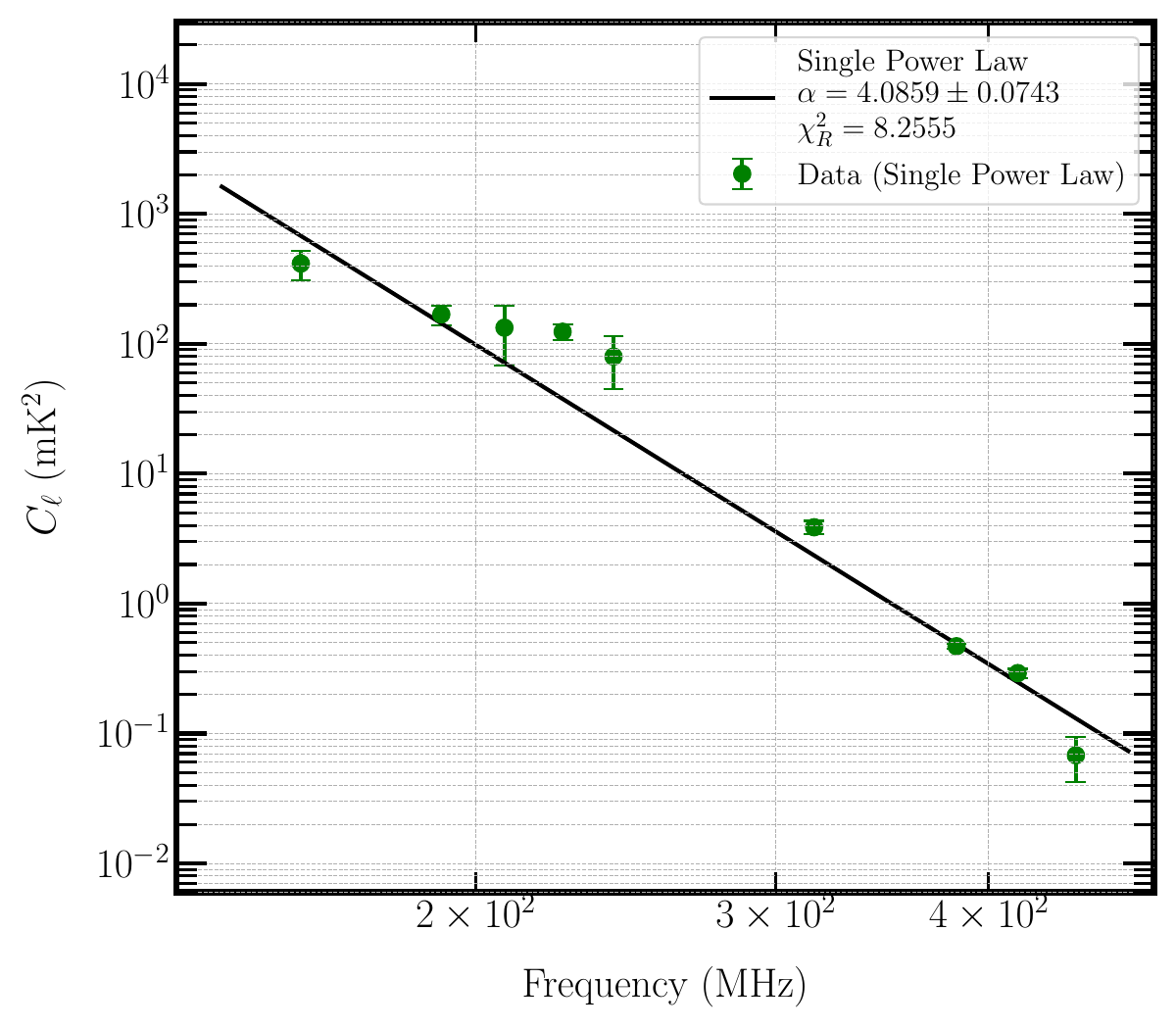}
\end{minipage}\hfill
\begin{minipage}{0.45\textwidth}
    \centering
    \includegraphics[width=\linewidth]{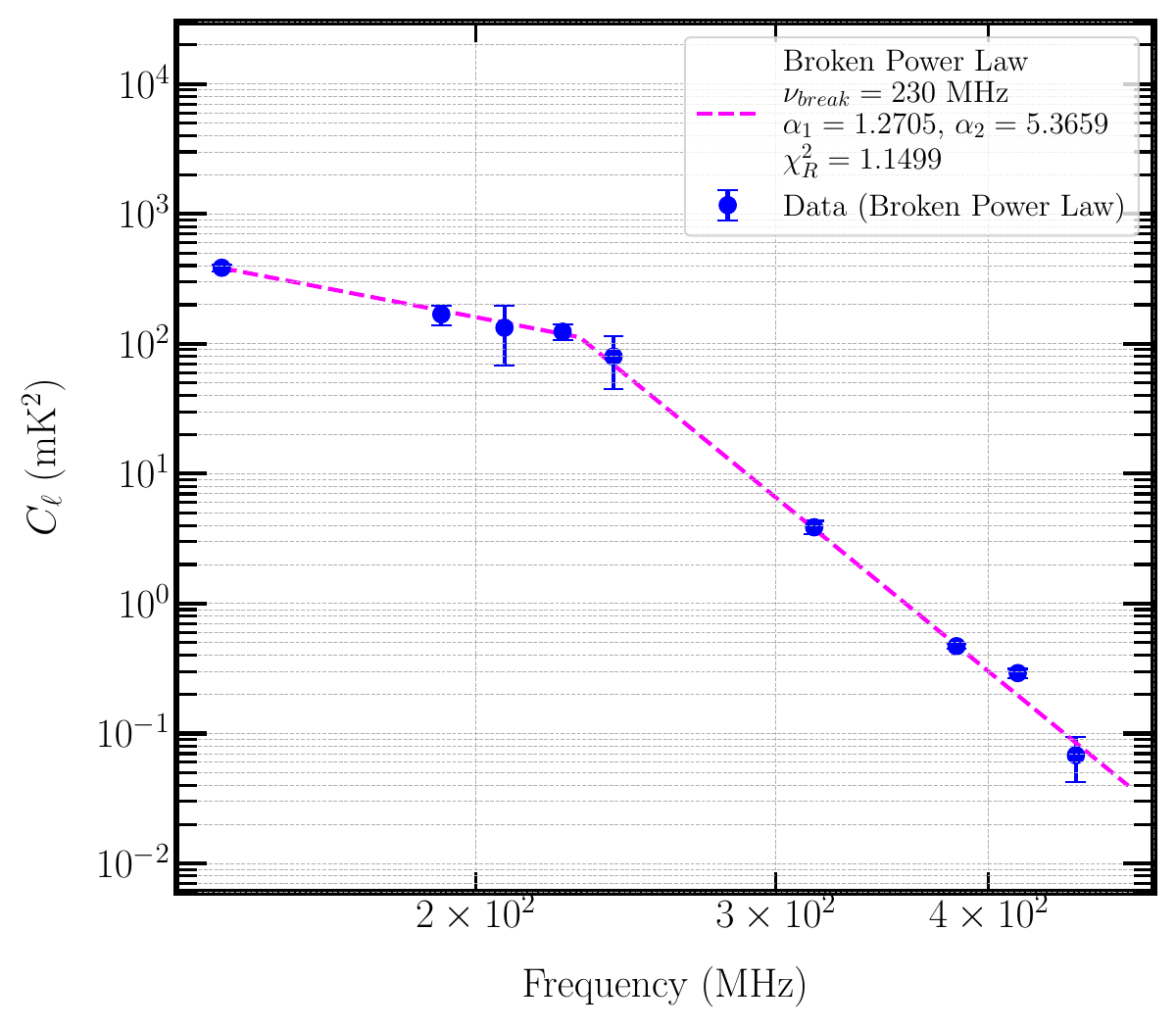}
\end{minipage}
\caption{Angular Power Spectrum of the DGSE as a function of frequency ($\nu$) using i-APS technique is shown for the entire bandwidth in Band-2 ($120-250$\,MHz) and Band-3 ($300-500$\,MHz), modelled using both single and broken power laws. The $C_{\ell}$ values are normalized at two different multipoles: $\ell$ = 600 (top panel) and $\ell$ = 900 (bottom panel). The best-fitting single power-law model ($C_{\ell} = A{\nu}^{-2{\alpha}}$) is shown as a solid black line with the data points in green (top and bottom left panels), and the corresponding mean spectral index ${\alpha}$ is also indicated in figure legends. The broken power-law (Equation \ref{eqn:7}) is shown in the top and bottom right panels, where the best-fitted spectral indices (${\alpha}_{1}$, ${\alpha}_{2}$) are shown with a dashed magenta line and blue data points. The break frequency ${\nu}_\text{break}$, mean spectral indices, and reduced chi-square ${\chi}_R^2$ values for the broken power-law model are provided in Table \ref{table:tab8} and noted in the figure legends. The green and blue points correspond to the best subsets of $C_{\ell}$ values used for fitting the single and broken power-law models, respectively, in each frequency sub-band.}
\label{fig:Fig11}
\end{figure*}
%%%%%%%%%%%%%%%%%%%%%%%%%%%%%%%%%%%%%%%%%%%%%%%%%%%%%%%%%%%%%%%%%%%%%%%%%%%%%%
The synchrotron spectrum is typically flatter below the spectral break frequency ($\nu_\text{break}$), indicating a population of electrons with minimal energy loss. Above $\nu_\text{break}$, the spectrum steepens due to radiative losses, particularly synchrotron and inverse Compton cooling, suggesting a transition in the energy distribution of the emitting electrons \citep{1962AZh....39..393K, 1965ARA&A...3..297G, 2011A&A...534A..54S}. \\
The ELAIS-N1 is located at significantly high galactic latitude ($b_\text{ELAIS-N1} = +44^{\circ}$) and the residuals (images and visibilities) of the field consist of galactic synchrotron emission and unresolved extragalactic sources. 
While these extragalactic sources, like faint radio galaxies or galaxy clusters, are part of the overall foreground, they are generally too sparse or compact to significantly alter the large-scale APS of the DGSE \citep{2011MNRAS.415.3641V, 2021MNRAS.505.4178V}. In contrast, the Galactic structures such as spur or loops \citep{2013JCAP...06..041M, 2015MNRAS.451.4311R} can contribute, causing a localised peak in $\beta$. Additionally, a change in the energy spectrum of CRe results in synchrotron emission. The spectral index ${\alpha}$ is related to the energy distribution of CRe, described by ${j_e}(E) = E^{-s}$ \citep[where ${\alpha} = (s-3)/2$;][]{1965ARA&A...3..297G}. \cite{1998ApJ...507..327P, 2011A&A...534A..54S, 2012JCAP...01..049B, 2013JCAP...03..036D} have discussed that there is a significant chance of a break in the energy spectrum of electron. \\
In this work, after analysing the DGSE in terms of angular scale, we have studied its properties in the frequency domain (Equation \ref{eqn:6}). We have studied the wideband data of Band-2 and Band-3 in the ELAIS-N1 field to investigate the broken power-law behaviour of DSE within the frequency range of $120-250$\,MHz and $300-500$\,MHz. We utilised the C$_{\ell}$ for i-APS across each sub-band at the central frequencies specified in Table \ref{table:tab7}. It also highlights the $\ell$ range where the DGSE exhibits a dominant power-law behaviour in both bands. Here, we have fitted the multi-frequency APS (MF-APS) of DGSE to study its spectral properties in the form of $C_{\ell} = A{\nu}^{-2{\alpha}}$ for the whole frequency range in Figure \ref{fig:Fig11}. We have chosen three multipole values, ${\ell}_0 =$ 600, 700 and 900, to estimate the best-fit parameters (A and ${\beta}$) for all sub-bands using the formula $C_{\ell} = A({\ell}_0/{\ell})^{{\beta}}$ and then perform the fitting. These ${\ell}_0$ values lie within the ${\ell}_\text{min}-{\ell}_\text{max}$ range where the DGSE dominates. We find that the spectral break remains consistent for all three choices of ${\ell}_0$. In Figure \ref{fig:Fig11}, the single power law has been shown with data points of green circles (top and bottom left panel), while the broken power law has been represented with data points of the blue circle (top and bottom right panel). The mean spectral index ${\alpha}$ for the entire band ($120-500$\,MHz) resulting in ${\chi}_R^2$ which is too high for the single power law for given normalised $C_{\ell}$ values (mentioned in the figure legends). Thus, we have investigated the possibility of a broken power law. We have fitted the break at the ${\nu}_\text{break}$ frequency for three different $\ell$ values. The model for broken power law is as follows:

%%%%%%%%%%%%%%%%%%%%%%%%%%%%%%%%%%%%%%%%%%%%%%%%%%%%%%%%%%%%%%%%%%%%%%%%%%%%%%%%%%%
\begin{equation}
\label{eqn:7}
\begin{aligned}
C_{\ell}(\nu) =
\begin{cases} 
A \left( \frac{\nu}{\nu_{\text{break}}} \right)^{-2\alpha_1} & \text{if } \nu \leq \nu_{\text{break}} \\
A \left( \frac{\nu}{\nu_{\text{break}}} \right)^{-2\alpha_2} & \text{if } \nu \geq \nu_{\text{break}}
\end{cases}
\end{aligned}
\end{equation}

%%%%%%%%%%%%%%%%%%%%%%%%%%%%%%%%%%%%%%%%%%%%%%%%%%%%%%%%%%%%%%%%%%%%%%%%%%%%%%%%%%%%%%%
\begin{table}
\caption{Best fitted multi-frequency angular power spectrum (MF-APS) in the form of a broken power-law of the whole band from $120-500$\,MHz (Band-2 and Band-3). The fitted mean spectral indices, break frequency and ${\chi}_R^2$ values of broken power law at three different multipoles ($\ell$) is provided in this table.}
\centering
\begin{tabular}{ccccc}
\hline
{${\nu}_\text{break}$} & & Broken Power-law  \\ 
(MHz) & ${\alpha}_1$ & ${\alpha}_2$ & ${\chi}_R^2$ \\
% ~& dd/mm/yy & (hours) & (\%) & (hours)  \\ 
\hline\hline
{225\,{$\pm$}\,8}$_{({\ell} = 600)}$  & $1.23\,{\pm}\,0.26$ & $ 5.82\,{\pm}\,0.26$ & 1.20\\ [0.5ex]
{233\,{$\pm$}\,6}$_{({\ell} = 700)}$  &  $1.05\,{\pm}\,0.30$ & $ 6.44\,{\pm}\,0.24$ & 1.55\\ [0.5ex]
{230\,{$\pm$}\,8}$_{({\ell} = 900)}$  &  $1.27\,{\pm}\,0.15$ & $ 5.37\,{\pm}\,0.27$ & 1.15\\ [0.5ex]
\hline
\end{tabular}
\label{table:tab8}
\end{table}
%%%%%%%%%%%%%%%%%%%%%%%%%%%%%%%%%%%%%%%%%%%%%%%%%%%%%%%%%%%%%%%%%%%%%%%%%%%%%%%%%%%%%%%
where ${\alpha_{1}}$ and ${\alpha_{2}}$ are the spectral indices before and after the break. The best-fitted spectral index values for the broken power law are mentioned in Table \ref{table:tab8} with their respective $\ell$ values. 
The high spectral curvature ${\Delta}{\alpha} = ({\alpha_{1} - \alpha_{2}})_{\ell}$ indicates that non-thermal synchrotron emission is in dominance with potent radiative losses \citep{2003A&A...404..133S}. Above the break ($\alpha_2$), the measured spectral index value falls between the Kardashev-Pacholczyk \citep[KP,][]{1962AZh....39..393K, 1970ranp.book.....P} and Jaffe-Perola \cite[JP,][]{1973A&A....26..423J} models \citep{1985ApJ...291...52M}. Here, we have calculated the synchrotron age $t_\text{syn} =$ 107, 105 and 106\,Myr \citep{1991ApJ...383..554C} using the assumed magnetic field B $=$ 10$\mu$G at three $\ell$ values, respectively. Thus, the synchrotron age at $120-500$\,MHz is $106\,{\pm}\,1$\,Myr with a break frequency $230\,{\pm}\,5$\,MHz. \\
The location of a spectral break or turnover in the APS of the DGSE is primarily influenced by intrinsic properties of the synchrotron-emitting medium, such as the energy distribution of CRe and variations in magnetic field structure \citep{1998ApJ...505..473P, 2010MNRAS.409.1647J, 2013JCAP...06..041M}. A low-energy cutoff in the CRe population can lead to spectral curvature at low frequencies \citep{1985ApJ...291...52M, 2013MNRAS.435.3353H, 2017ApJ...838...68K}. This result also indicates a weak magnetic field apart from low energy cutoff, since lower magnetic field strengths reduce radiative losses, shifting the break frequency downward and implying slower synchrotron ageing of the electron population \citep{1993MNRAS.261...57T,2016MNRAS.463.3143M}. In previous work using only uGMRT Band-3 ($300-500$\,MHz) data \citep{2019MNRAS.490..243C}, the limited frequency range could not reveal spectral behaviour of DGSE below 300\,MHz. The extended frequency coverage from Band-2 ($120-250$\,MHz) and Band-3 ($300-500$\,MHz) allows us to probe the synchrotron spectrum across the region where spectral curvature or break is found, thereby improving our sensitivity to deviations from a single power-law behavior. This low-frequency access allows the modelling to directly trace the behaviour of spectral curvature and flattening associated with slow ageing of synchrotron, resulting in a break at break frequency ${\nu_\text{break}} = 230\,{\pm}\,5$\,MHz. \\
To rule out the possibility that instrumental systematics between distinct bands are responsible for the different spectral slopes, we selected a few bright point sources located near the phase center. We examined whether their flux densities exhibit the same spectral break observed in the DGSE. If bandpass gain errors were present in different subbands, such effects would be imprinted in both the point-source fluxes and the DGSE. However, our analysis shows that these bright sources do not display any such spectral behaviour. We therefore conclude that no significant instrumental systematics are present, and the observed spectral break is the result of DGSE behaviour.

\section{Conclusions} \label{sec:conclusion}
This paper presents results from the observations of the ELAIS-N1 field at 183\,MHz, using the uGMRT during observing cycles 34 and 41. We discuss the analysis for calibration, imaging, catalogue and source count in detail. These observations cover an area of $5.86\,\text{deg}^2$. We have performed DD calibration on the 32 hours data and an image of $3.5^{\circ}\,{\times}\,3.5^{\circ}$ FoV was produced. This image reaches an off-source RMS noise of 237\,$\mu\text{Jy/beam}$ with a resolution of $11.45\arcsec$ and a dynamic range of ${\sim}\,5700$. A total of 1027 sources above $5\sigma_{\text{rms}}$ were identified to produce the final field catalogue. Low-frequency observations inherently face challenges such as RFI, systematics, and primary beam corrections. We compared the results with previous observations from the same sky region to ensure flux and positional accuracy. The measured flux values and positions of the detected sources are well-matched with those obtained in previous observations with GMRT and with recent LoTSS 150\,MHz observations \citep{2021A&A...648A...2S}. The median of the flux density ratio with LOFAR is found to be 1.07$^{+0.50}_{-0.31}$ whereas positional accuracy is  $\delta_\text{RA,median} = $ 0.637$^{+1.74}_{-1.47}$ arcsec and $\delta_\text{DEC,median} = $0.255$^{+0.85}_{-0.35}$ arcsec. Euclidean normalised differential source counts for the catalogue sources were also determined, correcting various errors and biases. The final source counts are consistent with simulations and previous observations for the same field and other sky patches. We have included a comparison with the LOFAR observations of the ELAIS-N1 field at 150\,MHz \citep{2021A&A...648A...2S} and found that our work does not indicate any ``drop'' or ``bump''  near sub-mJy levels as seen in LOFAR. This is mainly due to the lower number of sources in our catalogue compared to LOFAR at 150\,MHz, which is a direct consequence of our limited observation time, resulting in limited sensitivity. However, these are the first uGMRT observations at low frequency with wideband data of frequency range $120-250$\,MHz compared to LOFAR and provide complementary results to the LOFAR survey in the ELAIS-N1 field. In future work, extended uGMRT observations will enhance sensitivity, allowing for a deeper investigation of the ``drop and bump'' feature at sub-mJy flux density. \\
Furthermore, we have characterised foregrounds, which pose a significant challenge in recovering the redshifted $21$-cm signal. The foregrounds exhibit spatial fluctuations across the sky, which necessitate low-frequency observations to generate accurate and reliable foreground models. Using uGMRT Band-2 ($120-250$\,MHz) and Band-3 ($300-500$\,MHz), we have investigated the foreground smoothness in terms of power spectrum using image- and visibility-based estimators (i-APS and t-APS). We determined the APS, $C_{\ell}$ as a function of $\ell$ for sub-bands of both the Band-2 and Band-3 of the ELAIS-N1 field. We found that $C_{\ell}$ values at ${\ell}_0 = 1000$, ranged between $90-300\,$mK$^2$ for sub-bands in Band-2 while from $2.186-0.038$\,mK$^2$ for i-APS and from $1.087-0.144$\,mK$^2$ for t-APS in Band-3 (Table \ref{table:tab7}), which shows an increase in amplitude as we approached lower frequencies with both the estimators. We fit a power law of the form $A{\ell}^{- {\beta}}$ to the obtained APS and found that the best-fitted values for the power-law index (${\beta}$) in the range $1.7-2.4$ for t-APS while $1.9-2.1$ for i-APS for Band-2 (Table \ref{table:tab7}) except at 225\,MHz. 
For Band-3 the value of ${\beta}$ is $1.5-2.3$ for t-APS and $1.7-3.0$ for i-APS, respectively. In addition, we fitted a broken power law of DGSE in terms of MF-APS and found breaks near Band-2 (Figure \ref{fig:Fig11} and Table \ref{table:tab8}) at 225, 233 and 230\,MHz (${\nu}_\text{break}$) for three different ${\ell}$ with an accuracy of ${\chi}_R^2 = 1.20, 1.55\,\text{and}\,1.15$. \\
Combined uGMRT Band-2 and Band-3 (frequency range $120-500$\,MHz), $C_{\ell}$ values with respect to three different multipoles show a prominent break in the APS of DGSE at break frequency $230\,{\pm}\,5$\,MHz. This result supersedes previous results of a break at 405\,MHz \citep{2019MNRAS.490..243C} from Band-3 ($300-500$\,MHz), which arose from fitting a spectrum with missing data points to reveal the spectral curvature at lower frequencies. The break in this work shows similarity with radio source spectra \citep{2016MNRAS.463.3143M}, spectral energy distribution of radio galaxy in the frequency range $153-231$\,MHz with turnover at 230\,MHz \citep{2017ApJ...838...68K}, all-sky low-frequency maps \citep{2008MNRAS.388..247D} and modelling of CRe populations \citep{1998ApJ...505..473P, 2011A&A...534A..54S}.

\section*{Acknowledgements}

RS, AD and SM acknowledge the use of computing infrastructure for this work, which is hosted at the DAASE, IIT Indore and was procured through the funding via, Department of Science and Technology, Government of India sponsored DST-FIST grant no. SR/FST/PSII/2021/162 (C) awarded to DAASE, IIT Indore. NR acknowledges support from the United States-India Educational Foundation through the Fulbright Program. The authors thank the staff of GMRT, which is facilitated by the National Center for Radio Astrophysics (NCRA) of the Tata Institute of Fundamental Research (TIFR) to provide the observations. RS acknowledges the support of CSIR-UGC for the UGC-JRF fellowship. AD and SM acknowledge financial support through the project titled ``Observing the Cosmic Dawn in Multicolor using Next Generation Telescopes'' funded by the Science and Engineering Research Board (SERB), Department of Science and Technology, Government of India through the Core Research Grant No. CRG/2021/004025. This research work used APLpy \footnote{\url{https://pypi.org/project/aplpy/}}, an open source graphing package for PYTHON \citep{2012ascl.soft08017R, 2019zndo...2567476R, 2022ApJ...935..167A}, and Astropy\footnote{\url{https://www.astropy.org/}}, which is a PYTHON package for Astronomy \citep{2013A&A...558A..33A, 2018AJ....156..123A}. 

%%%%%%%%%%%%%%%%%%%%%%%%%%%%%%%%%%%%%%%%%%%%%%%%%%
\section*{Data Availability}
The complete catalogue (full version of Table \ref{table:tab3}, including all columns with their respective uncertainties) of 32 hours observations of uGMRT Band-2 is available in the electronic version of this paper.

%%%%%%%%%%%%%%%%%%%% REFERENCES %%%%%%%%%%%%%%%%%%

% The best way to enter references is to use BibTeX:

\bibliographystyle{mnras}
\bibliography{example} % if your bibtex file is called example.bib

\begin{thebibliography}{}
\makeatletter
\relax
\def\mn@urlcharsother{\let\do\@makeother \do\$\do\&\do\#\do\^\do\_\do\%\do\~}
\def\mn@doi{\begingroup\mn@urlcharsother \@ifnextchar [ {\mn@doi@} {\mn@doi@[]}}
\def\mn@doi@[#1]#2{\def\@tempa{#1}\ifx\@tempa\@empty \href {http://dx.doi.org/#2} {doi:#2}\else \href {http://dx.doi.org/#2} {#1}\fi \endgroup}
\def\mn@eprint#1#2{\mn@eprint@#1:#2::\@nil}
\def\mn@eprint@arXiv#1{\href {http://arxiv.org/abs/#1} {{\tt arXiv:#1}}}
\def\mn@eprint@dblp#1{\href {http://dblp.uni-trier.de/rec/bibtex/#1.xml} {dblp:#1}}
\def\mn@eprint@#1:#2:#3:#4\@nil{\def\@tempa {#1}\def\@tempb {#2}\def\@tempc {#3}\ifx \@tempc \@empty \let \@tempc \@tempb \let \@tempb \@tempa \fi \ifx \@tempb \@empty \def\@tempb {arXiv}\fi \@ifundefined {mn@eprint@\@tempb}{\@tempb:\@tempc}{\expandafter \expandafter \csname mn@eprint@\@tempb\endcsname \expandafter{\@tempc}}}

\bibitem[\protect\citeauthoryear{{Al Yazeedi}, {Katkov}, {Gelfand}, {Wylezalek}, {Zakamska}  \& {Liu}}{{Al Yazeedi} et~al.}{2021}]{2021ApJ...916..102A}
{Al Yazeedi} A.,  {Katkov} I.~Y.,  {Gelfand} J.~D.,  {Wylezalek} D.,  {Zakamska} N.~L.,   {Liu} W.,  2021, \mn@doi [\apj] {10.3847/1538-4357/abf5e1}, \href {https://ui.adsabs.harvard.edu/abs/2021ApJ...916..102A} {916, 102}

\bibitem[\protect\citeauthoryear{{Ali}, {Bharadwaj}  \& {Chengalur}}{{Ali} et~al.}{2008}]{2008MNRAS.385.2166A}
{Ali} S.~S.,  {Bharadwaj} S.,   {Chengalur} J.~N.,  2008, \mn@doi [\mnras] {10.1111/j.1365-2966.2008.12984.x}, \href {https://ui.adsabs.harvard.edu/abs/2008MNRAS.385.2166A} {385, 2166}

\bibitem[\protect\citeauthoryear{{Astropy Collaboration} et~al.,}{{Astropy Collaboration} et~al.}{2013}]{2013A&A...558A..33A}
{Astropy Collaboration} et~al., 2013, \mn@doi [\aap] {10.1051/0004-6361/201322068}, \href {https://ui.adsabs.harvard.edu/abs/2013A&A...558A..33A} {558, A33}

\bibitem[\protect\citeauthoryear{{Astropy Collaboration} et~al.,}{{Astropy Collaboration} et~al.}{2018}]{2018AJ....156..123A}
{Astropy Collaboration} et~al., 2018, \mn@doi [\aj] {10.3847/1538-3881/aabc4f}, \href {https://ui.adsabs.harvard.edu/abs/2018AJ....156..123A} {156, 123}

\bibitem[\protect\citeauthoryear{{Astropy Collaboration} et~al.,}{{Astropy Collaboration} et~al.}{2022}]{2022ApJ...935..167A}
{Astropy Collaboration} et~al., 2022, \mn@doi [\apj] {10.3847/1538-4357/ac7c74}, \href {https://ui.adsabs.harvard.edu/abs/2022ApJ...935..167A} {935, 167}

\bibitem[\protect\citeauthoryear{{Barry}, {Line}, {Lynch}, {Kriele}  \& {Cook}}{{Barry} et~al.}{2024}]{2024ApJ...964..158B}
{Barry} N.,  {Line} J.~L.~B.,  {Lynch} C.~R.,  {Kriele} M.,   {Cook} J.,  2024, \mn@doi [\apj] {10.3847/1538-4357/ad2e9b}, \href {https://ui.adsabs.harvard.edu/abs/2024ApJ...964..158B} {964, 158}

\bibitem[\protect\citeauthoryear{{Best} et~al.,}{{Best} et~al.}{2023}]{2023MNRAS.523.1729B}
{Best} P.~N.,  et~al., 2023, \mn@doi [\mnras] {10.1093/mnras/stad1308}, \href {https://ui.adsabs.harvard.edu/abs/2023MNRAS.523.1729B} {523, 1729}

\bibitem[\protect\citeauthoryear{{Bonaldi}, {Hartley}, {Ronconi}, {De Zotti}  \& {Bonato}}{{Bonaldi} et~al.}{2023}]{2023MNRAS.524..993B}
{Bonaldi} A.,  {Hartley} P.,  {Ronconi} T.,  {De Zotti} G.,   {Bonato} M.,  2023, \mn@doi [\mnras] {10.1093/mnras/stad1913}, \href {https://ui.adsabs.harvard.edu/abs/2023MNRAS.524..993B} {524, 993}

\bibitem[\protect\citeauthoryear{{Bonzini} et~al.,}{{Bonzini} et~al.}{2015}]{2015MNRAS.453.1079B}
{Bonzini} M.,  et~al., 2015, \mn@doi [\mnras] {10.1093/mnras/stv1675}, \href {https://ui.adsabs.harvard.edu/abs/2015MNRAS.453.1079B} {453, 1079}

\bibitem[\protect\citeauthoryear{{Bowman} et~al.,}{{Bowman} et~al.}{2013}]{2013PASA...30...31B}
{Bowman} J.~D.,  et~al., 2013, \mn@doi [\pasa] {10.1017/pas.2013.009}, \href {https://ui.adsabs.harvard.edu/abs/2013PASA...30...31B} {30, e031}

\bibitem[\protect\citeauthoryear{Bridle \& Greisen}{Bridle \& Greisen}{1994}]{Bridle_Greisen_1994}
Bridle A.~H.,  Greisen E.~W.,  1994, Technical report 87, The NRAO AIPS Project — A Summary.
National Radio Astronomy Observatory

\bibitem[\protect\citeauthoryear{{Briggs}}{{Briggs}}{1995}]{1995PhDT.......238B}
{Briggs} D.~S.,  1995, PhD thesis, New Mexico Institute of Mining and Technology

\bibitem[\protect\citeauthoryear{{Bringmann}, {Donato}  \& {Lineros}}{{Bringmann} et~al.}{2012}]{2012JCAP...01..049B}
{Bringmann} T.,  {Donato} F.,   {Lineros} R.~A.,  2012, \mn@doi [\jcap] {10.1088/1475-7516/2012/01/049}, \href {https://ui.adsabs.harvard.edu/abs/2012JCAP...01..049B} {2012, 049}

\bibitem[\protect\citeauthoryear{{Brunetti} \& {Jones}}{{Brunetti} \& {Jones}}{2014}]{2014IJMPD..2330007B}
{Brunetti} G.,  {Jones} T.~W.,  2014, \mn@doi [International Journal of Modern Physics D] {10.1142/S0218271814300079}, \href {https://ui.adsabs.harvard.edu/abs/2014IJMPD..2330007B} {23, 1430007}

\bibitem[\protect\citeauthoryear{Buch, Naik, Nalawade, Bhatporia, Gupta  \& Ajithkumar}{Buch et~al.}{2019}]{doi:10.1142/S2251171719400063}
Buch K.~D.,  Naik K.,  Nalawade S.,  Bhatporia S.,  Gupta Y.,   Ajithkumar B.,  2019, \mn@doi [Journal of Astronomical Instrumentation] {10.1142/S2251171719400063}, 08, 1940006

\bibitem[\protect\citeauthoryear{{Buch}, {Kale}, {Naik}, {Aragade}, {Muley}, {Kudale}  \& {Ajith Kumar}}{{Buch} et~al.}{2022}]{2022JAI....1150008B}
{Buch} K.~D.,  {Kale} R.,  {Naik} K.~D.,  {Aragade} R.,  {Muley} M.,  {Kudale} S.,   {Ajith Kumar} B.,  2022, \mn@doi [Journal of Astronomical Instrumentation] {10.1142/S2251171722500088}, \href {https://ui.adsabs.harvard.edu/abs/2022JAI....1150008B} {11, 2250008}

\bibitem[\protect\citeauthoryear{{Carilli}, {Perley}, {Dreher}  \& {Leahy}}{{Carilli} et~al.}{1991}]{1991ApJ...383..554C}
{Carilli} C.~L.,  {Perley} R.~A.,  {Dreher} J.~W.,   {Leahy} J.~P.,  1991, \mn@doi [\apj] {10.1086/170813}, \href {https://ui.adsabs.harvard.edu/abs/1991ApJ...383..554C} {383, 554}

\bibitem[\protect\citeauthoryear{{Ceraj} et~al.,}{{Ceraj} et~al.}{2018}]{2018A&A...620A.192C}
{Ceraj} L.,  et~al., 2018, \mn@doi [\aap] {10.1051/0004-6361/201833935}, \href {https://ui.adsabs.harvard.edu/abs/2018A&A...620A.192C} {620, A192}

\bibitem[\protect\citeauthoryear{{Chakraborty} et~al.,}{{Chakraborty} et~al.}{2019a}]{2019MNRAS.487.4102C}
{Chakraborty} A.,  et~al., 2019a, \mn@doi [\mnras] {10.1093/mnras/stz1580}, \href {https://ui.adsabs.harvard.edu/abs/2019MNRAS.487.4102C} {487, 4102}

\bibitem[\protect\citeauthoryear{{Chakraborty} et~al.,}{{Chakraborty} et~al.}{2019b}]{2019MNRAS.490..243C}
{Chakraborty} A.,  et~al., 2019b, \mn@doi [\mnras] {10.1093/mnras/stz2533}, \href {https://ui.adsabs.harvard.edu/abs/2019MNRAS.490..243C} {490, 243}

\bibitem[\protect\citeauthoryear{{Choudhuri}, {Bharadwaj}, {Roy}, {Ghosh}  \& {Ali}}{{Choudhuri} et~al.}{2016a}]{2016MNRAS.459..151C}
{Choudhuri} S.,  {Bharadwaj} S.,  {Roy} N.,  {Ghosh} A.,   {Ali} S.~S.,  2016a, \mn@doi [\mnras] {10.1093/mnras/stw607}, \href {https://ui.adsabs.harvard.edu/abs/2016MNRAS.459..151C} {459, 151}

\bibitem[\protect\citeauthoryear{{Choudhuri}, {Bharadwaj}, {Chatterjee}, {Ali}, {Roy}  \& {Ghosh}}{{Choudhuri} et~al.}{2016b}]{2016MNRAS.463.4093C}
{Choudhuri} S.,  {Bharadwaj} S.,  {Chatterjee} S.,  {Ali} S.~S.,  {Roy} N.,   {Ghosh} A.,  2016b, \mn@doi [\mnras] {10.1093/mnras/stw2254}, \href {https://ui.adsabs.harvard.edu/abs/2016MNRAS.463.4093C} {463, 4093}

\bibitem[\protect\citeauthoryear{{Choudhuri}, {Bharadwaj}, {Ali}, {Roy}, {Intema}  \& {Ghosh}}{{Choudhuri} et~al.}{2017}]{2017MNRAS.470L..11C}
{Choudhuri} S.,  {Bharadwaj} S.,  {Ali} S.~S.,  {Roy} N.,  {Intema} H.~T.,   {Ghosh} A.,  2017, \mn@doi [\mnras] {10.1093/mnrasl/slx066}, \href {https://ui.adsabs.harvard.edu/abs/2017MNRAS.470L..11C} {470, L11}

\bibitem[\protect\citeauthoryear{{Cochrane} et~al.,}{{Cochrane} et~al.}{2023}]{2023MNRAS.523.6082C}
{Cochrane} R.~K.,  et~al., 2023, \mn@doi [\mnras] {10.1093/mnras/stad1602}, \href {https://ui.adsabs.harvard.edu/abs/2023MNRAS.523.6082C} {523, 6082}

\bibitem[\protect\citeauthoryear{{Condon}}{{Condon}}{1992}]{1992ARA&A..30..575C}
{Condon} J.~J.,  1992, \mn@doi [\araa] {10.1146/annurev.aa.30.090192.003043}, \href {https://ui.adsabs.harvard.edu/abs/1992ARA&A..30..575C} {30, 575}

\bibitem[\protect\citeauthoryear{{Condon} et~al.,}{{Condon} et~al.}{2012}]{2012ApJ...758...23C}
{Condon} J.~J.,  et~al., 2012, \mn@doi [\apj] {10.1088/0004-637X/758/1/23}, \href {https://ui.adsabs.harvard.edu/abs/2012ApJ...758...23C} {758, 23}

\bibitem[\protect\citeauthoryear{{Cooray} \& {Furlanetto}}{{Cooray} \& {Furlanetto}}{2004}]{2004ApJ...606L...5C}
{Cooray} A.,  {Furlanetto} S.~R.,  2004, \mn@doi [\apjl] {10.1086/421241}, \href {https://ui.adsabs.harvard.edu/abs/2004ApJ...606L...5C} {606, L5}

\bibitem[\protect\citeauthoryear{{Cotton}}{{Cotton}}{2008}]{2008PASP..120..439C}
{Cotton} W.~D.,  2008, \mn@doi [\pasp] {10.1086/586754}, \href {https://ui.adsabs.harvard.edu/abs/2008PASP..120..439C} {120, 439}

\bibitem[\protect\citeauthoryear{{Croston} et~al.,}{{Croston} et~al.}{2019}]{2019A&A...622A..10C}
{Croston} J.~H.,  et~al., 2019, \mn@doi [\aap] {10.1051/0004-6361/201834019}, \href {https://ui.adsabs.harvard.edu/abs/2019A&A...622A..10C} {622, A10}

\bibitem[\protect\citeauthoryear{{Datta}, {Bowman}  \& {Carilli}}{{Datta} et~al.}{2010}]{2010ApJ...724..526D}
{Datta} A.,  {Bowman} J.~D.,   {Carilli} C.~L.,  2010, \mn@doi [\apj] {10.1088/0004-637X/724/1/526}, \href {https://ui.adsabs.harvard.edu/abs/2010ApJ...724..526D} {724, 526}

\bibitem[\protect\citeauthoryear{{Delvecchio} et~al.,}{{Delvecchio} et~al.}{2017}]{2017A&A...602A...3D}
{Delvecchio} I.,  et~al., 2017, \mn@doi [\aap] {10.1051/0004-6361/201629367}, \href {https://ui.adsabs.harvard.edu/abs/2017A&A...602A...3D} {602, A3}

\bibitem[\protect\citeauthoryear{Dewdney, Braun  \& Turner}{Dewdney et~al.}{2017}]{8105425}
Dewdney P.~E.,  Braun R.,   Turner W.,  2017, in 2017 XXXIInd General Assembly and Scientific Symposium of the International Union of Radio Science (URSI GASS). pp~1--4, \mn@doi{10.23919/URSIGASS.2017.8105425}

\bibitem[\protect\citeauthoryear{{Di Bernardo}, {Evoli}, {Gaggero}, {Grasso}  \& {Maccione}}{{Di Bernardo} et~al.}{2013}]{2013JCAP...03..036D}
{Di Bernardo} G.,  {Evoli} C.,  {Gaggero} D.,  {Grasso} D.,   {Maccione} L.,  2013, \mn@doi [\jcap] {10.1088/1475-7516/2013/03/036}, \href {https://ui.adsabs.harvard.edu/abs/2013JCAP...03..036D} {2013, 036}

\bibitem[\protect\citeauthoryear{{Di Matteo}, {Perna}, {Abel}  \& {Rees}}{{Di Matteo} et~al.}{2002}]{2002ApJ...564..576D}
{Di Matteo} T.,  {Perna} R.,  {Abel} T.,   {Rees} M.~J.,  2002, \mn@doi [\apj] {10.1086/324293}, \href {https://ui.adsabs.harvard.edu/abs/2002ApJ...564..576D} {564, 576}

\bibitem[\protect\citeauthoryear{{Di Matteo}, {Ciardi}  \& {Miniati}}{{Di Matteo} et~al.}{2004}]{2004MNRAS.355.1053D}
{Di Matteo} T.,  {Ciardi} B.,   {Miniati} F.,  2004, \mn@doi [\mnras] {10.1111/j.1365-2966.2004.08443.x}, \href {https://ui.adsabs.harvard.edu/abs/2004MNRAS.355.1053D} {355, 1053}

\bibitem[\protect\citeauthoryear{{Donoso}, {Best}  \& {Kauffmann}}{{Donoso} et~al.}{2009}]{2009MNRAS.392..617D}
{Donoso} E.,  {Best} P.~N.,   {Kauffmann} G.,  2009, \mn@doi [\mnras] {10.1111/j.1365-2966.2008.14068.x}, \href {https://ui.adsabs.harvard.edu/abs/2009MNRAS.392..617D} {392, 617}

\bibitem[\protect\citeauthoryear{{Dutta} \& {Nandakumar}}{{Dutta} \& {Nandakumar}}{2019}]{2019RAA....19...60D}
{Dutta} P.,  {Nandakumar} M.,  2019, \mn@doi [Research in Astronomy and Astrophysics] {10.1088/1674-4527/19/4/60}, \href {https://ui.adsabs.harvard.edu/abs/2019RAA....19...60D} {19, 060}

\bibitem[\protect\citeauthoryear{{Eddington}}{{Eddington}}{1913}]{1913MNRAS..73..359E}
{Eddington} A.~S.,  1913, \mn@doi [\mnras] {10.1093/mnras/73.5.359}, \href {https://ui.adsabs.harvard.edu/abs/1913MNRAS..73..359E} {73, 359}

\bibitem[\protect\citeauthoryear{{Franzen} et~al.,}{{Franzen} et~al.}{2015}]{2015MNRAS.453.4020F}
{Franzen} T.~M.~O.,  et~al., 2015, \mn@doi [\mnras] {10.1093/mnras/stv1866}, \href {https://ui.adsabs.harvard.edu/abs/2015MNRAS.453.4020F} {453, 4020}

\bibitem[\protect\citeauthoryear{{Franzen} et~al.,}{{Franzen} et~al.}{2016}]{2016MNRAS.459.3314F}
{Franzen} T.~M.~O.,  et~al., 2016, \mn@doi [\mnras] {10.1093/mnras/stw823}, \href {https://ui.adsabs.harvard.edu/abs/2016MNRAS.459.3314F} {459, 3314}

\bibitem[\protect\citeauthoryear{{Franzen}, {Vernstrom}, {Jackson}, {Hurley-Walker}, {Ekers}, {Heald}, {Seymour}  \& {White}}{{Franzen} et~al.}{2019}]{2019PASA...36....4F}
{Franzen} T.~M.~O.,  {Vernstrom} T.,  {Jackson} C.~A.,  {Hurley-Walker} N.,  {Ekers} R.~D.,  {Heald} G.,  {Seymour} N.,   {White} S.~V.,  2019, \mn@doi [\pasa] {10.1017/pasa.2018.52}, \href {https://ui.adsabs.harvard.edu/abs/2019PASA...36....4F} {36, e004}

\bibitem[\protect\citeauthoryear{{Garn}, {Green}, {Riley}  \& {Alexander}}{{Garn} et~al.}{2008}]{2008MNRAS.383...75G}
{Garn} T.,  {Green} D.~A.,  {Riley} J.~M.,   {Alexander} P.,  2008, \mn@doi [\mnras] {10.1111/j.1365-2966.2007.12562.x}, \href {https://ui.adsabs.harvard.edu/abs/2008MNRAS.383...75G} {383, 75}

\bibitem[\protect\citeauthoryear{{Gehlot} et~al.,}{{Gehlot} et~al.}{2022}]{2022A&A...662A..97G}
{Gehlot} B.~K.,  et~al., 2022, \mn@doi [\aap] {10.1051/0004-6361/202142939}, \href {https://ui.adsabs.harvard.edu/abs/2022A&A...662A..97G} {662, A97}

\bibitem[\protect\citeauthoryear{{Ghosh}, {Prasad}, {Bharadwaj}, {Ali}  \& {Chengalur}}{{Ghosh} et~al.}{2012}]{2012MNRAS.426.3295G}
{Ghosh} A.,  {Prasad} J.,  {Bharadwaj} S.,  {Ali} S.~S.,   {Chengalur} J.~N.,  2012, \mn@doi [\mnras] {10.1111/j.1365-2966.2012.21889.x}, \href {https://ui.adsabs.harvard.edu/abs/2012MNRAS.426.3295G} {426, 3295}

\bibitem[\protect\citeauthoryear{{Ginzburg} \& {Syrovatskii}}{{Ginzburg} \& {Syrovatskii}}{1965}]{1965ARA&A...3..297G}
{Ginzburg} V.~L.,  {Syrovatskii} S.~I.,  1965, \mn@doi [\araa] {10.1146/annurev.aa.03.090165.001501}, \href {https://ui.adsabs.harvard.edu/abs/1965ARA&A...3..297G} {3, 297}

\bibitem[\protect\citeauthoryear{Greisen}{Greisen}{1998}]{greisen1998recent}
Greisen E.~W.,  1998, Astronomical Data Analysis Software and Systems VII, 145, 204

\bibitem[\protect\citeauthoryear{{Greisen}}{{Greisen}}{2003}]{2003ASSL..285..109G}
{Greisen} E.~W.,  2003, in {Heck} A.,  ed.,  Astrophysics and Space Science Library Vol. 285, Information Handling in Astronomy - Historical Vistas. p.~109, \mn@doi{10.1007/0-306-48080-8_7}

\bibitem[\protect\citeauthoryear{{Gupta} et~al.,}{{Gupta} et~al.}{2017}]{2017CSci..113..707G}
{Gupta} Y.,  et~al., 2017, \mn@doi [Current Science] {10.18520/cs/v113/i04/707-714}, \href {https://ui.adsabs.harvard.edu/abs/2017CSci..113..707G} {113, 707}

\bibitem[\protect\citeauthoryear{{G{\"u}rkan} et~al.,}{{G{\"u}rkan} et~al.}{2019}]{2019A&A...622A..11G}
{G{\"u}rkan} G.,  et~al., 2019, \mn@doi [\aap] {10.1051/0004-6361/201833892}, \href {https://ui.adsabs.harvard.edu/abs/2019A&A...622A..11G} {622, A11}

\bibitem[\protect\citeauthoryear{{Hale} et~al.,}{{Hale} et~al.}{2019}]{2019A&A...622A...4H}
{Hale} C.~L.,  et~al., 2019, \mn@doi [\aap] {10.1051/0004-6361/201833906}, \href {https://ui.adsabs.harvard.edu/abs/2019A&A...622A...4H} {622, A4}

\bibitem[\protect\citeauthoryear{{Hancock}, {Murphy}, {Gaensler}, {Hopkins}  \& {Curran}}{{Hancock} et~al.}{2012}]{2012ascl.soft12009H}
{Hancock} P.~J.,  {Murphy} T.,  {Gaensler} B.~M.,  {Hopkins} A.,   {Curran} J.~R.,  2012, {Aegean: Compact source finding in radio images}, Astrophysics Source Code Library, record ascl:1212.009 (\mn@eprint {ascl} {1212.009})

\bibitem[\protect\citeauthoryear{{Hancock}, {Trott}  \& {Hurley-Walker}}{{Hancock} et~al.}{2018}]{2018PASA...35...11H}
{Hancock} P.~J.,  {Trott} C.~M.,   {Hurley-Walker} N.,  2018, \mn@doi [\pasa] {10.1017/pasa.2018.3}, \href {https://ui.adsabs.harvard.edu/abs/2018PASA...35...11H} {35, e011}

\bibitem[\protect\citeauthoryear{{Harwood}, {Hardcastle}, {Croston}  \& {Goodger}}{{Harwood} et~al.}{2013}]{2013MNRAS.435.3353H}
{Harwood} J.~J.,  {Hardcastle} M.~J.,  {Croston} J.~H.,   {Goodger} J.~L.,  2013, \mn@doi [\mnras] {10.1093/mnras/stt1526}, \href {https://ui.adsabs.harvard.edu/abs/2013MNRAS.435.3353H} {435, 3353}

\bibitem[\protect\citeauthoryear{{Helfand}, {White}  \& {Becker}}{{Helfand} et~al.}{2015}]{2015ApJ...801...26H}
{Helfand} D.~J.,  {White} R.~L.,   {Becker} R.~H.,  2015, \mn@doi [\apj] {10.1088/0004-637X/801/1/26}, \href {https://ui.adsabs.harvard.edu/abs/2015ApJ...801...26H} {801, 26}

\bibitem[\protect\citeauthoryear{{Iacobelli} et~al.,}{{Iacobelli} et~al.}{2013}]{2013A&A...558A..72I}
{Iacobelli} M.,  et~al., 2013, \mn@doi [\aap] {10.1051/0004-6361/201322013}, \href {https://ui.adsabs.harvard.edu/abs/2013A&A...558A..72I} {558, A72}

\bibitem[\protect\citeauthoryear{{Intema}}{{Intema}}{2014a}]{2014ascl.soft08006I}
{Intema} H.~T.,  2014a, {SPAM: Source Peeling and Atmospheric Modeling}, Astrophysics Source Code Library, record ascl:1408.006 (\mn@eprint {ascl} {1408.006})

\bibitem[\protect\citeauthoryear{{Intema}}{{Intema}}{2014b}]{2014ASInC..13..469I}
{Intema} H.~T.,  2014b, in Astronomical Society of India Conference Series. p.~469 (\mn@eprint {arXiv} {1402.4889})

\bibitem[\protect\citeauthoryear{{Intema}, {van der Tol}, {Cotton}, {Cohen}, {van Bemmel}  \& {R{\"o}ttgering}}{{Intema} et~al.}{2009}]{2009A&A...501.1185I}
{Intema} H.~T.,  {van der Tol} S.,  {Cotton} W.~D.,  {Cohen} A.~S.,  {van Bemmel} I.~M.,   {R{\"o}ttgering} H.~J.~A.,  2009, \mn@doi [\aap] {10.1051/0004-6361/200811094}, \href {https://ui.adsabs.harvard.edu/abs/2009A&A...501.1185I} {501, 1185}

\bibitem[\protect\citeauthoryear{{Intema}, {van Weeren}, {R{\"o}ttgering}  \& {Lal}}{{Intema} et~al.}{2011}]{2011A&A...535A..38I}
{Intema} H.~T.,  {van Weeren} R.~J.,  {R{\"o}ttgering} H.~J.~A.,   {Lal} D.~V.,  2011, \mn@doi [\aap] {10.1051/0004-6361/201014253}, \href {https://ui.adsabs.harvard.edu/abs/2011A&A...535A..38I} {535, A38}

\bibitem[\protect\citeauthoryear{{Intema}, {Jagannathan}, {Mooley}  \& {Frail}}{{Intema} et~al.}{2017}]{2017A&A...598A..78I}
{Intema} H.~T.,  {Jagannathan} P.,  {Mooley} K.~P.,   {Frail} D.~A.,  2017, \mn@doi [\aap] {10.1051/0004-6361/201628536}, \href {https://ui.adsabs.harvard.edu/abs/2017A&A...598A..78I} {598, A78}

\bibitem[\protect\citeauthoryear{{Ishwara-Chandra}, {Taylor}, {Green}, {Stil}, {Vaccari}  \& {Ocran}}{{Ishwara-Chandra} et~al.}{2020}]{2020MNRAS.497.5383I}
{Ishwara-Chandra} C.~H.,  {Taylor} A.~R.,  {Green} D.~A.,  {Stil} J.~M.,  {Vaccari} M.,   {Ocran} E.~F.,  2020, \mn@doi [\mnras] {10.1093/mnras/staa2341}, \href {https://ui.adsabs.harvard.edu/abs/2020MNRAS.497.5383I} {497, 5383}

\bibitem[\protect\citeauthoryear{{Jaffe} \& {Perola}}{{Jaffe} \& {Perola}}{1973}]{1973A&A....26..423J}
{Jaffe} W.~J.,  {Perola} G.~C.,  1973, \aap, \href {https://ui.adsabs.harvard.edu/abs/1973A&A....26..423J} {26, 423}

\bibitem[\protect\citeauthoryear{{Jeli{\'c}} et~al.,}{{Jeli{\'c}} et~al.}{2008}]{2008MNRAS.389.1319J}
{Jeli{\'c}} V.,  et~al., 2008, \mn@doi [\mnras] {10.1111/j.1365-2966.2008.13634.x}, \href {https://ui.adsabs.harvard.edu/abs/2008MNRAS.389.1319J} {389, 1319}

\bibitem[\protect\citeauthoryear{{Jeli{\'c}}, {Zaroubi}, {Labropoulos}, {Bernardi}, {de Bruyn}  \& {Koopmans}}{{Jeli{\'c}} et~al.}{2010}]{2010MNRAS.409.1647J}
{Jeli{\'c}} V.,  {Zaroubi} S.,  {Labropoulos} P.,  {Bernardi} G.,  {de Bruyn} A.~G.,   {Koopmans} L. V.~E.,  2010, \mn@doi [\mnras] {10.1111/j.1365-2966.2010.17407.x}, \href {https://ui.adsabs.harvard.edu/abs/2010MNRAS.409.1647J} {409, 1647}

\bibitem[\protect\citeauthoryear{{Johnston} et~al.,}{{Johnston} et~al.}{2008}]{2008ExA....22..151J}
{Johnston} S.,  et~al., 2008, \mn@doi [Experimental Astronomy] {10.1007/s10686-008-9124-7}, \href {https://ui.adsabs.harvard.edu/abs/2008ExA....22..151J} {22, 151}

\bibitem[\protect\citeauthoryear{{Jonas} \& {MeerKAT Team}}{{Jonas} \& {MeerKAT Team}}{2016}]{2016mks..confE...1J}
{Jonas} J.,  {MeerKAT Team} 2016, in MeerKAT Science: On the Pathway to the SKA. p.~1, \mn@doi{10.22323/1.277.0001}

\bibitem[\protect\citeauthoryear{{Kapi{\'n}ska} et~al.,}{{Kapi{\'n}ska} et~al.}{2017}]{2017ApJ...838...68K}
{Kapi{\'n}ska} A.~D.,  et~al., 2017, \mn@doi [\apj] {10.3847/1538-4357/aa5f5d}, \href {https://ui.adsabs.harvard.edu/abs/2017ApJ...838...68K} {838, 68}

\bibitem[\protect\citeauthoryear{{Kardashev}}{{Kardashev}}{1962}]{1962AZh....39..393K}
{Kardashev} N.~S.,  1962, \azh, \href {https://ui.adsabs.harvard.edu/abs/1962AZh....39..393K} {39, 393}

\bibitem[\protect\citeauthoryear{Kettenis, van Langevelde, Reynolds  \& Cotton}{Kettenis et~al.}{2006}]{kettenis2006parseltongue}
Kettenis M.,  van Langevelde H.,  Reynolds C.,   Cotton B.,  2006, in Astronomical Data Analysis Software and Systems XV. p.~497

\bibitem[\protect\citeauthoryear{{Klein}, {Lisenfeld}  \& {Verley}}{{Klein} et~al.}{2018}]{2018A&A...611A..55K}
{Klein} U.,  {Lisenfeld} U.,   {Verley} S.,  2018, \mn@doi [\aap] {10.1051/0004-6361/201731673}, \href {https://ui.adsabs.harvard.edu/abs/2018A&A...611A..55K} {611, A55}

\bibitem[\protect\citeauthoryear{{Kondapally} et~al.,}{{Kondapally} et~al.}{2021}]{2021A&A...648A...3K}
{Kondapally} R.,  et~al., 2021, \mn@doi [\aap] {10.1051/0004-6361/202038813}, \href {https://ui.adsabs.harvard.edu/abs/2021A&A...648A...3K} {648, A3}

\bibitem[\protect\citeauthoryear{{Koopmans} et~al.,}{{Koopmans} et~al.}{2015}]{2015aska.confE...1K}
{Koopmans} L.,  et~al., 2015, in Advancing Astrophysics with the Square Kilometre Array (AASKA14). p.~1 (\mn@eprint {arXiv} {1505.07568}), \mn@doi{10.22323/1.215.0001}

\bibitem[\protect\citeauthoryear{{Mahony} et~al.,}{{Mahony} et~al.}{2016}]{2016MNRAS.463.2997M}
{Mahony} E.~K.,  et~al., 2016, \mn@doi [\mnras] {10.1093/mnras/stw2225}, \href {https://ui.adsabs.harvard.edu/abs/2016MNRAS.463.2997M} {463, 2997}

\bibitem[\protect\citeauthoryear{{Mandal} et~al.,}{{Mandal} et~al.}{2021}]{2021A&A...648A...5M}
{Mandal} S.,  et~al., 2021, \mn@doi [\aap] {10.1051/0004-6361/202039998}, \href {https://ui.adsabs.harvard.edu/abs/2021A&A...648A...5M} {648, A5}

\bibitem[\protect\citeauthoryear{{Massardi}, {Bonaldi}, {Bonavera}, {L{\'o}pez-Caniego}, {de Zotti}  \& {Ekers}}{{Massardi} et~al.}{2011}]{2011MNRAS.415.1597M}
{Massardi} M.,  {Bonaldi} A.,  {Bonavera} L.,  {L{\'o}pez-Caniego} M.,  {de Zotti} G.,   {Ekers} R.~D.,  2011, \mn@doi [\mnras] {10.1111/j.1365-2966.2011.18802.x}, \href {https://ui.adsabs.harvard.edu/abs/2011MNRAS.415.1597M} {415, 1597}

\bibitem[\protect\citeauthoryear{{Mauch} et~al.,}{{Mauch} et~al.}{2020}]{2020ApJ...888...61M}
{Mauch} T.,  et~al., 2020, \mn@doi [\apj] {10.3847/1538-4357/ab5d2d}, \href {https://ui.adsabs.harvard.edu/abs/2020ApJ...888...61M} {888, 61}

\bibitem[\protect\citeauthoryear{{Mazumder}, {Chakraborty}, {Datta}, {Choudhuri}, {Roy}, {Wadadekar}  \& {Ishwara-Chandra}}{{Mazumder} et~al.}{2020}]{2020MNRAS.495.4071M}
{Mazumder} A.,  {Chakraborty} A.,  {Datta} A.,  {Choudhuri} S.,  {Roy} N.,  {Wadadekar} Y.,   {Ishwara-Chandra} C.~H.,  2020, \mn@doi [\mnras] {10.1093/mnras/staa1317}, \href {https://ui.adsabs.harvard.edu/abs/2020MNRAS.495.4071M} {495, 4071}

\bibitem[\protect\citeauthoryear{{McKean} et~al.,}{{McKean} et~al.}{2016}]{2016MNRAS.463.3143M}
{McKean} J.~P.,  et~al., 2016, \mn@doi [\mnras] {10.1093/mnras/stw2105}, \href {https://ui.adsabs.harvard.edu/abs/2016MNRAS.463.3143M} {463, 3143}

\bibitem[\protect\citeauthoryear{{Mellema}, {Koopmans}, {Shukla}, {Datta}, {Mesinger}  \& {Majumdar}}{{Mellema} et~al.}{2015}]{2015aska.confE..10M}
{Mellema} G.,  {Koopmans} L.,  {Shukla} H.,  {Datta} K.~K.,  {Mesinger} A.,   {Majumdar} S.,  2015, in Advancing Astrophysics with the Square Kilometre Array (AASKA14). p.~10 (\mn@eprint {arXiv} {1501.04203}), \mn@doi{10.22323/1.215.0010}

\bibitem[\protect\citeauthoryear{{Mertens} et~al.,}{{Mertens} et~al.}{2020}]{2020MNRAS.493.1662M}
{Mertens} F.~G.,  et~al., 2020, \mn@doi [\mnras] {10.1093/mnras/staa327}, \href {https://ui.adsabs.harvard.edu/abs/2020MNRAS.493.1662M} {493, 1662}

\bibitem[\protect\citeauthoryear{{Mertsch} \& {Sarkar}}{{Mertsch} \& {Sarkar}}{2013}]{2013JCAP...06..041M}
{Mertsch} P.,  {Sarkar} S.,  2013, \mn@doi [\jcap] {10.1088/1475-7516/2013/06/041}, \href {https://ui.adsabs.harvard.edu/abs/2013JCAP...06..041M} {2013, 041}

\bibitem[\protect\citeauthoryear{{Messias} et~al.,}{{Messias} et~al.}{2021}]{2021MNRAS.508.5259M}
{Messias} H.~G.,  et~al., 2021, \mn@doi [\mnras] {10.1093/mnras/stab1462}, \href {https://ui.adsabs.harvard.edu/abs/2021MNRAS.508.5259M} {508, 5259}

\bibitem[\protect\citeauthoryear{{Mohan} \& {Rafferty}}{{Mohan} \& {Rafferty}}{2015}]{2015ascl.soft02007M}
{Mohan} N.,  {Rafferty} D.,  2015, {PyBDSF: Python Blob Detection and Source Finder}, Astrophysics Source Code Library, record ascl:1502.007 (\mn@eprint {ascl} {1502.007})

\bibitem[\protect\citeauthoryear{{Myers} \& {Spangler}}{{Myers} \& {Spangler}}{1985}]{1985ApJ...291...52M}
{Myers} S.~T.,  {Spangler} S.~R.,  1985, \mn@doi [\apj] {10.1086/163040}, \href {https://ui.adsabs.harvard.edu/abs/1985ApJ...291...52M} {291, 52}

\bibitem[\protect\citeauthoryear{{Ocran}, {Taylor}, {Vaccari}, {Ishwara-Chandra}  \& {Prandoni}}{{Ocran} et~al.}{2020a}]{2020MNRAS.491.1127O}
{Ocran} E.~F.,  {Taylor} A.~R.,  {Vaccari} M.,  {Ishwara-Chandra} C.~H.,   {Prandoni} I.,  2020a, \mn@doi [\mnras] {10.1093/mnras/stz2954}, \href {https://ui.adsabs.harvard.edu/abs/2020MNRAS.491.1127O} {491, 1127}

\bibitem[\protect\citeauthoryear{{Ocran}, {Taylor}, {Vaccari}, {Ishwara-Chandra}, {Prandoni}, {Prescott}  \& {Mancuso}}{{Ocran} et~al.}{2020b}]{2020MNRAS.491.5911O}
{Ocran} E.~F.,  {Taylor} A.~R.,  {Vaccari} M.,  {Ishwara-Chandra} C.~H.,  {Prandoni} I.,  {Prescott} M.,   {Mancuso} C.,  2020b, \mn@doi [\mnras] {10.1093/mnras/stz3401}, \href {https://ui.adsabs.harvard.edu/abs/2020MNRAS.491.5911O} {491, 5911}

\bibitem[\protect\citeauthoryear{Ocran, Taylor, Vaccari, Ishwara-Chandra, Prandoni, Prescott  \& Mancuso}{Ocran et~al.}{2021}]{ocran2021evolution}
Ocran E.,  Taylor A.,  Vaccari M.,  Ishwara-Chandra C.,  Prandoni I.,  Prescott M.,   Mancuso C.,  2021, Monthly Notices of the Royal Astronomical Society, 500, 4685

\bibitem[\protect\citeauthoryear{{Offringa} \& {Smirnov}}{{Offringa} \& {Smirnov}}{2017}]{2017MNRAS.471..301O}
{Offringa} A.~R.,  {Smirnov} O.,  2017, \mn@doi [\mnras] {10.1093/mnras/stx1547}, \href {https://ui.adsabs.harvard.edu/abs/2017MNRAS.471..301O} {471, 301}

\bibitem[\protect\citeauthoryear{{Offringa} et~al.,}{{Offringa} et~al.}{2014}]{2014MNRAS.444..606O}
{Offringa} A.~R.,  et~al., 2014, \mn@doi [\mnras] {10.1093/mnras/stu1368}, \href {https://ui.adsabs.harvard.edu/abs/2014MNRAS.444..606O} {444, 606}

\bibitem[\protect\citeauthoryear{{Offringa}, {Singal}, {Heston}, {Horiuchi}  \& {Lucero}}{{Offringa} et~al.}{2022}]{2022MNRAS.509..114O}
{Offringa} A.~R.,  {Singal} J.,  {Heston} S.,  {Horiuchi} S.,   {Lucero} D.~M.,  2022, \mn@doi [\mnras] {10.1093/mnras/stab2865}, \href {https://ui.adsabs.harvard.edu/abs/2022MNRAS.509..114O} {509, 114}

\bibitem[\protect\citeauthoryear{{Oliver} et~al.,}{{Oliver} et~al.}{2000}]{2000MNRAS.316..749O}
{Oliver} S.,  et~al., 2000, \mn@doi [\mnras] {10.1046/j.1365-8711.2000.03550.x}, \href {https://ui.adsabs.harvard.edu/abs/2000MNRAS.316..749O} {316, 749}

\bibitem[\protect\citeauthoryear{{Pacholczyk}}{{Pacholczyk}}{1970}]{1970ranp.book.....P}
{Pacholczyk} A.~G.,  1970, {Radio astrophysics. Nonthermal processes in galactic and extragalactic sources}

\bibitem[\protect\citeauthoryear{{Padovani}, {Mainieri}, {Tozzi}, {Kellermann}, {Fomalont}, {Miller}, {Rosati}  \& {Shaver}}{{Padovani} et~al.}{2009}]{2009ApJ...694..235P}
{Padovani} P.,  {Mainieri} V.,  {Tozzi} P.,  {Kellermann} K.~I.,  {Fomalont} E.~B.,  {Miller} N.,  {Rosati} P.,   {Shaver} P.,  2009, \mn@doi [\apj] {10.1088/0004-637X/694/1/235}, \href {https://ui.adsabs.harvard.edu/abs/2009ApJ...694..235P} {694, 235}

\bibitem[\protect\citeauthoryear{{Padovani}, {Bonzini}, {Kellermann}, {Miller}, {Mainieri}  \& {Tozzi}}{{Padovani} et~al.}{2015}]{2015MNRAS.452.1263P}
{Padovani} P.,  {Bonzini} M.,  {Kellermann} K.~I.,  {Miller} N.,  {Mainieri} V.,   {Tozzi} P.,  2015, \mn@doi [\mnras] {10.1093/mnras/stv1375}, \href {https://ui.adsabs.harvard.edu/abs/2015MNRAS.452.1263P} {452, 1263}

\bibitem[\protect\citeauthoryear{{Perley}, {Chandler}, {Butler}  \& {Wrobel}}{{Perley} et~al.}{2011}]{2011ApJ...739L...1P}
{Perley} R.~A.,  {Chandler} C.~J.,  {Butler} B.~J.,   {Wrobel} J.~M.,  2011, \mn@doi [\apjl] {10.1088/2041-8205/739/1/L1}, \href {https://ui.adsabs.harvard.edu/abs/2011ApJ...739L...1P} {739, L1}

\bibitem[\protect\citeauthoryear{{Platania}, {Bensadoun}, {Bersanelli}, {De Amici}, {Kogut}, {Levin}, {Maino}  \& {Smoot}}{{Platania} et~al.}{1998}]{1998ApJ...505..473P}
{Platania} P.,  {Bensadoun} M.,  {Bersanelli} M.,  {De Amici} G.,  {Kogut} A.,  {Levin} S.,  {Maino} D.,   {Smoot} G.~F.,  1998, \mn@doi [\apj] {10.1086/306175}, \href {https://ui.adsabs.harvard.edu/abs/1998ApJ...505..473P} {505, 473}

\bibitem[\protect\citeauthoryear{{Pohl} \& {Esposito}}{{Pohl} \& {Esposito}}{1998}]{1998ApJ...507..327P}
{Pohl} M.,  {Esposito} J.~A.,  1998, \mn@doi [\apj] {10.1086/306298}, \href {https://ui.adsabs.harvard.edu/abs/1998ApJ...507..327P} {507, 327}

\bibitem[\protect\citeauthoryear{{Prandoni}, {Gregorini}, {Parma}, {de Ruiter}, {Vettolani}, {Wieringa}  \& {Ekers}}{{Prandoni} et~al.}{2000}]{2000A&AS..146...41P}
{Prandoni} I.,  {Gregorini} L.,  {Parma} P.,  {de Ruiter} H.~R.,  {Vettolani} G.,  {Wieringa} M.~H.,   {Ekers} R.~D.,  2000, \mn@doi [\aaps] {10.1051/aas:2000361}, \href {https://ui.adsabs.harvard.edu/abs/2000A&AS..146...41P} {146, 41}

\bibitem[\protect\citeauthoryear{{Prandoni}, {Parma}, {Wieringa}, {de Ruiter}, {Gregorini}, {Mignano}, {Vettolani}  \& {Ekers}}{{Prandoni} et~al.}{2006}]{2006A&A...457..517P}
{Prandoni} I.,  {Parma} P.,  {Wieringa} M.~H.,  {de Ruiter} H.~R.,  {Gregorini} L.,  {Mignano} A.,  {Vettolani} G.,   {Ekers} R.~D.,  2006, \mn@doi [\aap] {10.1051/0004-6361:20054273}, \href {https://ui.adsabs.harvard.edu/abs/2006A&A...457..517P} {457, 517}

\bibitem[\protect\citeauthoryear{{Reddy} et~al.,}{{Reddy} et~al.}{2017}]{2017JAI.....641011R}
{Reddy} S.~H.,  et~al., 2017, \mn@doi [Journal of Astronomical Instrumentation] {10.1142/S2251171716410117}, \href {https://ui.adsabs.harvard.edu/abs/2017JAI.....641011R} {6, 1641011}

\bibitem[\protect\citeauthoryear{{Remazeilles}, {Dickinson}, {Banday}, {Bigot-Sazy}  \& {Ghosh}}{{Remazeilles} et~al.}{2015}]{2015MNRAS.451.4311R}
{Remazeilles} M.,  {Dickinson} C.,  {Banday} A.~J.,  {Bigot-Sazy} M.~A.,   {Ghosh} T.,  2015, \mn@doi [\mnras] {10.1093/mnras/stv1274}, \href {https://ui.adsabs.harvard.edu/abs/2015MNRAS.451.4311R} {451, 4311}

\bibitem[\protect\citeauthoryear{{Robitaille}}{{Robitaille}}{2019}]{2019zndo...2567476R}
{Robitaille} T.,  2019, {APLpy v2.0: The Astronomical Plotting Library in Python}, Zenodo, \mn@doi{10.5281/zenodo.2567476}

\bibitem[\protect\citeauthoryear{{Robitaille} \& {Bressert}}{{Robitaille} \& {Bressert}}{2012}]{2012ascl.soft08017R}
{Robitaille} T.,  {Bressert} E.,  2012, {APLpy: Astronomical Plotting Library in Python}, Astrophysics Source Code Library, record ascl:1208.017 (\mn@eprint {ascl} {1208.017})

\bibitem[\protect\citeauthoryear{{Roy}, {Gupta}, {Pen}, {Peterson}, {Kudale}  \& {Kodilkar}}{{Roy} et~al.}{2010}]{2010ExA....28...25R}
{Roy} J.,  {Gupta} Y.,  {Pen} U.-L.,  {Peterson} J.~B.,  {Kudale} S.,   {Kodilkar} J.,  2010, \mn@doi [Experimental Astronomy] {10.1007/s10686-010-9187-0}, \href {https://ui.adsabs.harvard.edu/abs/2010ExA....28...25R} {28, 25}

\bibitem[\protect\citeauthoryear{{Sabater} et~al.,}{{Sabater} et~al.}{2021}]{2021A&A...648A...2S}
{Sabater} J.,  et~al., 2021, \mn@doi [\aap] {10.1051/0004-6361/202038828}, \href {https://ui.adsabs.harvard.edu/abs/2021A&A...648A...2S} {648, A2}

\bibitem[\protect\citeauthoryear{{Santos}, {Cooray}  \& {Knox}}{{Santos} et~al.}{2005}]{2005ApJ...625..575S}
{Santos} M.~G.,  {Cooray} A.,   {Knox} L.,  2005, \mn@doi [\apj] {10.1086/429857}, \href {https://ui.adsabs.harvard.edu/abs/2005ApJ...625..575S} {625, 575}

\bibitem[\protect\citeauthoryear{{Scaife} \& {Heald}}{{Scaife} \& {Heald}}{2012}]{2012MNRAS.423L..30S}
{Scaife} A. M.~M.,  {Heald} G.~H.,  2012, \mn@doi [\mnras] {10.1111/j.1745-3933.2012.01251.x}, \href {https://ui.adsabs.harvard.edu/abs/2012MNRAS.423L..30S} {423, L30}

\bibitem[\protect\citeauthoryear{{Shao}, {Wagg}, {Wang}, {Carilli}, {Riechers}, {Intema}, {Weiss}  \& {Menten}}{{Shao} et~al.}{2020}]{2020A&A...641A..85S}
{Shao} Y.,  {Wagg} J.,  {Wang} R.,  {Carilli} C.~L.,  {Riechers} D.~A.,  {Intema} H.~T.,  {Weiss} A.,   {Menten} K.~M.,  2020, \mn@doi [\aap] {10.1051/0004-6361/202038469}, \href {https://ui.adsabs.harvard.edu/abs/2020A&A...641A..85S} {641, A85}

\bibitem[\protect\citeauthoryear{{Shaver}, {Windhorst}, {Madau}  \& {de Bruyn}}{{Shaver} et~al.}{1999}]{1999A&A...345..380S}
{Shaver} P.~A.,  {Windhorst} R.~A.,  {Madau} P.,   {de Bruyn} A.~G.,  1999, \mn@doi [\aap] {10.48550/arXiv.astro-ph/9901320}, \href {https://ui.adsabs.harvard.edu/abs/1999A&A...345..380S} {345, 380}

\bibitem[\protect\citeauthoryear{{Shimwell} et~al.,}{{Shimwell} et~al.}{2025}]{2025A&A...695A..80S}
{Shimwell} T.~W.,  et~al., 2025, \mn@doi [\aap] {10.1051/0004-6361/202452930}, \href {https://ui.adsabs.harvard.edu/abs/2025A&A...695A..80S} {695, A80}

\bibitem[\protect\citeauthoryear{{Sinha} \& {Datta}}{{Sinha} \& {Datta}}{2023}]{2023MNRAS.525.5311S}
{Sinha} A.,  {Datta} A.,  2023, \mn@doi [\mnras] {10.1093/mnras/stad2544}, \href {https://ui.adsabs.harvard.edu/abs/2023MNRAS.525.5311S} {525, 5311}

\bibitem[\protect\citeauthoryear{{Sinha}, {Basu}, {Datta}  \& {Chakraborty}}{{Sinha} et~al.}{2022}]{2022MNRAS.514.4343S}
{Sinha} A.,  {Basu} A.,  {Datta} A.,   {Chakraborty} A.,  2022, \mn@doi [\mnras] {10.1093/mnras/stac1504}, \href {https://ui.adsabs.harvard.edu/abs/2022MNRAS.514.4343S} {514, 4343}

\bibitem[\protect\citeauthoryear{{Sinha}, {Mangla}  \& {Datta}}{{Sinha} et~al.}{2023}]{2023JApA...44...88S}
{Sinha} A.,  {Mangla} S.,   {Datta} A.,  2023, \mn@doi [Journal of Astrophysics and Astronomy] {10.1007/s12036-023-09978-0}, \href {https://ui.adsabs.harvard.edu/abs/2023JApA...44...88S} {44, 88}

\bibitem[\protect\citeauthoryear{{Sirothia}, {Dennefeld}, {Saikia}, {Dole}, {Ricquebourg}  \& {Roland}}{{Sirothia} et~al.}{2009}]{2009MNRAS.395..269S}
{Sirothia} S.~K.,  {Dennefeld} M.,  {Saikia} D.~J.,  {Dole} H.,  {Ricquebourg} F.,   {Roland} J.,  2009, \mn@doi [\mnras] {10.1111/j.1365-2966.2009.14317.x}, \href {https://ui.adsabs.harvard.edu/abs/2009MNRAS.395..269S} {395, 269}

\bibitem[\protect\citeauthoryear{{Smol{\v{c}}i{\'c}} et~al.,}{{Smol{\v{c}}i{\'c}} et~al.}{2017}]{2017A&A...602A...1S}
{Smol{\v{c}}i{\'c}} V.,  et~al., 2017, \mn@doi [\aap] {10.1051/0004-6361/201628704}, \href {https://ui.adsabs.harvard.edu/abs/2017A&A...602A...1S} {602, A1}

\bibitem[\protect\citeauthoryear{{Sohn}, {Klein}  \& {Mack}}{{Sohn} et~al.}{2003}]{2003A&A...404..133S}
{Sohn} B.~W.,  {Klein} U.,   {Mack} K.~H.,  2003, \mn@doi [\aap] {10.1051/0004-6361:20030435}, \href {https://ui.adsabs.harvard.edu/abs/2003A&A...404..133S} {404, 133}

\bibitem[\protect\citeauthoryear{{Strong}, {Orlando}  \& {Jaffe}}{{Strong} et~al.}{2011}]{2011A&A...534A..54S}
{Strong} A.~W.,  {Orlando} E.,   {Jaffe} T.~R.,  2011, \mn@doi [\aap] {10.1051/0004-6361/201116828}, \href {https://ui.adsabs.harvard.edu/abs/2011A&A...534A..54S} {534, A54}

\bibitem[\protect\citeauthoryear{{Swarup}, {Ananthakrishnan}, {Kapahi}, {Rao}, {Subrahmanya}  \& {Kulkarni}}{{Swarup} et~al.}{1991}]{1991CSci...60...95S}
{Swarup} G.,  {Ananthakrishnan} S.,  {Kapahi} V.~K.,  {Rao} A.~P.,  {Subrahmanya} C.~R.,   {Kulkarni} V.~K.,  1991, Current Science, \href {https://ui.adsabs.harvard.edu/abs/1991CSci...60...95S} {60, 95}

\bibitem[\protect\citeauthoryear{{Takeuchi}, {Yoshida}, {Cortese}, {Wong}, {Catinella}  \& {Cooray}}{{Takeuchi} et~al.}{2022}]{2022arXiv220400831T}
{Takeuchi} T.~T.,  {Yoshida} S.~A.,  {Cortese} L.,  {Wong} O.~I.,  {Catinella} B.,   {Cooray} S.,  2022, \mn@doi [arXiv e-prints] {10.48550/arXiv.2204.00831}, \href {https://ui.adsabs.harvard.edu/abs/2022arXiv220400831T} {p. arXiv:2204.00831}

\bibitem[\protect\citeauthoryear{{Taylor} \& {Jagannathan}}{{Taylor} \& {Jagannathan}}{2016}]{2016MNRAS.459L..36T}
{Taylor} A.~R.,  {Jagannathan} P.,  2016, \mn@doi [\mnras] {10.1093/mnrasl/slw038}, \href {https://ui.adsabs.harvard.edu/abs/2016MNRAS.459L..36T} {459, L36}

\bibitem[\protect\citeauthoryear{{Taylor} et~al.,}{{Taylor} et~al.}{2024}]{2024MNRAS.528.2511T}
{Taylor} A.~R.,  et~al., 2024, \mn@doi [\mnras] {10.1093/mnras/stae169}, \href {https://ui.adsabs.harvard.edu/abs/2024MNRAS.528.2511T} {528, 2511}

\bibitem[\protect\citeauthoryear{{Tingay} et~al.,}{{Tingay} et~al.}{2013}]{2013PASA...30....7T}
{Tingay} S.~J.,  et~al., 2013, \mn@doi [\pasa] {10.1017/pasa.2012.007}, \href {https://ui.adsabs.harvard.edu/abs/2013PASA...30....7T} {30, e007}

\bibitem[\protect\citeauthoryear{{Tribble}}{{Tribble}}{1993}]{1993MNRAS.261...57T}
{Tribble} P.~C.,  1993, \mn@doi [\mnras] {10.1093/mnras/261.1.57}, \href {https://ui.adsabs.harvard.edu/abs/1993MNRAS.261...57T} {261, 57}

\bibitem[\protect\citeauthoryear{{Vernstrom}, {Scott}  \& {Wall}}{{Vernstrom} et~al.}{2011}]{2011MNRAS.415.3641V}
{Vernstrom} T.,  {Scott} D.,   {Wall} J.~V.,  2011, \mn@doi [\mnras] {10.1111/j.1365-2966.2011.18990.x}, \href {https://ui.adsabs.harvard.edu/abs/2011MNRAS.415.3641V} {415, 3641}

\bibitem[\protect\citeauthoryear{{Vernstrom}, {Heald}, {Vazza}, {Galvin}, {West}, {Locatelli}, {Fornengo}  \& {Pinetti}}{{Vernstrom} et~al.}{2021}]{2021MNRAS.505.4178V}
{Vernstrom} T.,  {Heald} G.,  {Vazza} F.,  {Galvin} T.~J.,  {West} J.~L.,  {Locatelli} N.,  {Fornengo} N.,   {Pinetti} E.,  2021, \mn@doi [\mnras] {10.1093/mnras/stab1301}, \href {https://ui.adsabs.harvard.edu/abs/2021MNRAS.505.4178V} {505, 4178}

\bibitem[\protect\citeauthoryear{{White}, {Becker}, {Helfand}  \& {Gregg}}{{White} et~al.}{1997}]{1997ApJ...475..479W}
{White} R.~L.,  {Becker} R.~H.,  {Helfand} D.~J.,   {Gregg} M.~D.,  1997, \mn@doi [\apj] {10.1086/303564}, \href {https://ui.adsabs.harvard.edu/abs/1997ApJ...475..479W} {475, 479}

\bibitem[\protect\citeauthoryear{{White}, {Jarvis}, {H{\"a}u{\ss}ler}  \& {Maddox}}{{White} et~al.}{2015}]{2015MNRAS.448.2665W}
{White} S.~V.,  {Jarvis} M.~J.,  {H{\"a}u{\ss}ler} B.,   {Maddox} N.,  2015, \mn@doi [\mnras] {10.1093/mnras/stv134}, \href {https://ui.adsabs.harvard.edu/abs/2015MNRAS.448.2665W} {448, 2665}

\bibitem[\protect\citeauthoryear{{Whittam}, {Green}, {Jarvis}  \& {Riley}}{{Whittam} et~al.}{2017}]{2017MNRAS.464.3357W}
{Whittam} I.~H.,  {Green} D.~A.,  {Jarvis} M.~J.,   {Riley} J.~M.,  2017, \mn@doi [\mnras] {10.1093/mnras/stw2638}, \href {https://ui.adsabs.harvard.edu/abs/2017MNRAS.464.3357W} {464, 3357}

\bibitem[\protect\citeauthoryear{{Whittam} et~al.,}{{Whittam} et~al.}{2022}]{2022MNRAS.516..245W}
{Whittam} I.~H.,  et~al., 2022, \mn@doi [\mnras] {10.1093/mnras/stac2140}, \href {https://ui.adsabs.harvard.edu/abs/2022MNRAS.516..245W} {516, 245}

\bibitem[\protect\citeauthoryear{{Williams}, {Intema}  \& {R{\"o}ttgering}}{{Williams} et~al.}{2013}]{2013A&A...549A..55W}
{Williams} W.~L.,  {Intema} H.~T.,   {R{\"o}ttgering} H.~J.~A.,  2013, \mn@doi [\aap] {10.1051/0004-6361/201220235}, \href {https://ui.adsabs.harvard.edu/abs/2013A&A...549A..55W} {549, A55}

\bibitem[\protect\citeauthoryear{{Williams} et~al.,}{{Williams} et~al.}{2016}]{2016MNRAS.460.2385W}
{Williams} W.~L.,  et~al., 2016, \mn@doi [\mnras] {10.1093/mnras/stw1056}, \href {https://ui.adsabs.harvard.edu/abs/2016MNRAS.460.2385W} {460, 2385}

\bibitem[\protect\citeauthoryear{{Wilman} et~al.,}{{Wilman} et~al.}{2008}]{2008MNRAS.388.1335W}
{Wilman} R.~J.,  et~al., 2008, \mn@doi [\mnras] {10.1111/j.1365-2966.2008.13486.x}, \href {https://ui.adsabs.harvard.edu/abs/2008MNRAS.388.1335W} {388, 1335}

\bibitem[\protect\citeauthoryear{{Wilson} et~al.,}{{Wilson} et~al.}{2011}]{2011MNRAS.416..832W}
{Wilson} W.~E.,  et~al., 2011, \mn@doi [\mnras] {10.1111/j.1365-2966.2011.19054.x}, \href {https://ui.adsabs.harvard.edu/abs/2011MNRAS.416..832W} {416, 832}

\bibitem[\protect\citeauthoryear{{Windhorst}, {Miley}, {Owen}, {Kron}  \& {Koo}}{{Windhorst} et~al.}{1985}]{1985ApJ...289..494W}
{Windhorst} R.~A.,  {Miley} G.~K.,  {Owen} F.~N.,  {Kron} R.~G.,   {Koo} D.~C.,  1985, \mn@doi [\apj] {10.1086/162911}, \href {https://ui.adsabs.harvard.edu/abs/1985ApJ...289..494W} {289, 494}

\bibitem[\protect\citeauthoryear{{Wootten} \& {Thompson}}{{Wootten} \& {Thompson}}{2009}]{2009IEEEP..97.1463W}
{Wootten} A.,  {Thompson} A.~R.,  2009, \mn@doi [IEEE Proceedings] {10.1109/JPROC.2009.2020572}, \href {https://ui.adsabs.harvard.edu/abs/2009IEEEP..97.1463W} {97, 1463}

\bibitem[\protect\citeauthoryear{{de Oliveira-Costa}, {Tegmark}, {Gaensler}, {Jonas}, {Landecker}  \& {Reich}}{{de Oliveira-Costa} et~al.}{2008}]{2008MNRAS.388..247D}
{de Oliveira-Costa} A.,  {Tegmark} M.,  {Gaensler} B.~M.,  {Jonas} J.,  {Landecker} T.~L.,   {Reich} P.,  2008, \mn@doi [\mnras] {10.1111/j.1365-2966.2008.13376.x}, \href {https://ui.adsabs.harvard.edu/abs/2008MNRAS.388..247D} {388, 247}

\bibitem[\protect\citeauthoryear{{van Haarlem} et~al.,}{{van Haarlem} et~al.}{2013}]{2013A&A...556A...2V}
{van Haarlem} M.~P.,  et~al., 2013, \mn@doi [\aap] {10.1051/0004-6361/201220873}, \href {https://ui.adsabs.harvard.edu/abs/2013A&A...556A...2V} {556, A2}

\bibitem[\protect\citeauthoryear{{van der Vlugt}, {Hodge}, {Algera}, {Smail}, {Leslie}, {Radcliffe}, {Riechers}  \& {R{\"o}ttgering}}{{van der Vlugt} et~al.}{2022}]{2022ApJ...941...10V}
{van der Vlugt} D.,  {Hodge} J.~A.,  {Algera} H.~S.~B.,  {Smail} I.,  {Leslie} S.~K.,  {Radcliffe} J.~F.,  {Riechers} D.~A.,   {R{\"o}ttgering} H.,  2022, \mn@doi [\apj] {10.3847/1538-4357/ac99db}, \href {https://ui.adsabs.harvard.edu/abs/2022ApJ...941...10V} {941, 10}

\makeatother
\end{thebibliography}

% Alternatively you could enter them by hand, like this:
% This method is tedious and prone to error if you have lots of references
%\begin{thebibliography}{99}
%\bibitem[\protect\citeauthoryear{Author}{2012}]{Author2012}
%Author A.~N., 2013, Journal of Improbable Astronomy, 1, 1
%\bibitem[\protect\citeauthoryear{Others}{2013}]{Others2013}
%Others S., 2012, Journal of Interesting Stuff, 17, 198
%\end{thebibliography}

%%%%%%%%%%%%%%%%%%%%%%%%%%%%%%%%%%%%%%%%%%%%%%%%%%

%%%%%%%%%%%%%%%%% APPENDICES %%%%%%%%%%%%%%%%%%%%%

% \appendix

% \section{Some extra material}

% If you want to present additional material which would interrupt the flow of the main paper,
% it can be placed in an Appendix which appears after the list of references.

%%%%%%%%%%%%%%%%%%%%%%%%%%%%%%%%%%%%%%%%%%%%%%%%%%

% Don't change these lines
\bsp	% typesetting comment
\label{lastpage}
\end{document}